\documentclass[12pt,a4paper]{article}

\usepackage{amsfonts} 
\usepackage{mathtools} 
\usepackage{amssymb} 
\usepackage{subfigure} 
\usepackage{graphicx}
\usepackage{xcolor}
\usepackage{cite} 
\usepackage[normalem]{ulem}
\usepackage{mathrsfs}

\graphicspath{{Figs/}}

\newcommand{\abs}[1]{|#1|} 
\newcommand{\ABS}[1]{\left|#1\right|} 
\newcommand{\re}[1]{\text{Re}(#1)} 
\newcommand{\im}[1]{\text{Im}(#1)} 

\newcommand{\der}{\mathcal D}

\newcommand{\refEQ}[1]{Eq.\,\eqref{#1}} 
\newcommand{\refEQS}[1]{Eqs.\,\eqref{#1}} 
\newcommand{\Hd}[1]{\Phi_{#1}}

\newcommand{\Hdt}[1]{\tilde\Phi_{#1}}
\newcommand{\nHH}{\mathrm{H}^0}
\newcommand{\nHR}{\mathrm{R}^0}
\newcommand{\nHI}{\mathrm{I}^0}
\newcommand{\nh}{\mathrm{h}}
\newcommand{\nH}{\mathrm{H}}
\newcommand{\nA}{\mathrm{A}}
\newcommand{\cH}{\mathrm{H}^\pm}
\newcommand{\cHm}{\mathrm{H}^-}
\newcommand{\cHp}{\mathrm{H}^+}

\newcommand{\cb}{c_\beta}
\renewcommand{\sb}{s_\beta}
\newcommand{\ca}{c_\alpha}
\newcommand{\sa}{s_\alpha}
\newcommand{\cab}{c_{\beta\alpha}}
\newcommand{\sab}{s_{\beta\alpha}}

\newcommand{\ROT}[2]{\mathcal R_{\mathrm{[#1]}}(#2)}

\newcommand{\RS}[1]{\mathcal R_{#1}}
\newcommand{\Rb}{\mathcal R_{\beta}}
\newcommand{\CKM}{V}
\newcommand{\CKMd}{V^\dagger}
\newcommand{\V}[1]{{\CKM_{#1}^{\phantom{\ast}}}}
\newcommand{\Vc}[1]{{\CKM_{#1}^\ast}}

\newcommand{\UqX}[2]{U_{{\rm #1_{#2}}}^{\phantom{\dagger}}}
\newcommand{\UqXd}[2]{U_{{\rm #1_{#2}}}^{\dagger}}

\newcommand{\UdR}{\UqX{d}{R}}
\newcommand{\UdRd}{\UqXd{d}{R}}
\newcommand{\UdL}{\UqX{d}{L}}
\newcommand{\UdLd}{\UqXd{d}{L}}
\newcommand{\UuR}{\UqX{u}{R}}
\newcommand{\UuRd}{\UqXd{u}{R}}
\newcommand{\UuL}{\UqX{u}{L}}
\newcommand{\UuLd}{\UqXd{u}{L}}
\newcommand{\UlR}{\UqX{\ell}{R}}
\newcommand{\UlRd}{\UqXd{\ell}{R}}

\newcommand{\UlLd}{\UqXd{\ell}{L}}

\newcommand{\tr}[1]{\text{tr}\left(#1\right)}
\newcommand{\id}{\mathbf{1}}

\newcommand{\Yubase}{\Delta}
\newcommand{\Yu}[1]{\Yubase^{\phantom{\dagger}}_{#1}}
\newcommand{\Yud}[1]{\Yubase^{\dagger}_{#1}}
\newcommand{\YuH}[1]{{\large\Yubase}_{#1}}
\newcommand{\Ydbase}{\Gamma}
\newcommand{\Yd}[1]{\Ydbase^{\phantom{\dagger}}_{#1}}
\newcommand{\Ydd}[1]{\Ydbase^{\dagger}_{#1}}
\newcommand{\YdH}[1]{{\Large \Ydbase}_{#1}}
\newcommand{\Ylbase}{\Pi}
\newcommand{\Yl}[1]{\Ylbase^{\phantom{\dagger}}_{#1}}
\newcommand{\Yld}[1]{\Ylbase^{\dagger}_{#1}}
\newcommand{\YlH}[1]{{\Large \Ylbase}_{#1}}
\newcommand{\DYf}[2]{f^{\rm [#1]}_{#2}}
\newcommand{\DYg}[2]{g^{\rm [#1]}_{#2}}
\newcommand{\basewM}{\mathcal M^0}
\newcommand{\basewN}{\mathcal N^0}
\newcommand{\wMq}[1]{\basewM_{\mathrm{#1}}}\newcommand{\wNq}[1]{\basewN_{\mathrm{#1}}}

\newcommand{\wMd}{\wMq{d}}\newcommand{\wNd}{\wNq{d}}
\newcommand{\wMu}{\wMq{u}}\newcommand{\wNu}{\wNq{u}}
\newcommand{\wMl}{\wMq{\ell}}\newcommand{\wNl}{\wNq{\ell}}
\newcommand{\basemM}{\mathcal M}
\newcommand{\basemN}{\mathcal N}\newcommand{\basemNd}{\mathcal N^{\dagger}}
\newcommand{\mMq}[1]{\basemM_{\mathrm{#1}}}\newcommand{\mNq}[1]{\basemN_{\mathrm{#1}}}
\newcommand{\mNqd}[1]{\basemNd_{\mathrm{#1}}}
\newcommand{\mMd}{\mMq{d}}\newcommand{\mNd}{\mNq{d}}\newcommand{\mNdd}{\mNqd{d}}
\newcommand{\mMu}{\mMq{u}}\newcommand{\mNu}{\mNq{u}}\newcommand{\mNud}{\mNqd{u}}
\newcommand{\mMl}{\mMq{\ell}}\newcommand{\mNl}{\mNq{\ell}}\newcommand{\mNld}{\mNqd{\ell}}
\newcommand{\basewYD}{Y^{0\phantom{\dagger}}}\newcommand{\basewYDd}{Y^{0\dagger}}
\newcommand{\wYDq}[2]{\basewYD_{{\rm [#1]}#2}}\newcommand{\wYDqd}[2]{\basewYDd_{{\rm [#1]}#2}}
\newcommand{\wYDd}[1]{\wYDq{d}{#1}}\newcommand{\wYDdd}[1]{\wYDqd{d}{#1}}
\newcommand{\wYDu}[1]{\wYDq{u}{#1}}\newcommand{\wYDud}[1]{\wYDqd{u}{#1}}
\newcommand{\basemYD}{Y^{\phantom{\dagger}}}\newcommand{\basemYDd}{Y^{\dagger}}
\newcommand{\mYDq}[2]{\basemYD_{{\rm [#1]}#2}}\newcommand{\mYDqd}[2]{\basemYDd_{{\rm [#1]}#2}}
\newcommand{\mYDd}[1]{\mYDq{d}{#1}}\newcommand{\mYDdd}[1]{\mYDqd{d}{#1}}
\newcommand{\mYDu}[1]{\mYDq{u}{#1}}\newcommand{\mYDud}[1]{\mYDqd{u}{#1}}
\newcommand{\baseyD}[2]{y^{{\rm #1}\phantom{\ast}}_{#2}}\newcommand{\baseyDc}[2]{y^{{\rm #1}\ast}_{#2}}
\newcommand{\myDq}[3]{\baseyD{#1}{{ #2},#3}}\newcommand{\myDqc}[3]{\baseyDc{#1}{{ #2},#3}}
\newcommand{\myDd}[2]{\myDq{d}{#1}{#2}}\newcommand{\myDdc}[2]{\myDqc{d}{#1}{#2}}
\newcommand{\myDu}[2]{\myDq{u}{#1}{#2}}\newcommand{\myDuc}[2]{\myDqc{u}{#1}{#2}}
\newcommand{\baseYR}{{\rm W}}\newcommand{\baseYRc}{{\rm W}^{\ast}}\newcommand{\baseYRd}{{\rm W}^\dagger}
\newcommand{\YR}[1]{\baseYR^{\phantom{\ast}}_{#1}}\newcommand{\YRc}[1]{\baseYRc_{#1}}
\newcommand{\baseHNq}[1]{\mathcal H_{\mathrm{#1}}}\newcommand{\baseAHNq}[1]{\mathcal A_{\mathrm{#1}}}
\newcommand{\HNd}{\baseHNq{d}}\newcommand{\AHNd}{\baseAHNq{d}}
\newcommand{\HNu}{\baseHNq{u}}\newcommand{\AHNu}{\baseAHNq{u}}
\newcommand{\HNl}{\baseHNq{\ell}}\newcommand{\AHNl}{\baseAHNq{\ell}}
\newcommand{\nq}[2]{n_{#2}}\newcommand{\nqc}[2]{\nq{#1}{#2}^\ast}
\newcommand{\ND}[1]{\nq{d}{#1}}
\newcommand{\NU}[1]{\nq{u}{#1}}
\newcommand{\NL}[1]{\nq{\ell}{#1}}

\newcommand{\COMM}[2]{\left[#1\,,\,#2\right]}
\newcommand{\comm}[2]{[#1,#2]}

\begin{document}

\begin{titlepage}

\hfill\begin{minipage}[r]{0.3\textwidth}\begin{flushright}  IFIC/18-004\\    CFTP/18-004 \end{flushright} \end{minipage}

\begin{center}

\vspace{1.0cm}

{\large \bf {Flavour Conservation in Two Higgs Doublet Models}}

\vspace{1.0cm}

Francisco J. Botella  $^{a,}$\footnote{\texttt{Francisco.J.Botella@uv.es}}, 
Fernando Cornet-Gomez  $^{a,}$\footnote{\texttt{Fernando.Cornet@ific.uv.es}}, 
Miguel Nebot $^{b,}$\footnote{\texttt{miguel.r.nebot.gomez@tecnico.ulisboa.pt}}

\vspace{1.0cm}

\emph{$^a$ Departament de F\' \i sica Te\`orica and IFIC,\\
Universitat de Val\`encia-CSIC,\\ E-46100, Burjassot, Spain.} \\
\emph{$^b$ Centro de F\' \i sica Te\' orica de Part\' \i culas (CFTP),\\
Instituto Superior T\' ecnico (IST), U. de Lisboa (UL),\\ Av. Rovisco Pais, P-1049-001 Lisboa,
Portugal.}

\end{center}

\vspace{3cm}

\begin{abstract}
In extensions of the Standard Model with two Higgs doublets, flavour changing Yukawa couplings of the neutral scalars may be present at tree level. In this work we consider the most general scenario in which those flavour changing couplings are absent. We revise the conditions that the Yukawa coupling matrices must obey for such \emph{general flavour conservation} (gFC), and study the one loop renormalisation group evolution of such conditions in both the quark and lepton sectors. We show that gFC in the leptonic sector is one loop stable under the Renormalization Group Evolution (RGE) and in the quark sector we present some new Cabibbo like solution also one loop stable under RGE. At a phenomenological level, we obtain the regions for the different gFC parameters that are allowed by the existing experimental constraints related to the 125 GeV Higgs.
\end{abstract}

\end{titlepage}

%
%
\section{Introduction\label{SEC:INTRO}}
Two Higgs Doublet Models (2HDM) \cite{Lee:1973iz,Branco:2011iw,Ivanov:2017dad} are a simple and popular class of extensions of the Standard Model (SM). Besides the original motivation, in particular the possibility of having spontaneous CP violation \cite{Lee:1973iz}, extending the SM scalar sector with a second doublet allows a number of interesting phenomenological consequences. To name a few generic ones: the appearance of new fundamental scalar particles, non-standard properties of the ``quite Higgs-like'' scalar discovered at the LHC with a mass of 125 GeV \cite{Aad:2012tfa,Chatrchyan:2012xdj}, and, related to them, a number of potential deviations in low energy processes with respect to SM expectations. They have been the focus of intense scrutiny before and after the 2012 discovery  \cite{Haber:1978jt,Donoghue:1978cj,Abbott:1979dt,Hall:1981bc,Barger:1989fj,Atwood:1996vj,WahabElKaffas:2007xd,Aoki:2009ha,Mahmoudi:2009zx,Deschamps:2009rh,Crivellin:2013wna,Broggio:2014mna,Das:2015qva,Gaitan:2015hga,Altunkaynak:2015twa,Arhrib:2015maa,Kim:2015zla,Enomoto:2015wbn,Benbrik:2015evd,Cline:2015lqp,Han:2017pfo,Gori:2017qwg,Arbey:2017gmh}. Additional aspects, including dark matter candidates \cite{Deshpande:1977rw,LopezHonorez:2006gr} or sources of CP violation in addition to the Cabibbo-Kobayashi-Maskawa matrix \cite{Bona:2005vz,Charles:2004jd,Botella:2005fc}, of interest for baryogenesis \cite{Turok:1990zg,Guo:2016ixx,Fuyuto:2017ewj}, provide further interest in 2HDM.

In the SM, concentrating on quarks, a single Yukawa structure in each sector -- up and down -- is both responsible for: (i) the generation of mass upon spontaneous breaking of $SU(2)_{\rm L}\otimes U(1)_{\rm Y}$ into $U(1)_{\rm EM}$, and (ii) the couplings of the quarks to the only fundamental scalar leftover, the Higgs boson, after associating the three would-be Goldstone bosons to the longitudinal polarizations of the massive $Z$ and $W^\pm$ gauge bosons. As a consequence, there are no tree level Flavour Changing Neutral Couplings (FCNC) of the Higgs to quarks. With two independent Yukawa structures available in each sector, the situation is dramatically changed in the general 2HDM, and FCNC couplings of quarks do arise at tree level. To which extent they appear in the couplings of the different physical neutral scalars depends then on the details of the scalar potential \cite{Georgi:1978ri}: if the 125 GeV scalar is a mixture of the true-but-unphysical Higgs and the additional neutral scalars, FCNC ``leak'' into its couplings through that mixing. At the end of the day, as with many New Physics avenues, the presence of FCNC is a double edged feature: since the competing SM gauge mediated contributions to FCNC processes are loop induced, those transitions pose severe constraints while, on the same grounds, provide immediate opportunities to discover deviations from the SM picture.

The study of different ways to dispense without problematic too large FCNC couplings and the conditions for their appearance or absence, has drawn sustained attention over the years. As analysed in \cite{Glashow:1976nt,Paschos:1976ay}, the absence of FCNC is guaranteed by forcing each right-handed fermion type to couple to one and only one scalar doublet; this absence of FCNC, backed by a $\mathbb{Z}_2$ symmetry,  is a popular option, and several implementations of this \emph{Natural Flavour Conservation} (NFC) idea, namely 2HDM of types I, II, and of types X, Y (when the lepton sector is also considered) have been thoroughly explored. Additional $U(1)$ gauge symmetries have also been considered, for example, in \cite{Ko:2012hd,Campos:2017dgc}.
The general conditions for the absence of FCNC, that is, that the mass matrix and the remaining Yukawa coupling matrix can be diagonalised simultaneously, were identified early \cite{Gatto:1978dy,Gatto:1979mr,Sartori:1979gt,Grimus:1986mh}. The interplay of how a symmetry requirement could enforce that \emph{general} NFC and shed some light into the structure of the resulting CKM matrix was addressed in \cite{Barbieri:1978qh,Gatto:1978dy,Sartori:1979ms,Gatto:1979sh,Segre:1979gs,Segre:1979rt,Rothman:1980ev,Kang:1980yg,Leurer:1992wg} with interesting consequences.

On a different line of thought, stepping back from right out forbiddance, suppression of FCNC in other ``natural'' manners has also attracted significant interest, including suppression given by masses like in the Cheng-Sher ans\"atz \cite{Cheng:1987rs}, suppression obtained from broken/approximate flavour symmetries \cite{Antaramian:1992ya,Hall:1993ca,Wu:1994ja,Datta:2008qn}, and symmetry controlled FCNC scenarios \cite{Joshipura:1990xm,Joshipura:1990pi,Joshipura:2007cs,Joshipura:2010tz,Lavoura:1994ty}. Among the later, Branco-Grimus-Lavoura (BGL) models are worth mentioning in particular, since this suppression is simply given by products of CKM matrix elements \cite{Branco:1996bq,Botella:2009pq} (see also related extensions \cite{Botella:2011ne,Botella:2015hoa,Alves:2017xmk}). In a more recent popular scenario, the Aligned 2HDM \cite{Pich:2009sp}, the absence of FCNC is a priori achieved (and parametrised) with simple requirements on the Yukawa couplings (for an early mention of this kind of possibility, although in the context of real Yukawa couplings and spontaneous CP violation, see also \cite{Ecker:1989ay}). The possibility of having effective aligned scenarios has been studied in \cite{Serodio:2011hg,Varzielas:2011jr}. Radiative effects and the interplay of tree level FCNC with the Renormalization Group Evolution (RGE) have also been addressed by and large in the literature \cite{Wise:1980ux,Segre:1979gs,Hall:1981bc,Frere:1985bu,Cvetic:1997zd,Cvetic:1998uw,Ferreira:2010xe,Bijnens:2011gd,Botella:2015yfa}.

The aim of this work is to explore different facets of scenarios with general flavour conservation (gFC), i.e. \emph{generalised flavour alignment}, in 2HDM; in other words, analysing relevant aspects of the most general 2HDM scenarios where tree level FCNC are, a priori, absent. An analysis of FCNC induced in this context by the RGE has been recently presented in \cite{Penuelas:2017ikk}. On a purely phenomenological basis, a scenario of this type restricted to the lepton sector was also considered in \cite{Ahn:2010zza,Braeuninger:2010td}.

The paper is organised as follows. In section \ref{SEC:GEN}, we revisit some generalities of 2HDM, fix the notation for the discussion to follow, and recall the most relevant aspects of the conditions leading to gFC. They are then analysed attending to the Renormalization Group Evolution that they obey in section \ref{SEC:RGE}, leading to the full set of conditions required to have RGE-stable gFC. The well known type I and type II cases are briefly revisited in section \ref{sSEC:RGE:mqNqStable}; section \ref{sSEC:RGE:CabibboStable} is devoted to a particular solution which arises when the CKM matrix is reduced to a single Cabibbo-like mixing. The gFC stability of the lepton sector is discussed in section \ref{sSEC:RGE:Leptons}.
In section \ref{SEC:Phenomenology}, we discuss the most relevant experimental constraints on gFC arising from flavour conserving Higgs-related observables, leading to the analysis and results of section \ref{sSEC:Phen:Num}.\\ %
Appendix \ref{APP:RGEdetails} provides details omitted in the discussion of section \ref{SEC:RGE}.

%
\section{Yukawa Couplings and General Flavour Conservation\label{SEC:GEN}}
The Higgs doublets ($j=1,2$) of 2HDM are
\begin{equation}\label{eq:ScalarDoublet}
\Hd{j}=e^{i\theta_j}\begin{pmatrix}\varphi_j^+\\ (v_j+\rho_j+i\eta_j)/\sqrt{2}\end{pmatrix}
\end{equation}
where $v_j$, $\theta_j$ are real numbers, $\rho_j$, $\eta_j$, are neutral (hermitian) fields and $\varphi_j^\pm$ are charged fields. 
Equation \eqref{eq:ScalarDoublet} anticipates the assumption that the scalar potential $V(\Hd{1},\Hd{2})$ \cite{Branco:2011iw,Ivanov:2017dad} is such that $V(\langle\Hd{1}\rangle,\langle\Hd{2}\rangle)$ has an appropriate minimum at
\begin{equation}\label{eq:ScalarDoubletVEV}
\langle\Hd{j}\rangle=\begin{pmatrix}0\\ e^{i\theta_j}v_j/\sqrt{2}\end{pmatrix}\,.
\end{equation}
In the ``Higgs basis'' \cite{Georgi:1978ri,Donoghue:1978cj,Botella:1994cs}, only one linear combination of $\Hd{1}$ and $\Hd{2}$, $H_{1}$, has a non-vanishing vacuum expectation value,
\begin{equation}\label{eq:HiggsBasis:00}
\begin{pmatrix}H_{1}\\ H_{2}\end{pmatrix}=
\Rb
\begin{pmatrix}e^{-i\theta_1}\Hd{1}\\ e^{-i\theta_2}\Hd{2}\end{pmatrix},\quad
\langle H_{1}\rangle=\frac{v}{\sqrt 2}\begin{pmatrix}0\\ 1\end{pmatrix},\quad\langle H_{2}\rangle=\begin{pmatrix}0\\ 0\end{pmatrix},
\end{equation}
with $v=\sqrt{v_1^2+v_2^2}$, $\cb=\cos\beta\equiv v_1/v$, $\sb=\sin\beta=v_2/v$ and
\begin{equation}
\Rb\equiv\begin{pmatrix}\cb & \sb\\ -\sb & \phantom{-}\cb \end{pmatrix}\,,\ [\Rb]^{-1}=[\Rb]^{T}\,.
\end{equation}
The expansion of $H_{1}$, $H_{2}$ around that minimum of the potential reads
\begin{equation}\label{eq:HiggsBasis:01}
H_{1}=\begin{pmatrix} G^+\\ (v+\nHH+iG^0)/\sqrt{2}\end{pmatrix},\quad H_{2}=\begin{pmatrix} \cHp\\ (\nHR+i\nHI)/\sqrt{2}\end{pmatrix},
\end{equation}
where
\begin{equation}
\begin{pmatrix}G^+\\ \cHp\end{pmatrix}=\Rb\begin{pmatrix}\varphi_1^+\\ \varphi_2^+\end{pmatrix},\ 
\begin{pmatrix}G^0\\ \nHI\end{pmatrix}=\Rb\begin{pmatrix}\eta_1\\ \eta_2\end{pmatrix},\ 
\begin{pmatrix}\nHH\\ \nHR\end{pmatrix}=\Rb\begin{pmatrix}\rho_1\\ \rho_2\end{pmatrix}.
\end{equation}
The would-be Goldstone bosons $G^0$ and $G^\pm$ provide the longitudinal degrees of freedom of the $Z$ and $W^\pm$ gauge bosons; furthermore, while $\cH$ is already a physical charged scalar field, the physical neutral scalars $\{\nh,\nH,\nA\}$ are real linear combinations of $\{\nHH,\nHR,\nHI\}$,
\begin{equation}\label{eq:GenMix:Scalars:00}
\ROT{3}{\vec\alpha}\,\begin{pmatrix}\nh\\ \nH\\ \nA\end{pmatrix}=\begin{pmatrix}\nHH\\ \nHR\\ \nHI\end{pmatrix}\,,\ [\ROT{3}{\vec\alpha}]^{-1}=[\ROT{3}{\vec\alpha}]^{T}\,,
\end{equation}
with $[\ROT{3}{\vec\alpha}]$ a real orthogonal rotation described by three real mixing angles, $\vec \alpha=\{\alpha_{12},\alpha_{13},\alpha_{23}\}$ ($c_x\equiv \cos x$, $s_x\equiv \sin x$),
\begin{equation}\label{eq:Rot3:Scalars:00}
[\ROT{3}{\vec\alpha}]=
\begin{pmatrix} c_{\alpha_{12}} & -s_{\alpha_{12}} & 0\\ s_{\alpha_{12}} & c_{\alpha_{12}} & 0\\ 0 & 0 & 1\end{pmatrix}\,
\begin{pmatrix} c_{\alpha_{13}} & 0 & -s_{\alpha_{13}}\\ 0 & 1 & 0\\ s_{\alpha_{13}} & 0 & c_{\alpha_{13}}\end{pmatrix}\,
\begin{pmatrix} 1 & 0 & 0 \\ 0 & c_{\alpha_{23}} & -s_{\alpha_{23}}\\ 0 & s_{\alpha_{23}} & c_{\alpha_{23}} \end{pmatrix}\,.
\end{equation}
When there is no CP violation in the scalar potential, i.e. no mixing connecting the CP-even $\nHH$, $\nHR$, and the CP-odd $\nHI$, it is customary to introduce the mixing angle $\alpha$ 
\begin{equation}\label{eq:Rot2:Scalars:00}
\begin{pmatrix}\nh\\ \nH\end{pmatrix}=\begin{pmatrix}\phantom{-}\sa & \ca\\ -\ca & \sa \end{pmatrix}\begin{pmatrix}\rho_1\\ \rho_2\end{pmatrix}=\begin{pmatrix}\phantom{-}\sab & \cab\\ -\cab & \sab \end{pmatrix}\begin{pmatrix}\nHH\\ \nHR\end{pmatrix}
\end{equation}
where $\sab=\sin(\alpha+\beta)$ and $\cab=\cos(\alpha+\beta)$ (that is, $\alpha_{13}=\alpha_{23}=0$ and $\alpha_{12}=\pi/2-(\alpha+\beta)$ in \refEQ{eq:Rot3:Scalars:00}). Since a $\pm$ sign can be included in the definition of the scalar fields without changing their kinetic terms, different conventions for \refEQS{eq:Rot3:Scalars:00}--\eqref{eq:Rot2:Scalars:00} are used in the literature, which may be relevant when comparing expressions.\\ 
%
\subsection{Quark Yukawa Couplings in 2HDM\label{sSEC:Yukawa:Quarks}}
The Yukawa couplings of the quarks -- $SU(2)_L$ doublets $Q_L^0$ and singlets $d_R^0$, $u_R^0$ -- with the scalar doublets read
\begin{equation}\label{eq:Yukawa:00}%
\mathscr L_{\rm Y}^{[q]}=
-\bar Q_L^0 [\Yd{1}\Hd{1}+\Yd{2}\Hd{2}] d_R^0
-\bar Q_L^0 [\Yu{1}\Hdt{1}+\Yu{2}\Hdt{2}] u_R^0
+\text{H.c.}
\end{equation}
with $\Hdt{j}=i\sigma_2\Hd{j}^\ast$. Following \refEQS{eq:HiggsBasis:00}--\eqref{eq:HiggsBasis:01},
\begin{align}\label{eq:Yukawa:01}%
\mathscr L_{\rm Y}^{[q]}=&
-\frac{\sqrt 2}{v}\bar Q_L^0 [\wMd H_{1}+\wNd H_{2}] d_R^0
-\frac{\sqrt 2}{v}\bar Q_L^0 [\wMu\tilde H_{1}+\wNu\tilde H_{2}] u_R^0
+\text{H.c.}\\
=&\,\mathscr L_{\rm m}^{[q]}+\mathscr L_{\rm G}^{[q]}+\mathscr L_{\rm Ch}^{[q]} +\mathscr L_{\rm N}^{[q]}
\end{align}
with mass terms $\mathscr L_{\rm m}^{[q]}$, would-be Goldstone boson couplings $\mathscr L_{\rm G}^{[q]}$, and Yukawa couplings to charged and neutral scalars, $\mathscr L_{\rm Ch}^{[q]}$ and $\mathscr L_{\rm N}^{[q]}$:
\begin{align}
\mathscr L_{\rm m}^{[q]}\supset &-\bar d_L^0 \wMd d_R^0 -\bar u_L^0 \wMu u_R^0\,,\label{eq:Yukawas:Mass:01}\\
\mathscr L_{\rm G}^{[q]}\supset &-\frac{\sqrt 2}{v}\left[ G^+ \bar u_L^0 \wMd d_R^0 +i  G^0 \bar d_L^0 \wMd d_R^0 - G^- \bar d_L^0 \wMu u_R^0 -i  G^0 \bar u_L^0 \wMu u_R^0\right],\label{eq:Yukawas:Goldstone:01}\\
\mathscr L_{\rm Ch}^{[q]}\supset &-\frac{\sqrt 2}{v}\left[ \cHp \bar u_L^0 \wNd d_R^0 - \cHm \bar d_L^0 \wNu u_R^0 \right],\label{eq:Yukawas:Charged:01}\\
\mathscr L_{\rm N}^{[q]}\supset &-\frac{1}{v}\left[ \nHH\bar d_L^0 \wMd d_R^0 + (\nHR+i\nHI)\bar d_L^0 \wNd d_R^0 +\nHH\bar u_L^0 \wMu u_R^0 + (\nHR-i\nHI)\bar u_L^0 \wNu u_R^0\right].\label{eq:Yukawas:Neutral:01}
\end{align}
%
The mass matrices are
\begin{equation}\label{eq:MassMatrices:00}%
\wMd=\frac{v}{\sqrt 2}\left[e^{i\theta_1}\cb\Yd{1}+e^{i\theta_2}\sb\Yd{2}\right],\quad 
\wMu=\frac{v}{\sqrt 2}\left[e^{-i\theta_1}\cb\Yu{1}+e^{-i\theta_2}\sb\Yu{2}\right],
\end{equation}
and the second linear combinations of Yukawa matrices which encode the potential FCNC are
\begin{equation}\label{eq:NMatrices:00}%
\wNd=\frac{v}{\sqrt 2}\left[-e^{i\theta_1}\sb\Yd{1}+e^{i\theta_2}\cb\Yd{2}\right],\quad 
\wNu=\frac{v}{\sqrt 2}\left[-e^{-i\theta_1}\sb\Yu{1}+e^{-i\theta_2}\cb\Yu{2}\right]\,.
\end{equation}
For the usual bi-diagonalisation of the mass matrices $\wMd$, $\wMu$, the quark mass eigenstates (without ``0'' superscript) read
\begin{equation}\label{eq:QuarkMassEigenstates:00}
d_L=\UdLd d_L^0,\ d_R=\UdRd d_R^0,\ u_L=\UuLd u_L^0,\ u_R=\UuRd u_R^0,
\end{equation}
with
\begin{equation}\label{eq:MDiagonalisation:00}
\mMd=\UdLd\wMd\,\UdR=\text{diag}(m_d,m_s,m_b)\,,\ \mMu=\UuLd\wMu\,\UuR=\text{diag}(m_u,m_c,m_t)\,,
\end{equation}
\begin{equation}
\mNd=\UdLd\wNd\,\UdR\,,\ \mNu=\UuLd\wNu\,\UuR\,.
\end{equation}
The CKM matrix is $\CKM=\UuLd\UdL$. When both $\mNd$ and $\mNu$ are diagonal, tree-level FCNC are absent.\\ 
Expressing \refEQ{eq:Yukawa:01} in terms of quark and scalar mass eigenstates (as a shorthand we use $[\ROT{3}{\vec\alpha}]_{ij}=\RS{ij}$),
\begin{align}
\mathscr L_{\rm m}^{[q]}=&-\bar d_L \mMd d_R -\bar u_L \mMu u_R+\text{H.c.}\,,\label{eq:Yukawas:Mass:02}\\
\mathscr L_{\rm G}^{[q]}=&-\frac{\sqrt 2}{v}\left[ G^+ \bar u_L\CKM \mMd d_R +i  G^0 \bar d_L \mMd d_R + G^- \bar d_L\CKMd \mMu u_R -i  G^0 \bar u_L \mMu u_R\right]+\text{H.c.},\label{eq:Yukawas:Goldstone:02}
\end{align}
\begin{equation}%
\mathscr L_{\rm Ch}^{[q]}=
 -\frac{\sqrt 2}{v}\left\{ \cHp \left[\bar u_L\CKM \mNd d_R - \bar u_R\mNud \CKM d_L \right]+ \cHm \left[\bar d_R\mNdd \CKMd u_L - \bar d_L\CKMd \mNu u_R\right] \right\},\label{eq:Yukawas:Charged:03}
\end{equation}
\begin{align}%
\mathscr L_{\rm N}^{[q]}=
&-\frac{\nh}{v} \left\{ 
\bar d \left[\RS{11}\mMd+\RS{21}\HNd+i\RS{31}\AHNd\right] d
+ \bar d \left[\RS{21}\AHNd+i\RS{31}\HNd\right] \gamma_5 d
\right\}\nonumber\\
&-\frac{\nh}{v} \left\{ 
\bar u \left[\RS{11}\mMu+\RS{21}\HNu-i\RS{31}\AHNu\right] u
+ \bar u \left[\RS{21}\AHNu-i\RS{31}\HNu\right] \gamma_5 u
\right\}\nonumber\\
&-\frac{\nH}{v} \left\{
\bar d \left[\RS{12}\mMd+\RS{22}\HNd+i\RS{32}\AHNd\right] d
+\bar d \left[\RS{22}\AHNd+i\RS{32}\HNd\right] \gamma_5 d
\right\}\nonumber\\
&-\frac{\nH}{v} \left\{ 
\bar u \left[\RS{12}\mMu+\RS{22}\HNu-i\RS{32}\AHNu\right] u
+ \bar u \left[\RS{22}\AHNu-i\RS{32}\HNu\right] \gamma_5 u
\right\}\nonumber\\
&-\frac{\nA}{v} \left\{
\bar d \left[\RS{13}\mMd+\RS{23}\HNd+i\RS{33}\AHNd\right] d
+ \bar d \left[\RS{23}\AHNd+i\RS{33}\HNd\right] \gamma_5 d
\right\}\nonumber\\
&-\frac{\nA}{v} \left\{ 
\bar u \left[\RS{13}\mMu+\RS{23}\HNu-i\RS{33}\AHNu\right] u
+ \bar u \left[\RS{23}\AHNu-i\RS{33}\HNu\right] \gamma_5 u
\right\},\label{eq:Yukawas:Neutral:03}
\end{align}
where 
\begin{equation}
\baseHNq{q}\equiv\frac{\mNq{q}+\mNqd{q}}{2},\quad \baseAHNq{q}\equiv\frac{\mNq{q}-\mNqd{q}}{2},\quad q=u,d,
\end{equation}
are the hermitian and anti-hermitian combinations of $\mNq{q}$ and $\mNqd{q}$.\\ 
With no CP violation in the scalar sector, 
\begin{equation}\label{eq:ScalarCPconserving:00}
\RS{}=\begin{pmatrix}\sab & -\cab & 0\\ \cab & \phantom{-}\sab & 0\\ 0&0&1\end{pmatrix}\,,
\end{equation}
and \refEQ{eq:Yukawas:Neutral:03} reduces to
\begin{align}%
\mathscr L_{\rm N}^{[q]}=
&-\frac{\nh}{v} \left\{ \bar d \left[\sab\mMd+\cab\HNd\right] d + \cab\,\bar d \AHNd \gamma_5 d \right\}\nonumber\\
&-\frac{\nh}{v} \left\{ \bar u \left[\sab\mMu+\cab\HNu\right] u + \cab\,\bar u \AHNu \gamma_5 u \right\}\nonumber\\
&-\frac{\nH}{v} \left\{ \bar d \left[-\cab\mMd+\sab\HNd\right] d + \sab\,\bar d \AHNd \gamma_5 d \right\}\nonumber\\
&-\frac{\nH}{v} \left\{ \bar u \left[-\cab\mMu+\sab\HNu\right] u + \sab\,\bar u \AHNu \gamma_5 u \right\}\nonumber\\
&-i\frac{\nA}{v} \left\{ \bar d \AHNd d + \bar d \HNd \gamma_5 d \right\}
+i\frac{\nA}{v} \left\{ \bar u \AHNu u + \bar u \HNu \gamma_5 u \right\}.\label{eq:Yukawas:Neutral:04}
\end{align}

\subsection{Lepton Yukawa Couplings in 2HDM\label{sSEC:Yukawa:Leptons}}
The Yukawa couplings of the lepton $SU(2)_L$ doublets $L_L^0$ and singlets $\ell_R^0$ with the scalar doublets are
\begin{equation}\label{eq:YukawaLeptons:01}
\mathscr L_{\rm Y}^{[\ell]}=-\bar L_L^0(\Yl{1}\Hd{1}+\Yl{2}\Hd{2})\ell_R^0+\text{H.c.}=-\frac{\sqrt 2}{v}\bar L_L^0(\wMl H_1+\wNl H_2)\ell_R^0+\text{H.c.}
\end{equation}
where, similarly to the quark sector in the previous section,
\begin{equation}
\wMl=\frac{v}{\sqrt 2}\left[e^{i\theta_1}\cb\Yl{1}+e^{i\theta_2}\sb\Yl{2}\right],\quad 
\wNl=\frac{v}{\sqrt 2}\left[-e^{i\theta_1}\sb\Yl{1}+e^{i\theta_2}\cb\Yl{2}\right].
\end{equation}
The mass eigenstates, without ``0'' superscript, correspond to
\begin{equation}
\ell_L=\UlLd\ell_L^0,\quad \ell_R=\UlRd\ell_R^0,
\end{equation}
and
\begin{equation}
\mMl=\UlLd\,\wMl\,\UlR=\text{diag}(m_e,m_\mu,m_\tau),\quad \mNl=\UlLd\,\wNl\,\UlR\,.
\end{equation}
Notice that we do not include right-handed neutrinos $\nu_R^0$ and thus, unlike in the quark sector, there is only one set of Yukawa coupling matrices and we work in the massless neutrino approximation.
The leptonic analogs of the Yukawa couplings in \refEQS{eq:Yukawas:Charged:03}-\eqref{eq:Yukawas:Neutral:03} are
\begin{equation}%
\mathscr L_{\rm Ch}^{[\ell]}=
 -\frac{\sqrt 2}{v}\left\{ \cHp \bar \nu_L \mNl \ell_R + \cHm \bar \ell_R\mNld \nu_L \right\},
\label{eq:YukawasLeptons:Charged:01}
\end{equation}
\begin{align}%
\mathscr L_{\rm N}^{[\ell]}=
&-\frac{\nh}{v} \left\{ 
\bar \ell \left[\RS{11}\mMl+\RS{21}\HNl+i\RS{31}\AHNl\right] \ell
+ \bar \ell \left[\RS{21}\AHNl+i\RS{31}\HNl\right] \gamma_5 \ell
\right\}\nonumber\\
&-\frac{\nH}{v} \left\{
\bar \ell \left[\RS{12}\mMl+\RS{22}\HNl+i\RS{32}\AHNl\right] \ell
+\bar \ell \left[\RS{22}\AHNl+i\RS{32}\HNl\right] \gamma_5 \ell
\right\}\nonumber\\
&-\frac{\nA}{v} \left\{
\bar \ell \left[\RS{13}\mMl+\RS{23}\HNl+i\RS{33}\AHNl\right] \ell
+ \bar \ell \left[\RS{23}\AHNl+i\RS{33}\HNl\right] \gamma_5 \ell
\right\},\label{eq:YukawasLeptons:Neutral:01}
\end{align}
with
\begin{equation}
\HNl\equiv\frac{\mNl+\mNld}{2},\quad \AHNl\equiv\frac{\mNl-\mNld}{2}.
\end{equation}
\subsection{General Flavour Conservation\label{sSEC:FC}}
The necessary and sufficient conditions obeyed by the quark Yukawa coupling matrices $\Yd{\alpha}$, $\Yu{\alpha}$, $\alpha=1,2$, in order to have gFC \cite{Gatto:1978dy,Gatto:1979mr,Sartori:1979gt,Grimus:1986mh}, are that each of the sets 
\begin{equation}\label{eq:abelian:00b}
\{\Yd{\alpha}\Ydd{\beta}\},\ \{\Ydd{\alpha}\Yd{\beta}\},\ \{\Yu{\alpha}\Yud{\beta}\},\ \{\Yud{\alpha}\Yu{\beta}\},\quad \alpha,\beta=1,2,
\end{equation}
is \emph{abelian}, that is, their elements commute:
\begin{equation}\label{eq:abelian:01b}
\COMM{\Yd{\alpha}\Ydd{\beta}}{\Yd{\gamma}\Ydd{\delta}}=0,\ \COMM{\Ydd{\alpha}\Yd{\beta}}{\Ydd{\gamma}\Yd{\delta}}=0,\ \COMM{\Yu{\alpha}\Yud{\beta}}{\Yu{\gamma}\Yud{\delta}}=0,\ \COMM{\Yud{\alpha}\Yu{\beta}}{\Yud{\gamma}\Yu{\delta}}=0,
\end{equation}
with $\alpha,\beta,\gamma,\delta=1,2$. In that case, $\{\Yd{1},\Yd{2}\}$ are simultaneously bi-diagonalised, and $\{\Yu{1},\Yu{2}\}$ too.\\ %
A crucial corollary to these necessary and sufficient conditions is the fact that the simultaneous diagonalisability is intrinsic to the Yukawa coupling matrices themselves, independently of the spontaneous symmetry breaking vacuum characterised by the VEVs $v_1,v_2$. In other words, the property is independent of $\beta$ in \refEQS{eq:MassMatrices:00}, \eqref{eq:NMatrices:00}; the simultaneous bi-diagonalisability of $\{\wMq{q},\wNq{q}\}$ is equivalent to the simultaneous bi-diagonalisability of the Yukawa couplings matrices or of any other independent linear combinations of them. Of course, the actual values of the eigenvalues of both $\wMq{q}$ (the masses) and $\wNq{q}$ do depend on the particular linear combinations.\\ %
For leptons, similarly, $\{\Yl{\alpha}\Yld{\beta}\}$ and $\{\Yld{\alpha}\Yl{\beta}\}$ must be abelian in order to have gFC, and the previous corollary applies equally to them.\\ %
A very relevant consequence follows \cite{Barbieri:1978qh,Gatto:1978dy,Sartori:1979ms,Gatto:1979sh,Segre:1979gs,Segre:1979rt,Rothman:1980ev,Kang:1980yg,Leurer:1992wg}: if gFC is due to the Lagrangian in \refEQ{eq:Yukawa:00} being invariant under a (symmetry) transformation of quarks and scalars, the CKM mixing matrix cannot be related to the values of the masses; for example, predictions being made at the time (late 70's)\footnote{In the context of $SU(2)_L\otimes U(1)_Y$ gauge theories; the literature is richer in examples for $SU(2)_L\otimes SU(2)_R\otimes U(1)_Y$ scenarios.} for the Cabibbo angle, like $\tan\theta_c= m_d/m_s$ \cite{Pakvasa:1977in,Wyler:1978fj}, \emph{could not lead simultaneously to gFC}. Moreover, the resulting mixings are unrealistic (for example, no mixing or a permutation times a complex phase) and radiative corrections cannot be invoked to yield realistic mixings \cite{Segre:1979gs}.\\
The most general parameterisation of tree level couplings of fermions to scalars obeying gFC is, quite trivially,
\begin{equation}\label{eq:FC:Matrices:00}
\mNd=\begin{pmatrix}\ND{d}&0&0\\ 0&\ND{s}&0\\ 0&0&\ND{b}\end{pmatrix}\!,\ 
\mNu=\begin{pmatrix}\NU{u}&0&0\\ 0&\NU{c}&0\\ 0&0&\NU{t}\end{pmatrix}\!,\ 
\mNl=\begin{pmatrix}\NL{e}&0&0\\ 0&\NL{\mu}&0\\ 0&0&\NL{\tau}\end{pmatrix}\!,
\quad \nq{q}{j}\in\mathbb{C},
\end{equation}
which we use in the rest of the paper: in section \ref{SEC:RGE} for the study of the renormalization group evolution and in section \ref{SEC:Phenomenology} for a phenomenological analysis.\\ %
Notice that, while for the flavour changing couplings the simultaneous presence of scalar and pseudoscalar terms in fermion-scalar Yukawa interactions is not necessarily CP violating, in the diagonal, flavour conserving ones, on the contrary, it is CP violating (see for example \cite{Nebot:2015wsa}). With the flavour conserving matrices $\mNq{f}$ in \refEQ{eq:FC:Matrices:00}, the hermitian and antihermitian couplings in \refEQS{eq:Yukawas:Neutral:04} and \eqref{eq:YukawasLeptons:Neutral:01} are, respectively, their real and imaginary parts. For example, for a CP conserving scalar sector with non-zero mixing $\cab\neq 0$, if $\mNq{f}$ are not real, they constitute new sources of CP violation in neutral couplings. For the couplings to the charged scalar, without entering into details, if $\im{\nq{}{u_i}\nq{}{d_j}}\neq 0$, the combination of scalar and pseudoscalar terms in the coupling $\cHp\bar u_i\, d_j$ is CP violating.

%
\section{Renormalization Group Evolution and Flavour Conservation\label{SEC:RGE}}

\subsection{Evolution of the Quark Yukawa Coupling Matrices\label{sSEC:RGE:Yukawas}}

The one loop evolution of the Yukawa couplings under the renormalization group \cite{Cvetic:1997zd,Cvetic:1998uw,Grimus:2004yh,Ferreira:2010xe} is (with $\der\equiv 16\pi^2\frac{d}{d\ln\mu}$ and $\mu$ the energy scale):
\begin{multline}
\der\Yd{\alpha}=
a_d\Yd{\alpha}+\sum_{\rho=1}^{n=2}T^{d}_{\alpha,\rho}\Yd{\rho}+\sum_{\rho=1}^{n=2}\left(-2\Yu{\rho}\Yud{\alpha}\Yd{\rho}+\Yd{\alpha}\Ydd{\rho}\Yd{\rho}+\frac{1}{2}\Yu{\rho}\Yud{\rho}\Yd{\alpha}+\frac{1}{2}\Yd{\rho}\Ydd{\rho}\Yd{\alpha}\right)\\
\text{with }T^{d}_{\alpha,\rho}\equiv 3\,\tr{\Yd{\alpha}\Ydd{\rho}+\Yud{\alpha}\Yu{\rho}}+\tr{\Yl{\alpha}\Yld{\rho}},
\label{eq:RGE:Yd:00}
\end{multline}
\begin{multline}
\der\Yu{\alpha}=
a_u\Yu{\alpha}+\sum_{\rho=1}^{n=2}T^{u}_{\alpha,\rho}\Yu{\rho}+\sum_{\rho=1}^{n=2}\left(-2\Yd{\rho}\Ydd{\alpha}\Yu{\rho}+\Yu{\alpha}\Yud{\rho}\Yu{\rho}+\frac{1}{2}\Yd{\rho}\Ydd{\rho}\Yu{\alpha}+\frac{1}{2}\Yu{\rho}\Yud{\rho}\Yu{\alpha}\right)\\
\text{with }T^{u}_{\alpha,\rho}\equiv 3\,\tr{\Yu{\alpha}\Yud{\rho}+\Ydd{\alpha}\Yd{\rho}}+\tr{\Yld{\alpha}\Yl{\rho}}=T^{d\,\ast}_{\alpha,\rho}\,,
\label{eq:RGE:Yu:00}
\end{multline}
where
\begin{equation}
a_d=-8g_s^2-\frac{9}{4}g^2-\frac{5}{12}g^{\prime 2},\qquad a_u=a_d-g^{\prime 2},
\label{eq:RGEa:00}
\end{equation}
with $g_s$, $g$, $g^\prime$ the gauge coupling constants of $SU(3)_c$, $SU(2)_L$ and $U(1)_Y$, respectively.
Introducing
\begin{equation}
\YdH{L}=\sum_{\rho=1}^{n=2}\Yd{\rho}\Ydd{\rho}\,,\quad \YdH{R}=\sum_{\rho=1}^{n=2}\Ydd{\rho}\Yd{\rho}\,,\quad
\YuH{L}=\sum_{\rho=1}^{n=2}\Yu{\rho}\Yud{\rho}\,,\quad \text{and}\quad \YuH{R}=\sum_{\rho=1}^{n=2}\Yud{\rho}\Yu{\rho}\,,
\label{eq:RGEQuadraticLR:00}
\end{equation}
\refEQS{eq:RGE:Yd:00}--\eqref{eq:RGE:Yu:00} read
\begin{equation}
\der\Yd{\alpha}=a_d\Yd{\alpha}+\sum_{\rho=1}^{n=2}T^{d}_{\alpha,\rho}\Yd{\rho}+\Yd{\alpha}\YdH{R}+\frac{1}{2}\YdH{L}\Yd{\alpha}+\frac{1}{2}\YuH{L}\Yd{\alpha}-2\sum_{\rho=1}^{n=2}\Yu{\rho}\Yud{\alpha}\Yd{\rho}\,,
\label{eq:RGE:Yd:01}
\end{equation}
\begin{equation}
\der\Yu{\alpha}=a_u\Yu{\alpha}+\sum_{\rho=1}^{n=2}T^{u}_{\alpha,\rho}\Yu{\rho}+\Yu{\alpha}\YuH{R}+\frac{1}{2}\YuH{L}\Yu{\alpha}+\frac{1}{2}\YdH{L}\Yu{\alpha}-2\sum_{\rho=1}^{n=2}\Yd{\rho}\Ydd{\alpha}\Yu{\rho}\,.
\label{eq:RGE:Yu:01}
\end{equation}
Equations \eqref{eq:RGE:Yd:01}--\eqref{eq:RGE:Yu:01} are the starting point to analyse the one loop stability of the necessary and sufficient conditions for gFC. For that, one needs to know
\begin{equation}
\der\left(\COMM{\Yd{\alpha}\Ydd{\beta}}{\Yd{\gamma}\Ydd{\delta}}\right),\, \der\left(\COMM{\Ydd{\alpha}\Yd{\beta}}{\Ydd{\gamma}\Yd{\delta}}\right),\, \der\left(\COMM{\Yu{\alpha}\Yud{\beta}}{\Yu{\gamma}\Yud{\delta}}\right),\, \der\left(\COMM{\Yud{\alpha}\Yu{\beta}}{\Yud{\gamma}\Yu{\delta}}\right),
\label{eq:RGE:FC:00}
\end{equation}
under the assumption that \refEQ{eq:abelian:01b} holds. With that objective in mind, some simplifications are worth mentioning. Starting with $\Yd{\alpha}\Ydd{\beta}$, we first notice that
\begin{equation}
\der(\Yd{\alpha}\Ydd{\beta})=(\der\Yd{\alpha})\Ydd{\beta}+\Yd{\alpha}(\der\Yd{\beta})^\dagger=\DYf{d_L}{\alpha \beta}(\Ydbase)+\DYg{d_L}{\alpha \beta}(\Ydbase,\Yubase)\,,
\label{eq:RGEbilineardecomposition:00}
\end{equation}
with\footnote{The superscript ${\rm [d_L]}$ is chosen in correspondence with the $\Yd{\alpha}\Ydd{\beta}$ matrix combinations; similarly $\DYf{d_R}{\alpha \beta}$ and $\DYg{d_R}{\alpha \beta}$ will appear in $\der(\Ydd{\alpha}\Yd{\beta})$, and $\DYf{u_{L,R}}{\alpha \beta}$ in $\der(\Yu{\alpha}\Yud{\beta})$ and $\der(\Yud{\alpha}\Yu{\beta})$, but we concentrate for the moment on $\der(\Yd{\alpha}\Ydd{\beta})$.}
\begin{equation}
\DYf{d_L}{\alpha \beta}(\Ydbase)=2a_d\Yd{\alpha}\Ydd{\beta}+\sum_{\rho=1}^{n=2}\left[T^d_{\alpha,\rho}\Yd{\rho}\Ydd{\beta}+T^{d\ast}_{\beta,\rho}\Yd{\alpha}\Ydd{\rho}\right]+2\Yd{\alpha}\YdH{R}\Ydd{\beta}+\frac{1}{2}\YdH{L}\Yd{\alpha}\Ydd{\beta}+\frac{1}{2}\Yd{\alpha}\Ydd{\beta}\YdH{L}\,,
\label{eq:DerY:f:00}
\end{equation}
and
\begin{equation}
\DYg{d_L}{\alpha \beta}(\Ydbase,\Yubase)=\frac{1}{2}\YuH{L}\Yd{\alpha}\Ydd{\beta} +\frac{1}{2}\Yd{\alpha}\Ydd{\beta}\YuH{L} -2\sum_{\rho=1}^{n=2}\left[\Yu{\rho}\Yud{\alpha}\Yd{\rho}\Ydd{\beta}+\Yd{\alpha}\Ydd{\rho}\Yu{\beta}\Yud{\rho}\right].
\label{eq:DerY:g:00}
\end{equation}
The relevant property of the decomposition in \refEQ{eq:RGEbilineardecomposition:00} is that $\DYf{d_L}{\alpha \beta}$ depends only\footnote{Although $T^d_{\alpha,\rho}$ do depend on $\Yu{\alpha}$'s, there is no matrix depence, only $\mathbb{C}$ numbers; this also applies to the leptonic Yukawa couplings $\Yl{\alpha}$.}, in terms of matrices, on $\Ydbase\Ydbase^\dagger$ and $\Ydbase\Ydbase^\dagger\Ydbase\Ydbase^\dagger$ terms, while $\DYg{d_L}{\alpha \beta}$ collects the remaining dependence on $\Yubase$'s, which has terms $\Ydbase\Ydbase^\dagger\Yubase\Yubase^\dagger$ and $\Yubase\Yubase^\dagger\Ydbase\Ydbase^\dagger$.
Then,
\begin{multline}
\der\COMM{\Yd{\alpha}\Ydd{\beta}}{\Yd{\gamma}\Ydd{\delta}}=\COMM{\der(\Yd{\alpha}\Ydd{\beta})}{\Yd{\gamma}\Ydd{\delta}}+\COMM{\Yd{\alpha}\Ydd{\beta}}{\der(\Yd{\gamma}\Ydd{\delta})}=\\
\COMM{\DYf{d_L}{\alpha \beta}(\Ydbase)+\DYg{d_L}{\alpha \beta}(\Ydbase,\Yubase)}{\Yd{\gamma}\Ydd{\delta}}+\COMM{\Yd{\alpha}\Ydd{\beta}}{\DYf{d_L}{\gamma \delta}(\Ydbase)+\DYg{d_L}{\gamma \delta}(\Ydbase,\Yubase)}\,.
\label{eq:DerComm:00}
\end{multline}

\subsection{Evolution with gFC Matrices\label{sSEC:RGE:FC}}
It is clear that, if there is gFC, i.e. with \refEQ{eq:abelian:01b},
\begin{equation}
\COMM{\DYf{d_L}{\alpha \beta}(\Ydbase)}{\Yd{\gamma}\Ydd{\delta}}=\COMM{\Yd{\alpha}\Ydd{\beta}}{\DYf{d_L}{\gamma \delta}(\Ydbase)}=0\,,
\label{eq:DerComm:01}
\end{equation}
and thus
\begin{equation}
\der\COMM{\Yd{\alpha}\Ydd{\beta}}{\Yd{\gamma}\Ydd{\delta}}=\COMM{\DYg{d_L}{\alpha \beta}(\Ydbase,\Yubase)}{\Yd{\gamma}\Ydd{\delta}}+\COMM{\Yd{\alpha}\Ydd{\beta}}{\DYg{d_L}{\gamma \delta}(\Ydbase,\Yubase)}\,.
\label{eq:DerComm:02}
\end{equation}
After the simplication brought by \refEQ{eq:DerComm:01}, the next step is to trade \refEQ{eq:DerComm:02} for conditions expressed in terms of the physical parameters entering in the matrices $\wMd$, $\wNd$, $\wMu$, $\wNu$. It is convenient to introduce the following notation
\begin{equation}
\wYDd{1}=\wMd,\quad \wYDd{2}=\wNd,\quad \wYDu{1}=\wMu,\quad \wYDu{2}=\wNu,
\label{eq:MNY:00}
\end{equation}
which allows us to rewrite \refEQS{eq:MassMatrices:00}-\eqref{eq:NMatrices:00} compactly (with summation over repeated indices understood):
\begin{alignat}{2}
\frac{v}{\sqrt 2}\Yd{\alpha}&=\YR{\alpha i}\wYDd{i}\,,\qquad \wYDd{i}&=\frac{v}{\sqrt 2}\Yd{\alpha}\YRc{\alpha i}\,,\nonumber\\
\frac{v}{\sqrt 2}\Yu{\alpha}&=\YRc{\alpha i}\wYDu{i}\,,\qquad \wYDu{i}&=\frac{v}{\sqrt 2}\Yu{\alpha}\YR{\alpha i}\,,
\label{eq:YR:00}
\end{alignat}
where
\begin{equation}
\baseYR=\begin{pmatrix}e^{-i\theta_1} & 0\\ 0 & e^{-i\theta_2}\end{pmatrix} \begin{pmatrix} \cb & -\sb\\ \sb & \cb \end{pmatrix},\quad 
\baseYR\baseYRd=\baseYRd\baseYR=\mathbf{1}\,.
\label{eq:YR:01}
\end{equation}
For completeness, notice that
\begin{equation}
\frac{v}{\sqrt 2}\Ydd{\alpha}=\YRc{\alpha i}\wYDdd{i}\,,\ \frac{v}{\sqrt 2}\Yud{\alpha}=\YR{\alpha i}\wYDud{i}\,,\ 
\wYDdd{i}=\frac{v}{\sqrt 2}\Ydd{\alpha}\YR{\alpha i}\,,\ \wYDud{i}=\frac{v}{\sqrt 2}\Yud{\alpha}\YRc{\alpha i}\,,
\label{eq:YR:02}
\end{equation}
 i.e. the Hermitian conjugate ${}^\dagger$ (in the space of flavour indices) only gives a complex conjugate in $\baseYR$.
One can then write
\begin{equation}
\COMM{\Yd{\alpha}\Ydd{\beta}}{\Yd{\gamma}\Ydd{\delta}}=\frac{4}{v^4}\YR{\alpha i}\YRc{\beta j}\YR{\gamma k}\YRc{\delta l}\COMM{\wYDd{i}\wYDdd{j}}{\wYDd{k}\wYDdd{l}},
\label{eq:Yuk:YukD:00}
\end{equation}
and thus
\begin{multline}
\der\COMM{\Yd{\alpha}\Ydd{\beta}}{\Yd{\gamma}\Ydd{\delta}}=\der\left(\frac{4}{v^4}\YR{\alpha i}\YRc{\beta j}\YR{\gamma k}\YRc{\delta l}\right)\COMM{\wYDd{i}\wYDdd{j}}{\wYDd{k}\wYDdd{l}}\\
+\frac{4}{v^4}\YR{\alpha i}\YRc{\beta j}\YR{\gamma k}\YRc{\delta l}\ \der\COMM{\wYDd{i}\wYDdd{j}}{\wYDd{k}\wYDdd{l}}\,.
\label{eq:der:Yuk:YukD:00}
\end{multline}
With gFC, the first commutator vanishes, and we just have a linear combination of different $\der\comm{\wYDd{i}\wYDdd{j}}{\wYDd{k}\wYDdd{l}}$. One can indeed invert \refEQ{eq:der:Yuk:YukD:00},
\begin{equation}
\frac{4}{v^4}\der\COMM{\wYDd{i}\wYDdd{j}}{\wYDd{k}\wYDdd{l}}=\YRc{\alpha i}\YR{\beta j}\YRc{\gamma k}\YR{\delta l}\ \der\COMM{\Yd{\alpha}\Ydd{\beta}}{\Yd{\gamma}\Ydd{\delta}}\,
\label{eq:der:Yuk:YukD:01}
\end{equation}
and express the right-hand side of \refEQ{eq:der:Yuk:YukD:01} in terms of $\wYDd{i}$, $\wYDu{j}$:
\begin{align}
\frac{v^2}{2}&\der\COMM{\wYDd{i}\wYDdd{j}}{\wYDd{k}\wYDdd{l}}=\nonumber\\
&\quad \wYDd{i}\wYDdd{j}\wYDu{h}\wYDud{h}\wYDd{k}\wYDdd{l}-\wYDd{k}\wYDdd{l}\wYDu{h}\wYDud{h}\wYDd{i}\wYDdd{j}\nonumber\\
&-2\COMM{\wYDu{h}\wYDud{i}}{\wYDd{k}\wYDdd{l}}\wYDd{h}\wYDdd{j}-2\wYDd{i}\wYDdd{h}\COMM{\wYDu{j}\wYDud{h}}{\wYDd{k}\wYDdd{l}}\nonumber\\
&+2\COMM{\wYDu{h}\wYDud{k}}{\wYDd{i}\wYDdd{j}}\wYDd{h}\wYDdd{l}+2\wYDd{k}\wYDdd{h}\COMM{\wYDu{l}\wYDud{h}}{\wYDd{i}\wYDdd{j}}.
\label{eq:der:Yuk:YukD:03}
\end{align}
As expected from the discussion in section \ref{sSEC:FC}, having a gFC scenario is related to the Yukawa coupling matrices themselves, it does not hinge on the particular EW vacuum configuration that determines which particular combinations of them are the mass matrices $\wMd$, $\wMu$ and the matrices $\wNd$, $\wNu$ (the vacuum configuration is ``encoded'' in $\baseYR$, which does not appear in \refEQ{eq:der:Yuk:YukD:03}). The last step is to transform into the mass eigenstate basis with $\UdL$ in \refEQ{eq:QuarkMassEigenstates:00}:
\begin{align}
\frac{v^2}{2}&\UdLd\left(\der\COMM{\wYDd{i}\wYDdd{j}}{\wYDd{k}\wYDdd{l}}\right)\UdL=\nonumber\\
&\quad \mYDd{i}\mYDdd{j}\,\CKMd\mYDu{h}\mYDud{h}\CKM\,\mYDd{k}\mYDdd{l}-\mYDd{k}\mYDdd{l}\,\CKMd\mYDu{h}\mYDud{h}\CKM\,\mYDd{i}\mYDdd{j}\nonumber\\
&-2\COMM{\CKMd\mYDu{h}\mYDud{i}\CKM}{\mYDd{k}\mYDdd{l}}\mYDd{h}\mYDdd{j}-2\mYDd{i}\mYDdd{h}\COMM{\CKMd\mYDu{j}\mYDud{h}\CKM}{\mYDd{k}\mYDdd{l}}\nonumber\\
&+2\COMM{\CKMd\mYDu{h}\mYDud{k}\CKM}{\mYDd{i}\mYDdd{j}}\mYDd{h}\mYDdd{l}+2\mYDd{k}\mYDdd{h}\COMM{\CKMd\mYDu{l}\mYDud{h}\CKM}{\mYDd{i}\mYDdd{j}},
\label{eq:der:Yuk:YukD:05}
\end{align}
where the CKM matrix $\CKM=\UuLd\UdL$ appears together with the diagonal matrices
\begin{equation}
\mYDd{1}=\mMd,\ \mYDd{2}=\mNd,\ \mYDu{1}=\mMu,\ \mYDu{2}=\mNu\,.
\label{eq:DiagY:00}
\end{equation}
In this generic notation -- \refEQ{eq:MNY:00} --,
\begin{align}
\mYDd{i}=\text{diag}(\myDd{i}{j}),\quad 
&\{\myDd{1}{1},\myDd{1}{2},\myDd{1}{3}\}=\{m_d,m_s,m_b\},\nonumber\\
&\{\myDd{2}{1},\myDd{2}{2},\myDd{2}{3}\}=\{\nq{}{d},\nq{}{s},\nq{}{b}\},\nonumber\\
\mYDu{i}=\text{diag}(\myDu{i}{j}),\quad 
&\{\myDu{1}{1},\myDu{1}{2},\myDu{1}{3}\}=\{m_u,m_c,m_t\},\nonumber\\
&\{\myDu{2}{1},\myDu{2}{2},\myDu{2}{3}\}=\{\nq{}{u},\nq{}{c},\nq{}{t}\}.
\label{eq:DiagY:01}
\end{align}
The previous derivation concerns the set $\{\Yd{\alpha}\Ydd{\beta}\}$; the evolution equations for $\{\Ydd{\alpha}\Yd{\beta}\}$, $\{\Yu{\alpha}\Yud{\beta}\}$ and $\{\Yud{\alpha}\Yu{\beta}\}$ are given in appendix \ref{APP:RGEdetails}.\\
In order to have a gFC scenario stable under the one loop RGE, one needs that the simultaneous diagonalisability of $\{\wYDq{q}{1},\wYDq{q}{2}\}$ is preserved, that is
\begin{alignat}{2}
&\der\COMM{\wYDd{i}\wYDdd{j}}{\wYDd{k}\wYDdd{l}}=0\,,\quad &\der\COMM{\wYDu{i}\wYDud{j}}{\wYDu{k}\wYDud{l}}=0\,,\nonumber\\
&\der\COMM{\wYDdd{i}\wYDd{j}}{\wYDdd{k}\wYDd{l}}=0\,,\quad &\der\COMM{\wYDud{i}\wYDu{j}}{\wYDud{k}\wYDu{l}}=0\,.
\label{eq:Stability:00}
\end{alignat}
With \refEQS{eq:der:Yuk:YukD:05}--\eqref{eq:DiagY:01}, the conditions expressed by the matrix equations in \eqref{eq:Stability:00} are formulated in full generality, for fixed mass matrices $\mMd$, $\mMu$, and CKM mixings $\CKM$, in terms of the 6 complex parameters $\nq{q}{j}$ in \refEQ{eq:FC:Matrices:00}. For example, element $(a,b)$ of the first stability condition in \refEQ{eq:Stability:00}, for $i=j$, $k=l$, $\der\comm{\wYDd{i}\wYDdd{i}}{\wYDd{k}\wYDdd{k}}=0$, reads 
\begin{align}
0=\sum_{q=1}^3\sum_{h=1}^2\Vc{qa}\V{qb}\Big\{&
\abs{\myDu{h}{q}}^2 \left( \abs{\myDd{i}{a}}^2\abs{\myDd{k}{b}}^2 - \abs{\myDd{i}{b}}^2\abs{\myDd{k}{a}}^2 \right)\label{eq:RGEStab:Matrix:00}\\
&-2\left(\myDu{h}{q}\myDuc{i}{q}\myDd{h}{b}\myDdc{i}{b}+\myDu{i}{q}\myDuc{h}{q}\myDd{i}{a}\myDdc{h}{a}\right)\left(\abs{\myDd{k}{b}}^2-\abs{\myDd{k}{a}}^2\right)\nonumber\\
&+2\left(\myDu{h}{q}\myDuc{k}{q}\myDd{h}{b}\myDdc{k}{b}+\myDu{k}{q}\myDuc{h}{q}\myDd{k}{a}\myDdc{h}{a}\right)\left(\abs{\myDd{i}{b}}^2-\abs{\myDd{i}{a}}^2\right)
\Big\}.
\nonumber
\end{align}
The complete set of conditions is given in appendix \ref{APP:RGEdetails}. For each set in \refEQ{eq:Stability:00} there are six choices of $i,j,k,l=1,2$, in 2HDM, which give, at least, 3 independent complex equations each. It is clear that, in terms of the 6 complex parameters $\nq{}{j}$, the system is largely overconstrained. In section \ref{sSEC:RGE:mqNqStable} below, we check that the known stable solutions with $\mNq{f}\propto \mMq{f}$ are recovered. It is however beyond the scope of this work to address if other solutions could a priori exist for the general one loop RGE stability conditions of gFC.\\ %
The lepton sector is discussed in section \ref{sSEC:RGE:Leptons}. Finally, in section \ref{sSEC:RGE:CabibboStable}, we present some particular solutions which arise when the CKM matrix reduces to a Cabibbo-like block diagonal mixing. 

\subsection{Stable gFC with $\mNq{f}\propto \mMq{f}$\label{sSEC:RGE:mqNqStable}}
When one substitutes $\mNq{q}=\alpha_q\mMq{q}$, $\alpha_q\in\mathbb{C}$, in the conditions for one loop RGE stability of gFC given in appendix \ref{APP:RGEdetails}, solving them for $\alpha_u$, $\alpha_d$, reduces to finding solutions of
\begin{equation}
(1+\alpha_d\alpha_u)(\alpha_u^\ast-\alpha_d)=0\,,
\label{eq:FCstable:MqNq:00}
\end{equation}
that is $\alpha_u=-\alpha_d^{-1}$ or $\alpha_u^\ast=\alpha_d$. In both cases, there is a basis for the scalars \cite{Ferreira:2010xe}
\begin{equation}
\begin{pmatrix}H_1^\prime\\ H_2^\prime\end{pmatrix}=
\frac{1}{\sqrt{1+\abs{\alpha_d}^2}}
\begin{pmatrix}1 & \alpha_d\\ -\alpha_d^\ast & 1\end{pmatrix}
\begin{pmatrix}H_1 \\ H_2\end{pmatrix},
\label{eq:FCstable:MqNq:01}
\end{equation}
with $H_1$ and $H_2$ in \refEQ{eq:HiggsBasis:00}, such that in \refEQ{eq:Yukawa:01} $H_1^\prime$ couples only to $d_R^0$ while $H_2^\prime$ couples only to $u_R^0$ for $\alpha_u=-\alpha_d^{-1}$; for $\alpha_u=\alpha_d^\ast$, $H_1^\prime$ couples to $d_R^0$ and $u_R^0$, while $H_2^\prime$ does not. These cases are none other than the 2HDM of type II and I respectively. For the particular case $\alpha_u=\alpha_d^\ast=0$, the scalar doublet which has a zero vacuum expectation value has vanishing Yukawa couplings: this is the Inert 2HDM \cite{Deshpande:1977rw}.

\subsection{Stable gFC in the Lepton Sector\label{sSEC:RGE:Leptons}}
The one loop RGE of the lepton Yukawa couplings in \refEQ{eq:YukawaLeptons:01} reads \cite{Cheng:1973nv,Grimus:2004yh}
\begin{equation}
\der\Yl{\alpha}=
a_\ell\Yl{\alpha}+\sum_{\rho=1}^{n=2}T^{\ell}_{\alpha,\rho}\Yl{\rho}+\sum_{\rho=1}^{n=2}\left(\Yl{\alpha}\Yld{\rho}\Yl{\rho}+\frac{1}{2}\Yl{\rho}\Yld{\rho}\Yl{\alpha}\right)\quad
\text{with }T^{\ell}_{\alpha,\rho}=T^{d}_{\alpha,\rho}\,,
\label{eq:RGE:Yl:00}
\end{equation}
where $a_\ell=-\frac{9}{4}g^2-\frac{15}{4}g^{\prime 2}$. With $\YlH{L}=\sum_{\rho=1}^{n=2}\Yl{\rho}\Yld{\rho}$, $\YlH{R}=\sum_{\rho=1}^{n=2}\Yld{\rho}\Yl{\rho}$, 
\begin{equation}
\der\Yl{\alpha}=a_\ell\Yl{\alpha}+\sum_{\rho=1}^{n=2}T^{\ell}_{\alpha,\rho}\Yl{\rho}+\Yl{\alpha}\YlH{R}+\frac{1}{2}\YlH{L}\Yl{\alpha}\,.
\label{eq:RGE:Yl:01}
\end{equation}
The crucial difference in the leptonic sector is that, following \refEQ{eq:RGE:Yl:01},
\begin{multline}
\der(\Yl{\alpha}\Yld{\beta})=
2a_\ell\Yl{\alpha}\Yld{\beta}+\sum_{\rho=1}^{n=2}\left(T^{\ell}_{\alpha,\rho}\Yl{\rho}\Yld{\beta}+T^{\ell\ast}_{\beta,\rho}\Yl{\alpha}\Yld{\rho}\right)+2\Yl{\alpha}\YlH{R}\Yld{\beta}\\
 +\frac{1}{2}\left(\YlH{L}\Yl{\alpha}\Yld{\beta}+\Yl{\alpha}\Yld{\beta}\YlH{L}\right),
\end{multline}
and thus it is clear that, if $\{\Yl{\alpha}\Yld{\beta}\}_{\alpha,\beta=1,2}$ is abelian, then
\begin{equation}
\der\COMM{\Yl{\alpha}\Yld{\beta}}{\Yl{\gamma}\Yld{\delta}}=\COMM{\der(\Yl{\alpha}\Yld{\beta})}{\Yl{\gamma}\Yld{\delta}}+\COMM{\Yl{\alpha}\Yld{\beta}}{\der(\Yl{\gamma}\Yld{\delta})}=0\,.
\end{equation}
Similarly,
\begin{equation}
\der(\Yld{\alpha}\Yl{\beta})=
2a_\ell\Yld{\alpha}\Yl{\beta}+\sum_{\rho=1}^{n=2}\left(T^{\ell\ast}_{\alpha,\rho}\Yld{\rho}\Yl{\beta}+T^{\ell}_{\beta,\rho}\Yld{\alpha}\Yl{\rho}\right)+\YlH{R}\Yld{\alpha}\Yld{\beta}+\Yld{\alpha}\Yld{\beta}\YlH{R}+\Yld{\alpha}\YlH{L}\Yl{\beta}\,,
\end{equation}
and thus, if $\{\Yld{\alpha}\Yl{\beta}\}_{\alpha,\beta=1,2}$ is abelian, then
\begin{equation}\label{eq:leptonRGEstable:00}
\der\COMM{\Yld{\alpha}\Yl{\beta}}{\Yld{\gamma}\Yl{\delta}}=\COMM{\der(\Yld{\alpha}\Yl{\beta})}{\Yld{\gamma}\Yl{\delta}}+\COMM{\Yld{\alpha}\Yl{\beta}}{\der(\Yld{\gamma}\Yl{\delta})}=0\,.
\end{equation}
That is, if the Yukawa couplings of leptons are gFC, as in \refEQ{eq:FC:Matrices:00}, this is not altered by the RGE: general flavour alignment is one-loop stable in the lepton sector. This can be directly traced back to the absence of right-handed neutrinos and Yukawa couplings involving them in \refEQ{eq:YukawaLeptons:01}, in clear contrast with the quark sector.
This result represents a generalization of previous results restricted to the so called aligned case and pointed out in \cite{Botella:2015yfa}, in agreement with the findings of \cite{Braeuninger:2010td,Jung:2010ik}: \emph{at one loop level the charged lepton sector remains general Flavour Conserving in full generality without any additional constraint}.
To be specific and going to the simplest aligned cases, type I, II, X and Y models are defined in the quark sector by
\begin{equation}\label{eq:typesIIIXY:00}
\text{Type I,X}\ \left\{\begin{matrix}\mNd=\text{cot}\beta\,\mMd,\\ \mNu=\text{cot}\beta\,\mMu,\end{matrix}\right.\qquad 
\text{Type II,Y}\ \left\{\begin{matrix}\mNd=-\text{tan}\beta\,\mMd,\\ \mNu=\text{cot}\beta\,\mMu.\end{matrix}\right.
\end{equation}
The fact that the leptonic sector alignment was known to be stable under RGE implies that one could analyse the experimental data with previous equation together with the more general leptonic structure ($\Yl{2}=\xi_\ell e^{-i\theta}\Yl{1}$)
\begin{equation}\label{eq:typesIIIXY:01}
\mNl=\text{cot}\beta\left(\frac{-\tan\beta+\xi_\ell}{\text{cot}\beta+\xi_\ell}\right)\mMl
\end{equation}
in the framework of a model one loop stable under RGE. This would include in a single analysis both type I and X or type II and Y. Note that with the appropriate limits $\xi_\ell\to 0$ or $\xi_\ell\to\infty$ one recovers the four models.
Equation \eqref{eq:leptonRGEstable:00} implies the new more general result that the models implemented by \refEQ{eq:typesIIIXY:00} together with an \emph{arbitrary} diagonal $\mNl$ (not just with \refEQ{eq:typesIIIXY:01}) are one loop stable under RGE.

\subsection{Stable gFC with Cabibbo-like mixing\label{sSEC:RGE:CabibboStable}}
The CKM matrix has a hierarchical structure; keeping only the largest mixing, it has the form
\begin{equation}
\V{\theta_c}=\begin{pmatrix}\cos\theta_c & \sin\theta_c & 0\\ -\sin\theta_c & \cos\theta_c & 0\\ 0 & 0& 1\end{pmatrix}
\label{eq:CabibboMix:00}
\end{equation}
with $\theta_c\simeq 0.22$ the Cabibbo mixing angle. It is interesting to analyse the question of one loop RGE stability of gFC conditions with $\CKM\to\V{\theta_c}$ in  \refEQ{eq:CabibboMix:00}. First, it is interesting on its own to know if this simplified mixing allows for some stable gFC scenario; second, if that is the case, in terms of those $\mNq{q}$ matrices, the deviations of gFC produced by the RGE would be controlled by the initial deviations of the complete CKM matrix from $\V{\theta_c}$, the subleading mixings.\\ %
One should first notice that, since $\V{\theta_c}$ decouples the third quark generation, $\nq{}{b}$ and $\nq{}{t}$ are expected to remain free parameters. Then, since the only remaining stability conditions concern elements $(a,b)=(1,2)$ or $(2,1)$, all the mixing combinations $\Vc{qa}\V{qb}$, $\V{aq}\Vc{bq}$ equal either $\cos\theta_c\sin\theta_c$ or $-\cos\theta_c\sin\theta_c$, and thus the dependence of the stability conditions on $\theta_c$ disappears. \\ %
Two classes of stable gFC scenarios follow from the discussion in section \ref{sSEC:RGE:mqNqStable}. The first, with
\begin{equation}
\mNd=\text{diag}(\alpha\, m_d,\,\alpha\, m_s,\,\nq{}{b}),\quad \mNu=\text{diag}(\alpha^\ast m_u,\,\alpha^\ast m_c,\nq{}{t}\,),
\label{eq:CabibboStableSolTypeI:00}
\end{equation}
corresponds to a type I 2HDM for the first two generations, while $\nq{}{b}$ and $\nq{}{t}$ are free (and thus $\mMq{q}^{-1}\mNq{q}\neq\alpha_q\id$). Some particular limit -- the extreme chiral limit -- of \refEQ{eq:CabibboStableSolTypeI:00} was already obtained in \cite{Botella:2015yfa} to justify $\CKM\simeq \id$. The second is
\begin{equation}
\mNd=\text{diag}(\alpha\, m_d,\,\alpha\, m_s,\,\nq{}{b}),\quad \mNu=\text{diag}(-\alpha^{-1} m_u,\,-\alpha^{-1} m_c,\nq{}{t}\,),
\label{eq:CabibboStableSolTypeII:00}
\end{equation}
which corresponds instead to a type II 2HDM for the first two generations (with free $\nq{}{b}$, $\nq{}{t}$ and $\mMq{q}^{-1}\mNq{q}\neq\alpha_q\id$ too). In addition to \refEQS{eq:CabibboStableSolTypeI:00}--\eqref{eq:CabibboStableSolTypeII:00}, one can check that
\begin{equation}
\mNd=\text{diag}(e^{i\varphi_d}m_s,\,e^{i\varphi_d}m_d,\,\nq{}{b}),\quad \mNu=\text{diag}(e^{i\varphi_u}m_c,\,e^{i\varphi_u}m_u,\nq{}{t}\,),
\label{eq:CabibboStableSol:00}
\end{equation}
with arbitrary real $\varphi_d$, $\varphi_u$ (and again, arbitrary complex $\nq{}{b}$ and $\nq{}{t}$), gives indeed another stable gFC scenario where $\mNq{q}$ and  $\mMq{q}$ are not even proportional in the first two generations sector. 

%
\section{Phenomenology\label{SEC:Phenomenology}}

\subsection{General considerations\label{sSEC:Phen:Gen}}
The Yukawa interactions in \refEQS{eq:Yukawas:Neutral:03} or \eqref{eq:Yukawas:Neutral:04}, together with the absence of tree level FCNC parameterised in \refEQ{eq:FC:Matrices:00}, have interesting phenomenological consequences in different observables, since they may produce deviations from SM expectations. Those windows on New Physics in different observables are, of course, related: they are controlled by the parameters $\nq{}{j}$ in \refEQ{eq:FC:Matrices:00}, by the values of the masses $m_{\cH}$, $m_{\nH}$, $m_{\nA}$, and by the mixings in the scalar sector, $\RS{ij}$ in \refEQ{eq:Yukawas:Neutral:03}. In the following we consider for simplicity the CP conserving case in \refEQ{eq:ScalarCPconserving:00}. Our interest lies on the parameters $\nq{}{j}$ in \refEQ{eq:FC:Matrices:00}. Among the observables of interest, those that (i) involve the lowest number of new non-SM parameters and (ii) provide direct constraints from existing measurements, are the following.
\begin{itemize}
\item Observables probing the couplings of the 125 GeV Higgs-like scalar, that we identify with $\nh$, that is (i) production mechanisms and (ii) decay modes. In addition to the $\nq{}{j}$ parameters, they involve one extra parameter, the mixing $\cab$ if there is no CP violation in the scalar sector; in the general case, two independent mixings are involved.
\item Observables probing the couplings of the charged scalar $\cH$, in particular effects of $\cH$ in flavour changing processes where the SM contributions involve virtual $W^\pm$ exchange like (i) tree level decays, modifying for example the expected universality of weak interactions, and (ii) one loop FCNC processes like neutral meson mixings and rare decays. These observables, besides the $\nq{}{j}$ parameters, depend on the mass $m_{\cH}$ (and no dependence on the neutral scalar mixings).
\end{itemize}
We concentrate in the rest of this work on the flavour conserving observables related to $\nh$: besides probing the gFC matrices in \refEQ{eq:FC:Matrices:00}, the bounds they impose also apply to the same flavour conserving couplings of a general 2HDM.\\ %
Before addressing the different constraints related to experiment, one can formulate a first theoretical requirement on the perturbativity of the Yukawa couplings:
\begin{equation}\label{eq:perturbativity:00}
\frac{\abs{\nq{}{j}}}{v}\leq \mathcal O(1).
\end{equation}
The precise value adopted in \refEQ{eq:perturbativity:00}, for example $\mathcal O(1)\to 1$ or $\sqrt{4\pi}$, is not expected to be specially relevant: other phenomenological requirements will be, typically, more restrictive. There is, however, an exception: the ``decoupling limit'' \cite{Gunion:2002zf} of the 2HDM, in which $\sab\to 1$ ($\cab\to 0$) removes the non-SM effects from the $\nh$ couplings (while $m_{\cH}\gg v$ suppresses $\cH$ mediated non-SM effects), leaving the perturbativity requirement as the only effective constraint. One may further argue that having either $m_j\ll \abs{\nq{}{j}}$ or $m_j\gg \abs{\nq{}{j}}$, involves fine tuning between quantities of very different nature: both $m_j$ and $\nq{}{j}$ are linear combinations, controlled by $\beta$, of Yukawa couplings (times $v$), but $\beta$ originates in the scalar potential, meaning that very disparate values of $m_j$ and $\nq{}{j}$ involve significant cancellations in one or the other, unless $\beta\to 0$ or $\beta\to \pi/2$. For the sake of clarity, we will only consider \refEQ{eq:perturbativity:00} and ignore the previous concerns about eventual fine tuning.

\subsection{Production and decay of $\nh$\label{sSEC:Phen:h}}

For the observables related to $\nh$, one should consider constraints on $\nq{}{j}$ and $\cab$ arising from $\nh$ production and decay processes at the LHC \cite{Khachatryan:2016vau}. In connection to them, additional attention should be paid to the decays of $\nh$ into light fermions since enhanced decays into light fermions can increase the total width and modify the precise SM pattern of branching ratios. The cross sections for direct $q\bar q\to \nh$ production is also important, since large couplings of $\nh$ to light quarks, in combination with the luminosities given by the parton distribution functions, could significantly increase them.\\ %
Before addressing the Yukawa couplings themselves, we recall that, owing to the mixing in the scalar sector, the couplings $\nh VV$ ($V=W,Z$) are modified with respect to the SM as
\begin{equation}\label{eq:HiggsVectorFactor:00}
\nh VV,\qquad \text{SM}:\, m_V\ \mapsto\ \text{gFC-2HDM}:\, \sab m_V\,.
\end{equation}
These couplings are involved in vector boson fusion (VBF) and associated production mechanisms, and in decays $\nh\to VV^{*}$.\\ 
For the different couplings to fermions $\mathscr L_{\nh ff}=-\nh\bar f(a_f+ib_f\gamma_5)f$ in \refEQ{eq:Yukawas:Neutral:04}, we have a scalar term $a_f$, straightforward to compare with the SM one,
\begin{equation}\label{eq:HiggsScalarFactor:00}
a_f:\qquad \text{SM}:\, {m_j}/{v}\ \mapsto\ \text{gFC-2HDM}:\, (\sab m_j+\cab\re{n_j})/{v}\,,
\end{equation}
and a pseudoscalar term $b_f$ absent in the SM,
\begin{equation}\label{eq:HiggsPseudoScalarFactor:00}
b_f:\qquad \text{SM}:\, 0\ \mapsto\ \text{gFC-2HDM}:\, \cab\im{n_j}/{v}\,.
\end{equation}
We now discuss in turn decay and production processes.

\subsubsection{Decays of $\nh$\label{ssSEC:h:decay}}

The decay width $\nh\to \bar ff$, for a generic Yukawa interaction $\mathscr L_{\nh ff}=-\nh\bar f(a_f+ib_f\gamma_5)f$, is, at tree level,
\begin{equation}\label{eq:HiggsFermions:00}
\Gamma(\nh\to \bar ff) = N_c(f) \frac{m_{\nh}}{8\pi} \sqrt{1-4\frac{m_f^2}{m_{\nh}^2}} \left[ \left(1-4\frac{m_f^2}{m_{\nh}^2}\right)\abs{a_f}^2 + \abs{b_f}^2 \right],
\end{equation}
with $N_c=3$ for quarks and $N_c=1$ for leptons; neglecting $4m_f^2/m_{\nh}^2\ll 1$,
\begin{equation}\label{eq:HiggsFermions:00b}
\Gamma(\nh\to \bar ff) = N_c(f) \frac{m_{\nh}}{8\pi} \left[ \abs{a_f}^2 + \abs{b_f}^2 \right].
\end{equation}
With \refEQS{eq:HiggsScalarFactor:00}--\eqref{eq:HiggsPseudoScalarFactor:00},
\begin{multline}\label{eq:HiggsFermions:01}
\Gamma(\nh\to \bar ff)_{\text{SM}}:\, \frac{N_c(f)}{8\pi}\frac{m_{\nh}}{v^2}m_f^2\ \mapsto\\ 
\Gamma(\nh\to \bar ff)_{\text{gFC-2HDM}}:\, \frac{N_c(f)}{8\pi}\frac{m_{\nh}}{v^2}\left[\sab^2 m_f^2 + 2\sab\cab m_f\re{\nq{}{f}} + \cab^2\abs{\nq{}{f}}^2\right]\,.
\end{multline}
The decay $\nh\to \gamma\gamma$, central in the discovery of the Higgs, has an amplitude controlled in the SM by two interfering contributions, the one loop triangle diagrams with virtual $W$'s and top quarks. The former is modified according to \refEQ{eq:HiggsVectorFactor:00}. The later is the only relevant one involving quarks in the SM because of the large $\nh\bar tt$ coupling: $m_t/v$; this amplitude is modified according to \refEQ{eq:HiggsScalarFactor:00}. With a pseudoscalar coupling now present, \refEQ{eq:HiggsPseudoScalarFactor:00}, there is an additional contribution which, however, does not interfere with the SM-like top(scalar coupling)+$W$. Furthermore, there are other contributions that one may consider: one due to diagrams with virtual $\cH$'s, and the ones due to other fermions with enhanced couplings to $\nh$ due to sizable $\nq{}{j}$. For the charged scalar, they cannot be neglected if $\cH$ is relatively light, and thus, barring that possibility, we do not consider them. For the remaining fermions, the values of $\cab$ that $\nh\leftrightarrows WW$ decay and production require are typically small ($|\cab|\leq 0.1$), and thus the values of $\nq{}{j}$ that one would need for their contributions to be relevant would be at least $\nq{}{j}\sim m_t$: they would produce huge contributions to the width $\Gamma(\nh)$ or to $\bar qq\to\nh$ production cross sections (see the discussion in section \ref{ssSEC:h:production}), in addition to the perturbativity and fine tuning concerns on the Yukawa couplings already mentioned: we thus ignore them altogether, since they will be rendered negligible once other constraints are considered. The width of $\nh\to \gamma\gamma$ reads
\begin{multline}\label{eq:hgaga:00}
\Gamma(\nh\to\gamma\gamma)_{\text{gFC-2HDM}}=\frac{\alpha^2}{256\pi^3}\frac{m_{\nh}^3}{v^2}\times\\
\left(\ABS{\sum_f N_c(f)Q_f^2 \frac{a_f v}{m_f} A_F(x_f)+\sab A_V(x_W)+g_{\cH}A_S(x_{\cH})}^2+\ABS{\sum_f N_c(f)Q_f^2 \frac{b_f v}{m_f} \hat A_F(x_f)}^2\right),
\end{multline}
with $x_X=4m_X^2/m_{\nh}^2$. The sum over fermions $f$ includes up and down type quarks, with $Q_f=2/3$ and $-1/3$ respectively, and charged leptons with $Q_f=-1$. 
The contribution of the charged scalar $\cH$ corresponds to an interaction $\mathscr L_{\cHp\cHm\nh}=-g_{\cH}\frac{2m_{\cH}^2}{v}\cHp\cHm\nh$. $g_{\cH}$ depends on the details of the scalar potential that we do not address since this contribution can be safely neglected for $m_{\cH}>v$.\\ %
The decay into gluons $\nh\to gg$ proceeds through similar diagrams, with the ones mediated by leptons and by $W$ and $\cH$ bosons absent:
\begin{equation}\label{eq:hgg:00}
\Gamma(\nh\to gg)_{\text{gFC-2HDM}}=\frac{\alpha_S^2}{128\pi^3}\frac{m_{\nh}^3}{v^2}\left(\ABS{\sum_f \frac{a_fv}{m_f}A_F(x_f)}^2+\ABS{\sum_f \frac{b_fv}{m_f} \hat A_F(x_f)}^2\right).
\end{equation}
The couplings $a_f$ and $b_f$ in \refEQS{eq:hgaga:00}--\eqref{eq:hgg:00} appear divided by fermion mass $m_f$ since the $A_F$ and $\hat A_F$ functions are defined including the mass factor of the SM $\nh\bar ff$ vertex.
The loop functions are \cite{Gunion:1989we}
\begin{alignat}{3}
A_F(x)&=-2x\left[1+(1-x)f(x)\right], &\hat A_F(x)&=-2xf(x),\nonumber\\
A_V(x)&=2+3x+3x(2-x)f(x),\quad & A_S(x)&=x(1-xf(x)),
\end{alignat}
where
\begin{equation}
f(x)=\left\{\begin{matrix}\text{arcsin}^2(1/\sqrt{x}),& x\geq 1,\\ -\frac{1}{4}\left[\ln\left(\frac{1+\sqrt{1-x}}{1-\sqrt{1-x}}\right)-i\pi\right]^2,& x<1.\end{matrix}\right.
\end{equation}
The dominant contribution in $\nh\to\gamma\gamma$ comes from $A_V(x_W)=-8.339$. Other representative values of the functions are shown in Table \ref{TAB:TriangleLoopValuesFermion}. It is important to stress that, while QCD corrections to \refEQ{eq:hgaga:00} are small, that is not the case for \refEQ{eq:hgg:00} (see for example \cite{Spira:1995rr}): we account for them by using
\begin{equation}\label{eq:hgg:01}
\Gamma(\nh\to gg)_{\text{gFC-2HDM}}\to\frac{\Gamma(\nh\to gg)_{\text{gFC-2HDM}}}{\Gamma(\nh\to gg)_{\text{SM}}}\times \Gamma(\nh\to gg)_{\text{SM ref.}},
\end{equation}
with $\Gamma(\nh\to gg)_{\text{SM ref.}}=0.351$ MeV the SM reference value from Table \ref{TAB:BRsLHC:SM}, and $\frac{\Gamma(\nh\to gg)_{\text{gFC-2HDM}}}{\Gamma(\nh\to gg)_{\text{SM}}}$ computed according to \refEQ{eq:hgg:00} (for the SM denominator $\frac{a_f v}{m_f}=1$, $b_f=0$). For completeness, reference values of the SM Higgs decays \cite{Dittmaier:2011ti,Dittmaier:2012vm,Heinemeyer:2013tqa,deFlorian:2016spz} are reproduced in Table \ref{TAB:BRsLHC:SM}.\\ %
\begin{table}[h!tb]
\begin{center}
\begin{tabular}{|c||c|c|c|}\hline
$f$ & $t$  & $b$ & $\tau$\\ \hline
$A_F(x_f)$  & {\small $1.3796$} & {\small $-(4.37+4.75i)10^{-2}$} & {\small $-(2.30+2.09i)10^{-2}$}\\ \hline
$\hat A_F(x_f)$  & {\small $2.1010$} & {\small $-(4.78+4.76i)10^{-2}$} & {\small $-(2.46+2.09i)10^{-2}$}\\ \hline\hline
$f$ & $c$ & $s$ & $\mu$\\ \hline
$A_F(x_f)$ & {\small $-(4.87+3.29i)10^{-3}$} & {\small $-(8.99+3.89i)10^{-5}$} & {\small $-(2.53+1.20i)10^{-4}$}\\ \hline
$\hat A_F(x_f)$ & {\small $-(5.07+3.29i)10^{-3}$} & {\small $-(9.15+3.89i)10^{-5}$} & {\small $-(2.59+1.20i)10^{-4}$}\\ \hline
\end{tabular}
\caption{Values of $A_F$ and $\hat A_F$ for charged fermions of the 2$^{\text{nd}}$ and 3$^{\text{rd}}$ generations; running masses at $\mu=m_{\nh}$ \cite{Xing:2011aa} are used.\label{TAB:TriangleLoopValuesFermion}}
\end{center}
\end{table}
%
%
\begin{table}[h!tb]
\begin{center}
\begin{tabular}{|c||c|c|c|c|c|}\hline
\phantom{$\hat I$} Channels $\bar ff$\phantom{$\hat I$} & $\bar bb$  & $\bar \tau\tau$   & $\bar cc$   & $\bar \mu\mu$  & $\bar ss$\\ \hline
BR  & $0.577$ & $6.32\cdot 10^{-2}$ & $2.91\cdot 10^{-2}$ & $2.19\cdot 10^{-4}$ & $2.46\cdot 10^{-4}$\\ \hline\hline
\phantom{$\hat I$} Channels $VV$ \phantom{$\hat I$} & gg  & $WW^{(\ast)}$ & $ZZ^{(\ast)}$ & $\gamma\gamma$ & $\gamma Z$ \\ \hline
BR  & $8.57\cdot 10^{-2}$ & $0.215$ & $2.64\cdot 10^{-2}$ & $2.28\cdot 10^{-3}$ & $1.54\cdot 10^{-3}$\\ \hline
\end{tabular}
\caption{Reference SM Higgs decay branching ratios for $m_{\nh}=125$ GeV; the total width is $\Gamma(\nh)=4.1$ MeV.\label{TAB:BRsLHC:SM}}
\end{center}
\end{table}
\subsubsection{Production of $\nh$\label{ssSEC:h:production}}

In addition to the decay widths, production mechanisms are also modified. Besides VBF and associated production, already commented ( \refEQ{eq:HiggsVectorFactor:00}), the most relevant one is gluon-gluon fusion (ggF) $gg\to \nh$ \cite{Georgi:1977gs}. The elementary process is the reverse of the decay $\nh\to gg$, which is then convoluted with the gluon distribution functions in the proton (in the narrow width approximation production and decay are  related straightforwardly). As in the case of the decay, \refEQ{eq:hgg:01}, we incorporate QCD corrections by normalizing the SM prediction to the reference value in Table \ref{TAB:xSectionsLHC:SM}, which shows  reference cross sections for different production mechanisms \cite{Dittmaier:2011ti,Dittmaier:2012vm,Heinemeyer:2013tqa,deFlorian:2016spz}.\\ %
\begin{table}[h!tb]
\begin{center}
\begin{tabular}{c|c|c|c|c|c|c|}\cline{2-7}
       & ggF   & VBF   & WH     & ZH     & ttH    & bbH\\ \hline
\multicolumn{1}{|c|}{8 TeV}  & 19.27 & 1.578 & 0.7046 & 0.4153 & 0.1293 & 0.2035 \\ \hline
\multicolumn{1}{|c|}{13 TeV} & 43.92 & 3.748 & 1.380  & 0.8696 & 0.5085 & 0.5116 \\ \hline
\multicolumn{1}{|c|}{14 TeV} & 49.47 & 4.233 & 1.522  & 0.9690 & 0.6113 & 0.5805 \\ \hline
\end{tabular}
\caption{Reference SM production cross sections for $m_{\nh}=125$ GeV (in pb).\label{TAB:xSectionsLHC:SM}}
\end{center}
\end{table}

\begin{figure}[h!tb]
\begin{center}
\includegraphics[height=0.25\textwidth]{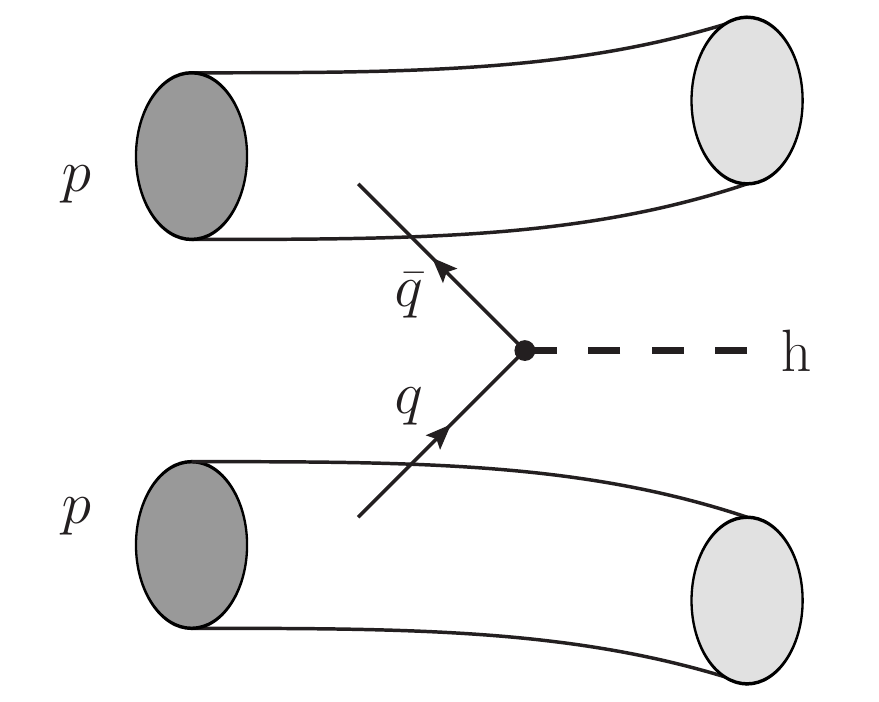}
\caption{$\bar qq\to \nh$ process.\label{FIG:qqh}}
\end{center}
\end{figure}

\noindent We now turn to the direct $\bar qq\to\nh$ production mechanism shown in Figure \ref{FIG:qqh}. The motivation to consider this production mechanism is that, when $\abs{\nq{}{q}}\gg m_q$, the corresponding cross section may become inappropriately large; one is considering light quarks $q\neq t,b$. Sensitivity to enhanced Yukawa couplings of light quarks at the LHC has also been discussed, for example, in \cite{Zhou:2015wra,Perez:2015lra,Yu:2016rvv,Cohen:2017rsk}.

 For a generic Yukawa interaction $\mathscr L_{\nh qq}=-\nh\bar q(a_q+ib_q\gamma_5)q$, the tree level cross section for direct production $pp(\bar qq)\to \nh$ is, in the narrow width approximation,
\begin{equation}\label{eq:pph:xSection:00}
\sigma[pp(\bar qq)\to\nh]=\left(\abs{a_q}^2+\abs{b_q}^2\right)\,\sigma_0(E)\,\mathcal L_{\bar qq}(E)\,,
\end{equation}
where 
\begin{align}
\sigma_0(E)&\equiv 2\frac{\pi}{8 N_c E^2}=\left(\frac{\text{TeV}}{E}\right)^2\,101.8\ \text{pb}\,,\nonumber\\
\mathcal L_{\bar qq}(E)&\equiv \int_{x_0}^1\!\!\!dx\,f^p_{\bar q}(x,Q^2)f^p_{q}(x_0/x,Q^2)\frac{1}{x}\,,
\label{eq:pph:xSection:01}
\end{align}
with $E$ the center of mass proton energy, $f^p_y$ the distribution function of parton $y$ in the proton, $x_0=\frac{m_{\nh}^2}{4E^2}$ and $Q$ is the factorization scale.  

\noindent In Table \ref{TAB:xSectionsLHC:qqh} we collect the values of $\sigma[pp(\bar qq)\to\nh]$ and $\mathcal L_{\bar qq}$ computed \cite{Martin:2009iq} for different quarks\footnote{\refEQ{eq:pph:xSection:00} is obtained using the tree level partonic cross section; furthermore, the results in table \ref{TAB:xSectionsLHC:qqh} are obtained multiplying these simple predictions by a common $\mathcal O(1)$ factor (one for each LHC energy case), chosen such that $\sigma[pp(\bar bb)\to\nh]$ in \refEQ{eq:pph:xSection:00} reproduces the improved reference values in \cite{Dittmaier:2011ti,Dittmaier:2012vm,Heinemeyer:2013tqa,deFlorian:2016spz}. We also take $Q=E/2$.} and setting $\abs{a_q}^2+\abs{b_q}^2=1$ in \refEQ{eq:pph:xSection:00}: this value of the couplings is obviously too large since it effectively corresponds, with respect to the SM, to the change $m_j/v\mapsto v/v$, but it allows for easy use. 
\begin{table}[h!tb]
\begin{center}
\begin{tabular}{c|c|c|c|c|c|c|c|c|c|c|c|c|c|}\cline{2-9}
\phantom{$\hat I$} & \multicolumn{2}{|c|}{$\bar dd$} & \multicolumn{2}{|c|}{$\bar uu$} & \multicolumn{2}{|c|}{$\bar ss$} & \multicolumn{2}{|c|}{$\bar cc$} \\ \cline{2-9}
 & $\mathcal L_{\bar dd}$ & $\sigma$ & $\mathcal L_{\bar uu}$ & $\sigma$ & $\mathcal L_{\bar ss}$ & $\sigma$ & $\mathcal L_{\bar cc}$ & $\sigma$ \\ \hline
\multicolumn{1}{|c|}{8  TeV} & 14.56 & 16.60 & 21.53 & 24.53 & 4.41 & 5.02 & 2.65 & 3.01\\ \hline
\multicolumn{1}{|c|}{13 TeV} & 74.57 & 29.17 & 105.28 & 41.18 & 27.70 & 10.84 & 17.92 & 7.01\\ \hline
\multicolumn{1}{|c|}{14 TeV} & 95.49 & 31.53 & 133.90 & 44.21 & 36.46 & 12.04 & 23.83 & 7.87\\ \hline
\end{tabular}
\caption{$\sigma[pp(\bar qq)\to\nh]$ $(\times 10^3)$ in pb and $\mathcal L_{\bar qq}(E^2)$ $(\times 10^3)$ for different $\bar qq$.\label{TAB:xSectionsLHC:qqh}}
\end{center}
\end{table}
Consider, for illustration, that for the LHC at 8 TeV $\sigma[pp(\bar uu)\to\nh]\sim 10$ pb: one can readily obtain
\begin{equation}\label{eq:uuh:example:00}
\sigma[pp(\bar uu)\to\nh]\sim 10\text{ pb}\ 
\Leftrightarrow\ \abs{a_q}^2+\abs{b_q}^2\sim 7.3\times 10^{-5}\,.
\end{equation}
Although considering $\sigma[pp(\bar uu)\to\nh]\sim 10$ pb may be unrealistic (the \emph{total} production cross section in Table \ref{TAB:xSectionsLHC:SM} for 8 TeV is $\sim 22$ pb), from \refEQ{eq:uuh:example:00}, $\Gamma(\nh\to u\bar u)\sim 1$ MeV: even if it is a significant contribution to the width $\Gamma(\nh)$, it might still be compatible with the overall pattern of Higgs signal strengths. To the knowledge of the authors, there are no dedicated analyses of $\bar qq\to\nh$ ($q\neq b,t$) from which experimental input can be used in this manner. However, it is reasonable to expect that this kind of production potentially ``contaminates'' the analyses of gluon-gluon fusion: in that case, one should add all $\sigma[pp(\bar qq)\to\nh]$ contributions for light $q$ to the gluon-gluon fusion cross section when analysing Higgs signal strengths. It is then clear that bounds more stringent than \refEQ{eq:uuh:example:00} would follow for the sum over all the different channels involved.  
The simple connection among the decays $\nh\to \bar qq$ and the $\bar qq\to\nh$ production mechanism -- in the narrow width approximation -- that follows from \refEQS{eq:HiggsFermions:01} and \eqref{eq:pph:xSection:00}, is
\begin{equation}\label{eq:xSectionVSwidth:00}
\frac{\sigma[pp(\bar qq)\to\nh]\,/\,1\text{pb}}{\Gamma(\nh\to\bar qq)\,/\,1\text{MeV}}=6.825\,\left(\frac{\text{TeV}}{E}\right)^2\left(\frac{\mathcal L_{\bar qq}(E)}{10^3}\right)\,,
\end{equation}
which allows for easy comparison of the relative strengths of the constraints imposed by $\bar qq\to\nh$ production and $\nh\to\bar qq$ decay for light quarks $q$. 

\subsubsection{Constraints\label{ssSEC:h:constraints}}
The main source of experimental constraints that we use is the combined analysis of LHC-Run I data from the ATLAS and CMS collaborations in  \cite{Khachatryan:2016vau}, which provides detailed information on (production) $\times$ (decay) of the 125 GeV Higgs $\nh$ for
\begin{itemize}
\item production: ggF, VBF, associated $W\nhÇ$, $Z\nh$, and $tt\nh$;
\item decay: $\nh\to\gamma\gamma$, $ZZ$, $WW$, $\tau\bar\tau$ and $b\bar b$.
\end{itemize}
Results from LHC-Run II in specific channels like (associated $V\nh$) $\times$ ($\nh\to b\bar b$) \cite{Aaboud:2017xsd,Sirunyan:2017elk}, or (ggF+VBF) $\times$ ($\nh\to\tau\bar\tau$) \cite{Sirunyan:2017khh} are starting to improve over \cite{Khachatryan:2016vau}. 
One should also consider off-shell (ggF+VBF)$\to\nh^{(\ast)}\to WW^{(\ast)}$ constraints on the total width $\Gamma(\nh)$ \cite{Kauer:2012hd}, even if they are still weak \cite{Aad:2015xua,Khachatryan:2016ctc}. 
Finally, dedicated studies like \cite{Khachatryan:2014aep,Aaboud:2017ojs} put useful bounds on $\nh\to\mu^+\mu^-,e^+e^-$.

\subsection{Electric dipole moments\label{sSEC:EDMs}}
\begin{figure}[h!tb]
\begin{center}
\includegraphics[height=0.225\textwidth]{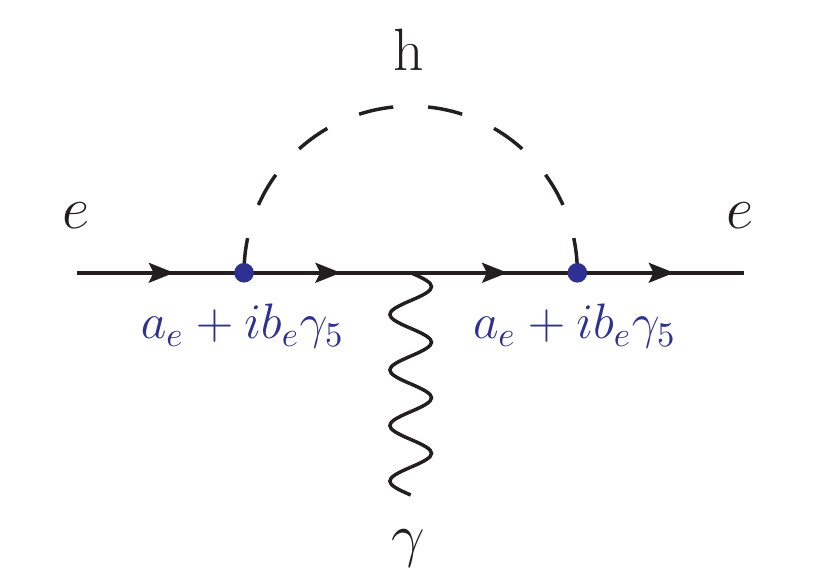}
\caption{$\nh$ mediated contribution to $d_e$ at one loop.\label{FIG:de}}
\end{center}
\end{figure}

As discussed at the end of section \ref{SEC:GEN}, non-real $\mNq{f}$ matrices are a source of CP violation in scalar-fermion interactions, which can induce electric dipole moments (EDMs). Consider for example an electron-Higgs coupling $\mathscr L_{\nh ee}=-\nh\,\bar e(a_e+ib_e\gamma_5)e$; the one loop diagram in Figure \ref{FIG:de} gives a contribution to the electron EDM $d_e$:
\begin{equation}\label{eq:de:00}
d_e=\frac{3m_e a_e b_e}{16\pi^2 m_{\nh}^2}\left(1+\mathcal O\left(\frac{m_e^2}{m_{\nh}^2}\right)\right).
\end{equation}
It is to be noticed that, for $a_e\sim b_e\sim m_e/v$, \refEQ{eq:de:00} gives $d_e\sim 10^{-34}$e$\cdot$cm. When $\abs{a_e}$, $\abs{b_e}\gg m_e/v$ are a priori allowed, up to the effect of other constraints, a significant enhancement in $d_e$ can be expected. For current experimental bounds $\abs{d_e}<10^{-27}$e $\cdot$cm, considering only this contribution gives
\begin{equation}\label{eq:de:01}
a_e b_e<8\times 10^{-5},
\end{equation}
or, with \refEQS{eq:HiggsScalarFactor:00}--\eqref{eq:HiggsPseudoScalarFactor:00} and neglecting $m_e$ with respect to $\cab\nq{}{e}$,
\begin{equation}\label{eq:de:02}
\cab^2\re{\nq{}{e}}\im{\nq{}{e}}< 5\text{ GeV}^2.
\end{equation}
Anticipating results from section \ref{sSEC:Phen:Num}, in particular Figure \ref{sFIG:ble:ale}, it is clear that the bounds imposed by the LHC results are more stringent than \refEQ{eq:de:02}. It should also be noticed that including contributions analog to Figure \ref{FIG:de} with $\nh\to\nH,\nA$, gives
\begin{equation}\label{eq:de:03}
\re{\nq{}{e}}\im{\nq{}{e}}\left(\cab^2+\sab^2\frac{m_{\nh}^2}{m_{\nH}^2}+\frac{m_{\nh}^2}{m_{\nA}^2}\right)< 5\text{ GeV}^2,
\end{equation}
and does not change this conclusion. Furthermore, one loop contributions with virtual $\cH$ and neutrinos are suppressed.\\ %
\noindent It is well known that two loop ``Barr-Zee'' \cite{Barr:1990vd,Leigh:1990kf,Gunion:1990ce,Chang:1990sf,Kao:1992jv,Ilisie:2015tra} contributions can be significant: studies such as \cite{Inoue:2014nva,Altmannshofer:2015qra} address such constraints on CP violating Higgs-fermion couplings. However, those contributions involve different $\nq{}{f}$ couplings simultaneously, together with the masses of the different scalars, preventing a simple translation into bounds on a single parameter. It is to be noticed too that cancellations among different diagrams in that class may occur \cite{Botella:2015hoa,Bian:2014zka}. Including such kind of analysis is beyond the scope of this work; in any case one should keep in mind that the analysis of EDMs may have some impact on the results of section \ref{sSEC:Phen:Num}. The previous discussion also applies to the EDMs of the $u$ and $d$ quarks and the experimental constraints that the neutron EDM bounds impose, including, in addition, the impact of QCD effects \cite{Jung:2013hka}.

\subsection{Analysis\label{sSEC:Phen:Num}}
\begin{figure}[h!tb]
\begin{center}
\subfigure[$u$\label{sFIG:nqu:cab}]{\includegraphics[height=0.3\textwidth]{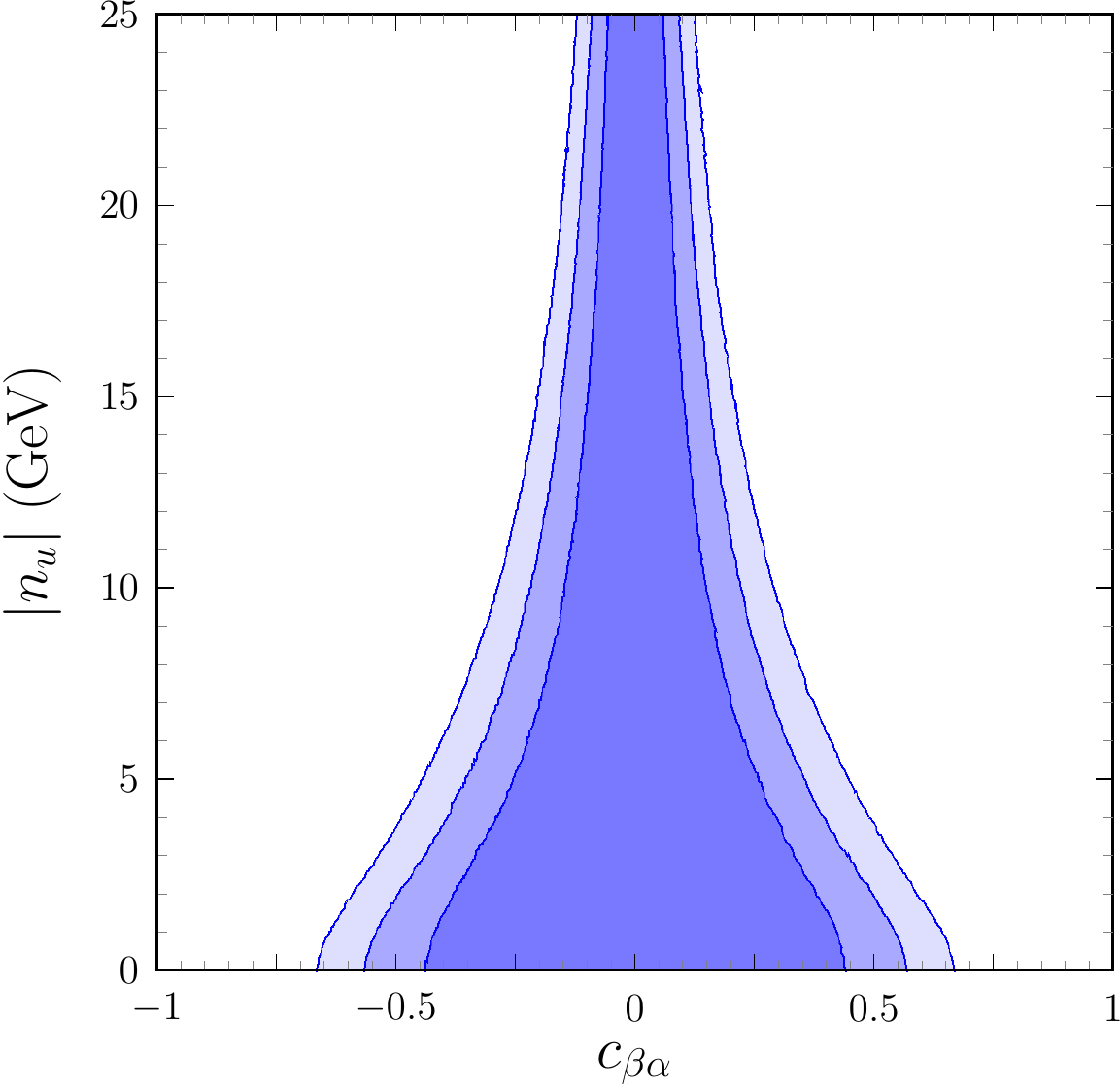}}\quad
\subfigure[$c$\label{sFIG:nqc:cab}]{\includegraphics[height=0.3\textwidth]{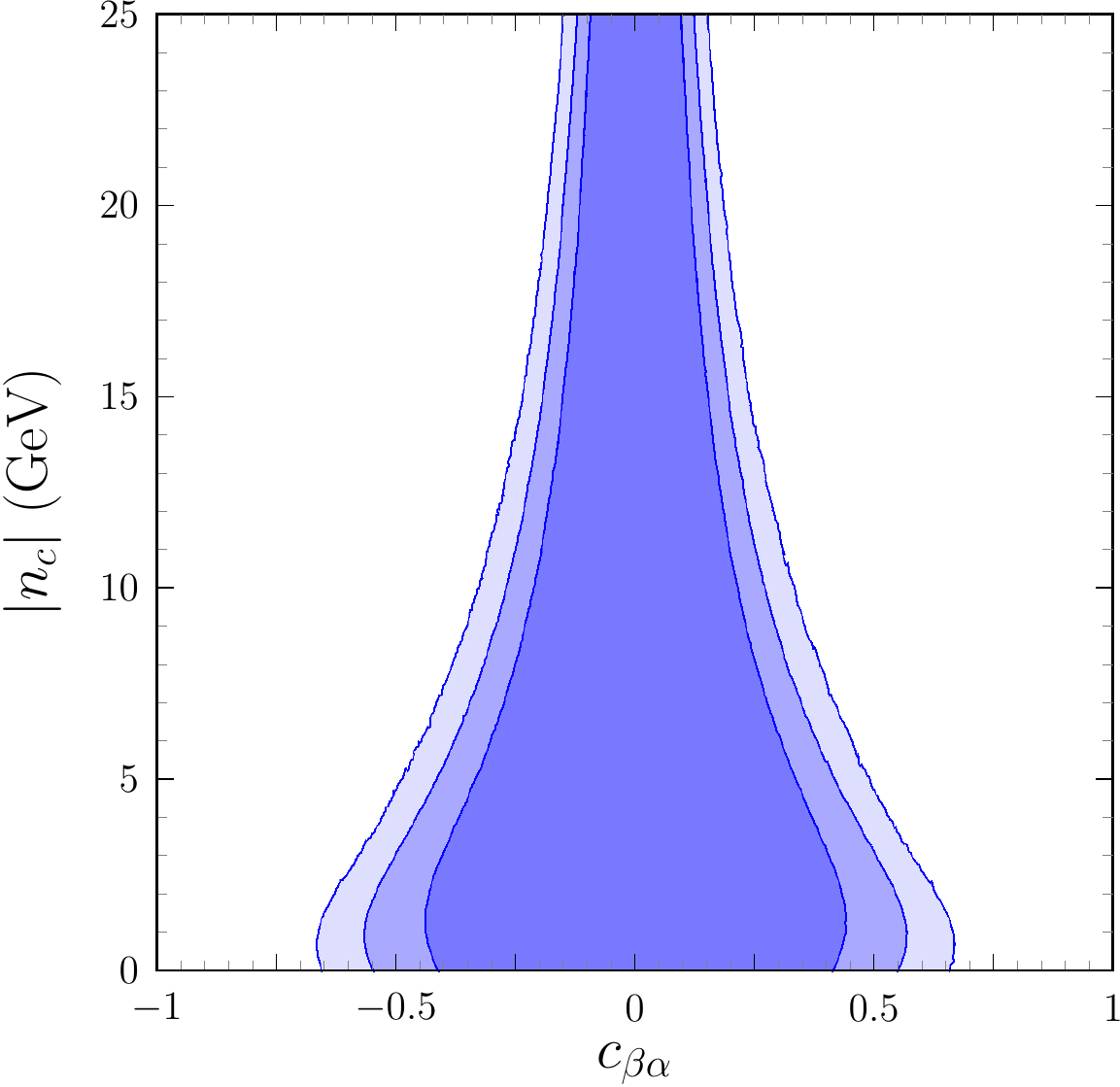}}\quad
\subfigure[$t$\label{sFIG:nqt:cab}]{\includegraphics[height=0.3\textwidth]{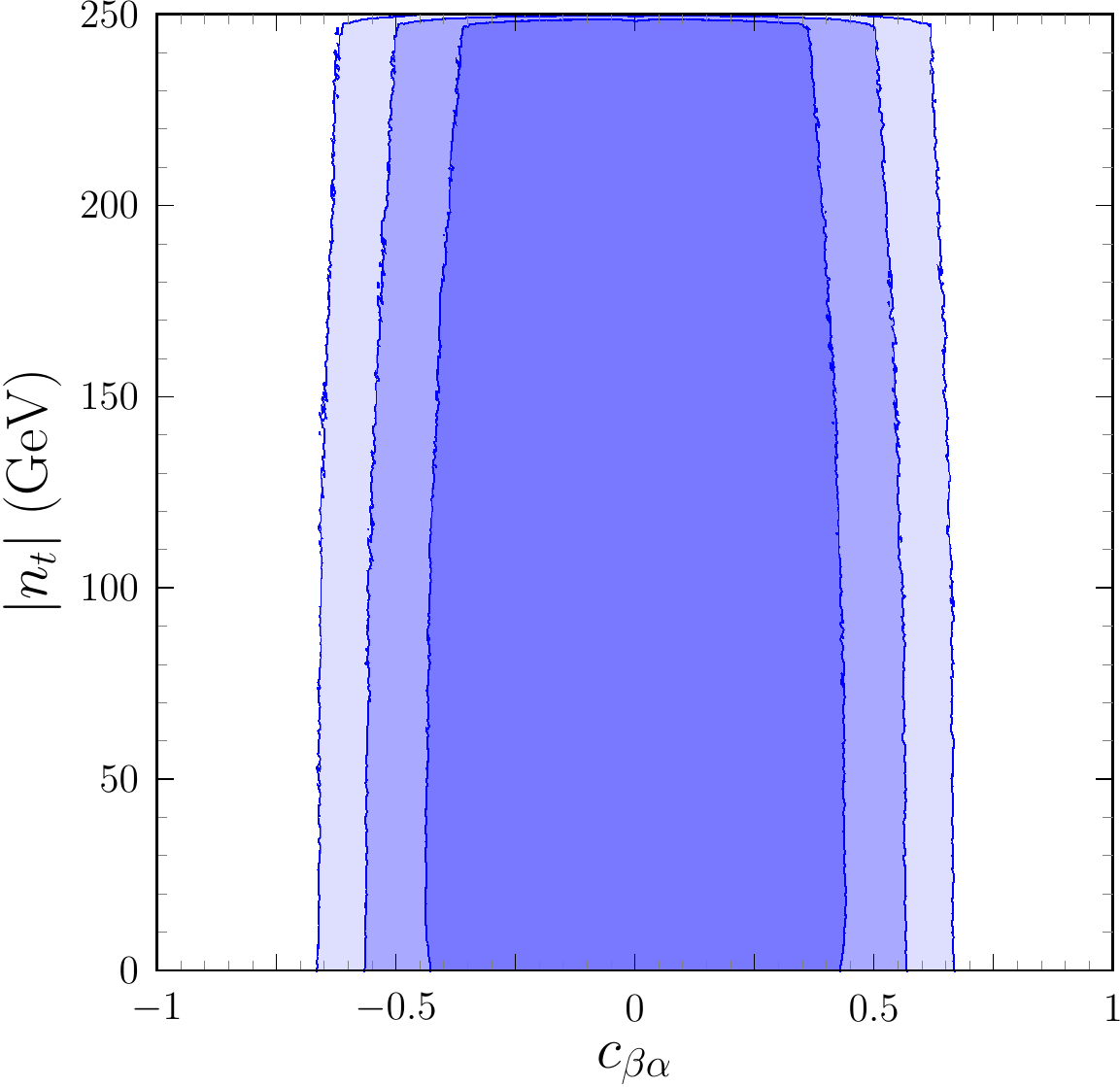}}\\
\subfigure[$d$\label{sFIG:nqd:cab}]{\includegraphics[height=0.3\textwidth]{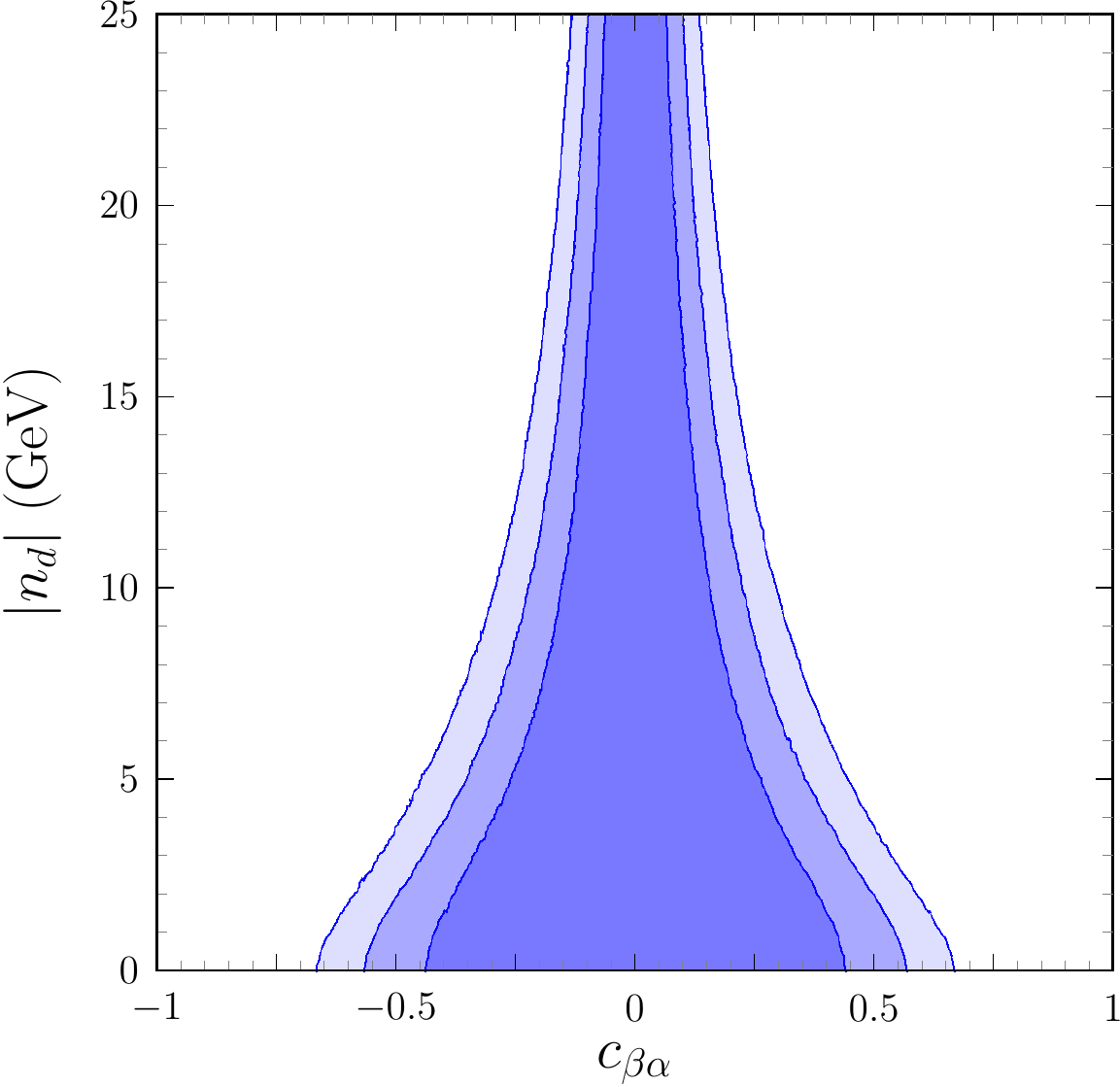}}\quad
\subfigure[$s$\label{sFIG:nqs:cab}]{\includegraphics[height=0.3\textwidth]{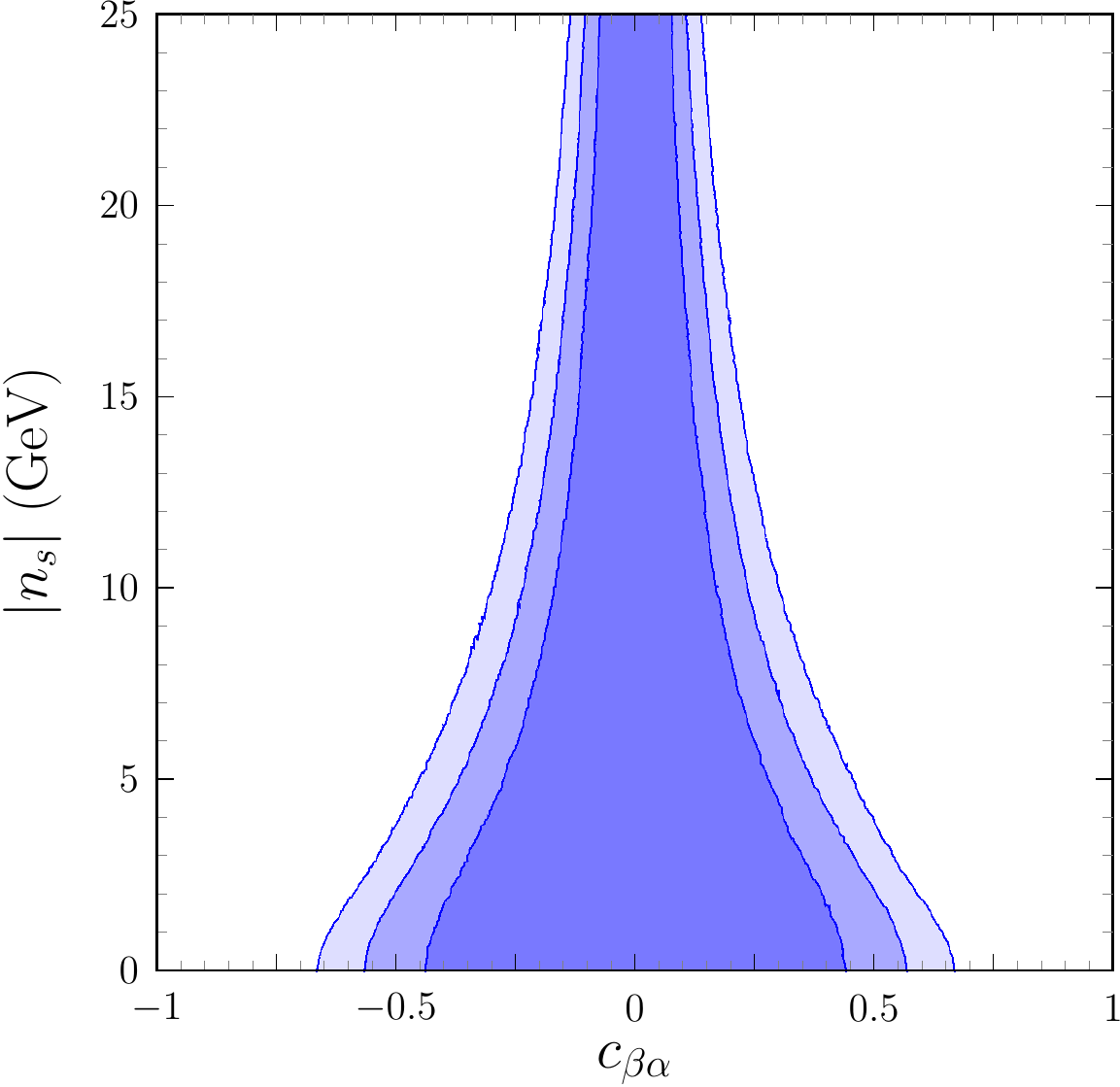}}\quad
\subfigure[$b$\label{sFIG:nqb:cab}]{\includegraphics[height=0.3\textwidth]{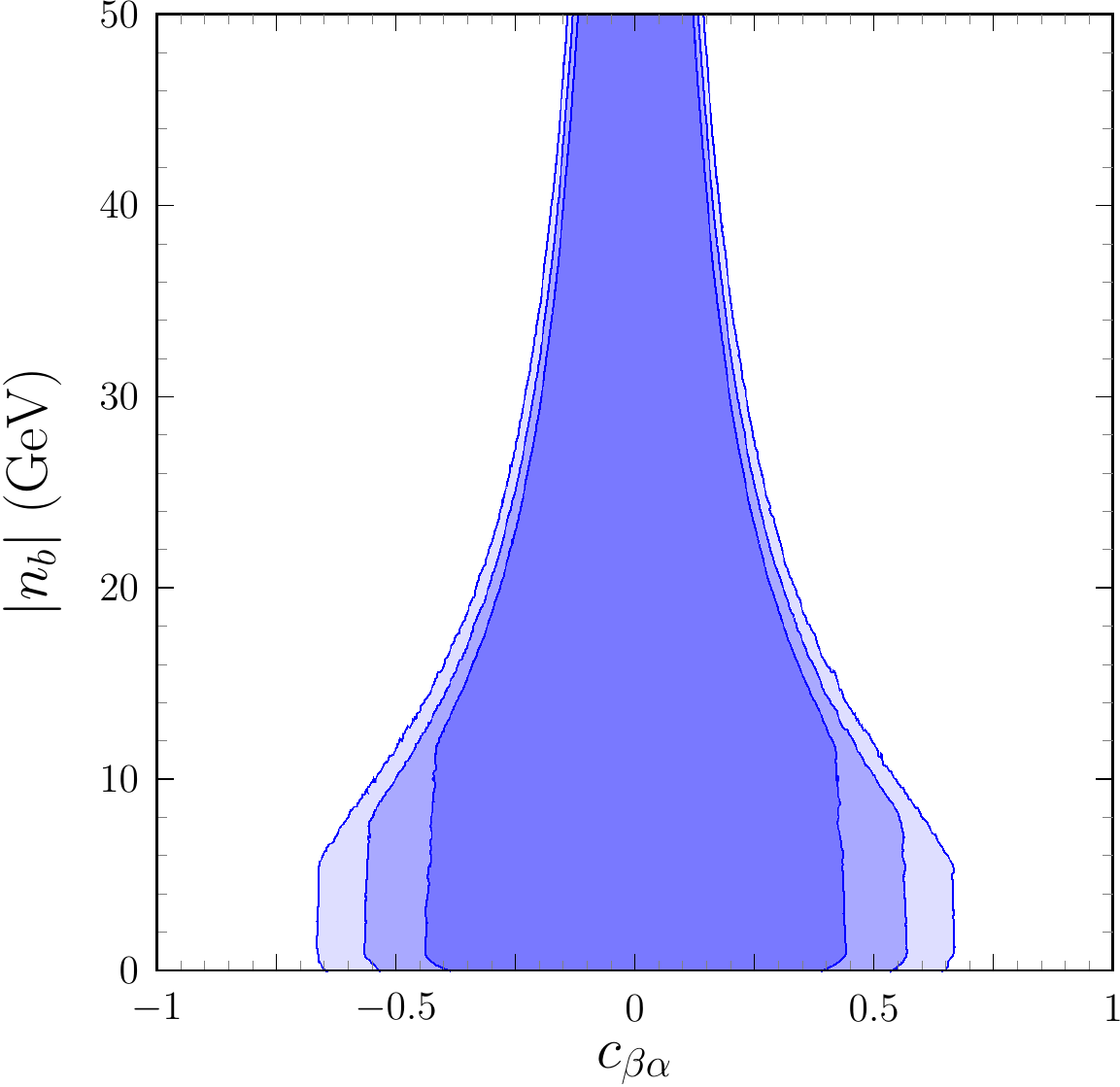}}\\
\subfigure[$e$\label{sFIG:nle:cab}]{\includegraphics[height=0.3\textwidth]{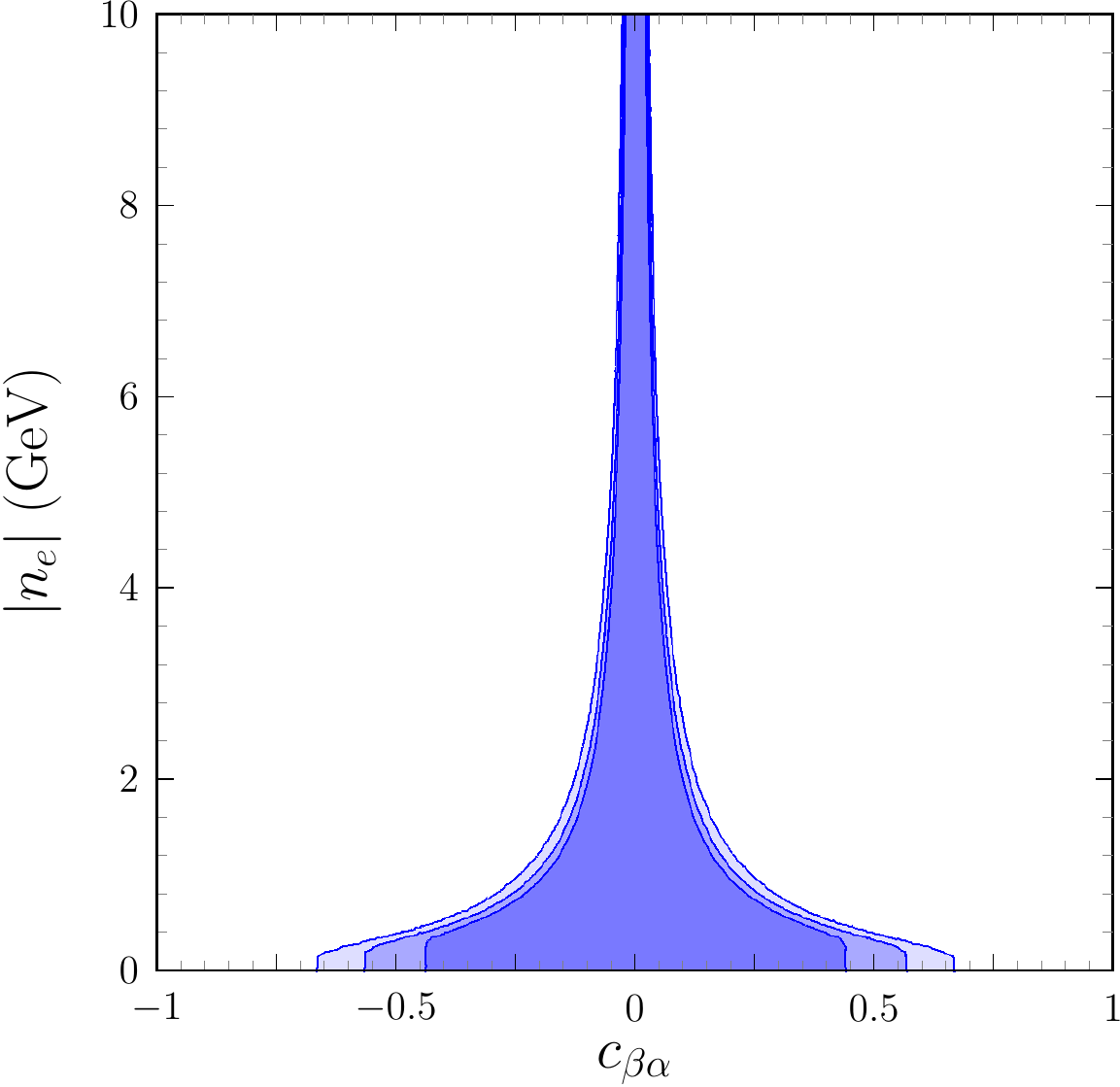}}\quad
\subfigure[$\mu$\label{sFIG:nlm:cab}]{\includegraphics[height=0.3\textwidth]{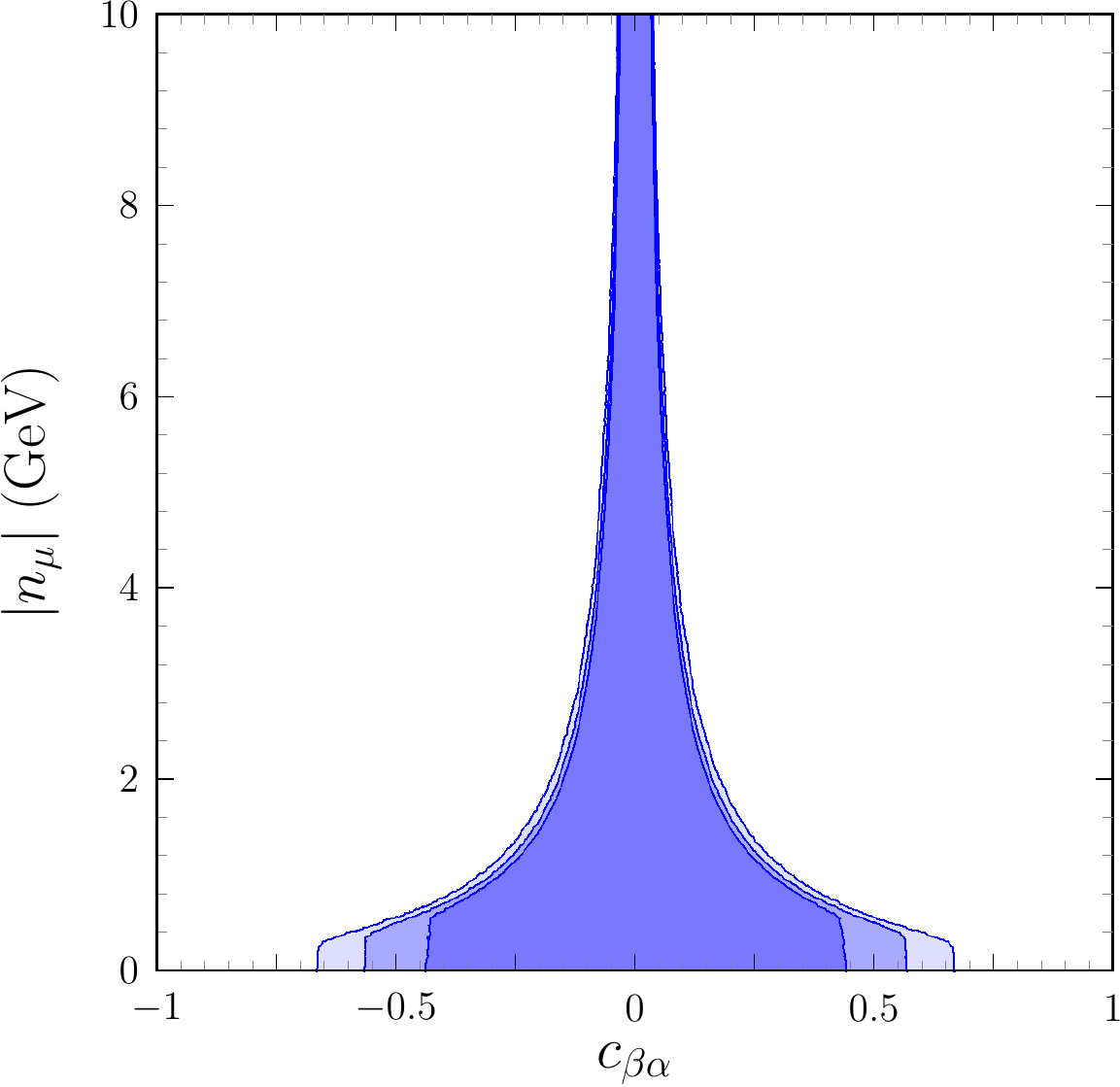}}\quad
\subfigure[$\tau$\label{sFIG:nlt:cab}]{\includegraphics[height=0.3\textwidth]{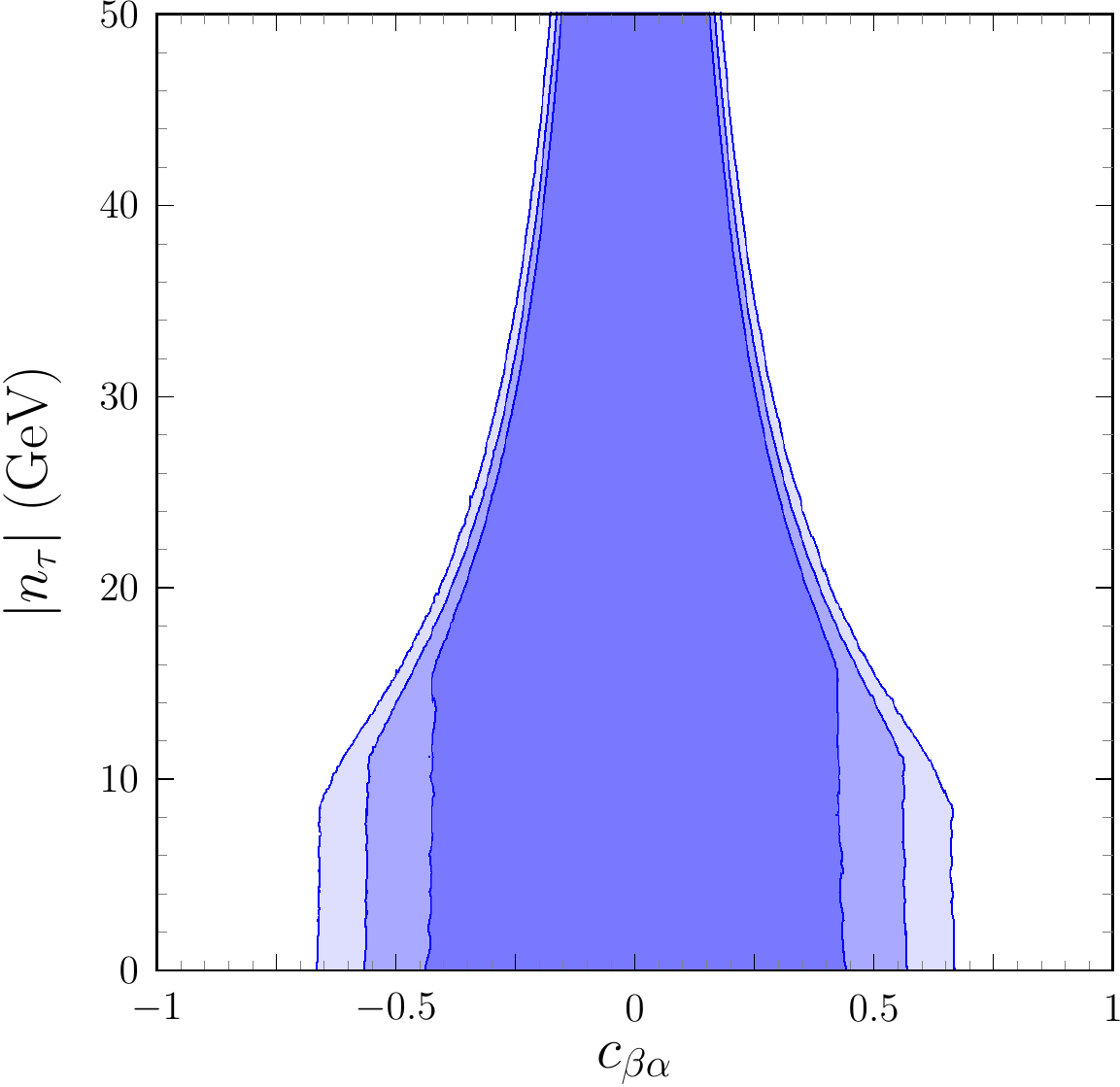}}
\caption{$|n_f|$ vs. $\cab$ for the different fermions $f$; darker to lighter regions correspond to 68, 95 and 99\% CL.\label{FIG:nf:cab}}
\end{center}
\end{figure}

\begin{figure}[h!tb]
\begin{center}
\subfigure[${\sigma[q\bar q\nh]}$ vs. $\Gamma(\nh)$, ${\sigma[q\bar q\nh]\subset\text{ ggF}}$.\label{sFIG:qqH:Gh:gFC-0}]{\includegraphics[height=0.43\textwidth]{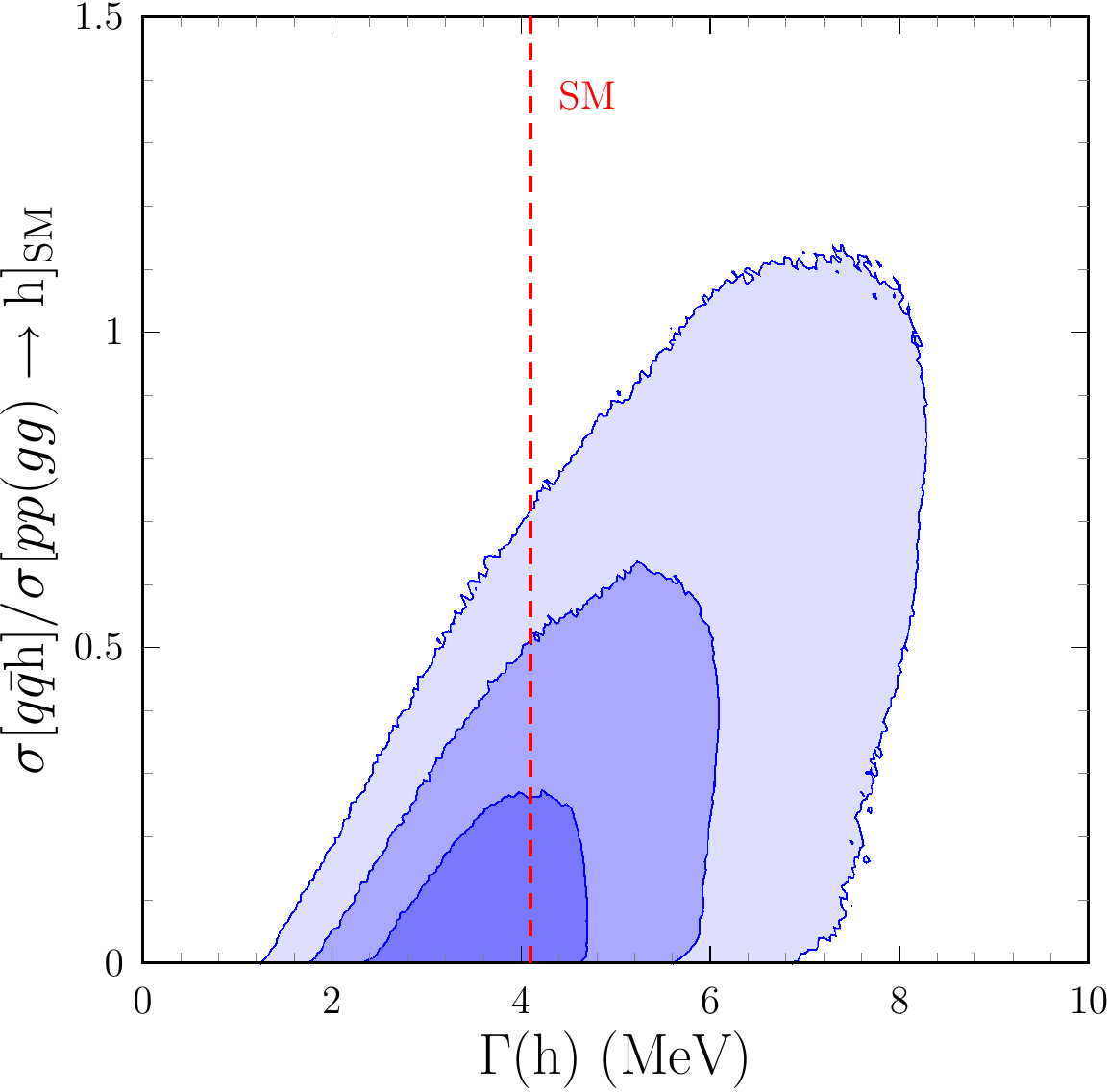}}\quad
\subfigure[${\sigma[q\bar q\nh]}$ vs. ${\sigma[pp(gg)\to\nh]}$, ${\sigma[q\bar q\nh]\subset\text{ ggF}}$\label{sFIG:qqH:ggF:gFC-0}]{\includegraphics[height=0.43\textwidth]{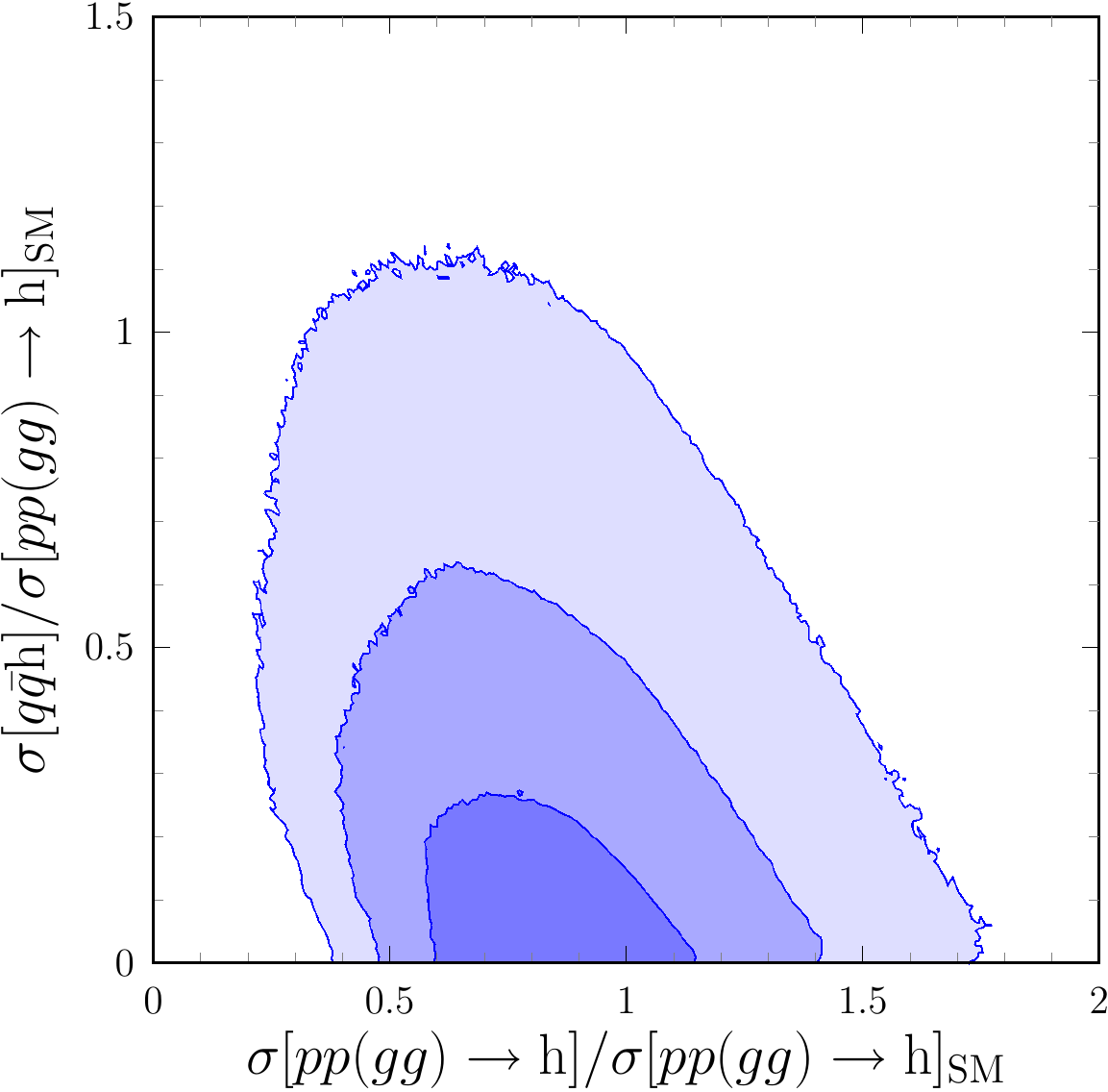}}\\ \vspace{0.1cm}
\subfigure[${\sigma[q\bar q\nh]}$ vs. $\Gamma(\nh)$, ${\sigma[q\bar q\nh]\not\subset\text{ ggF}}$\label{sFIG:qqH:Gh:gFC-1}]{\includegraphics[height=0.43\textwidth]{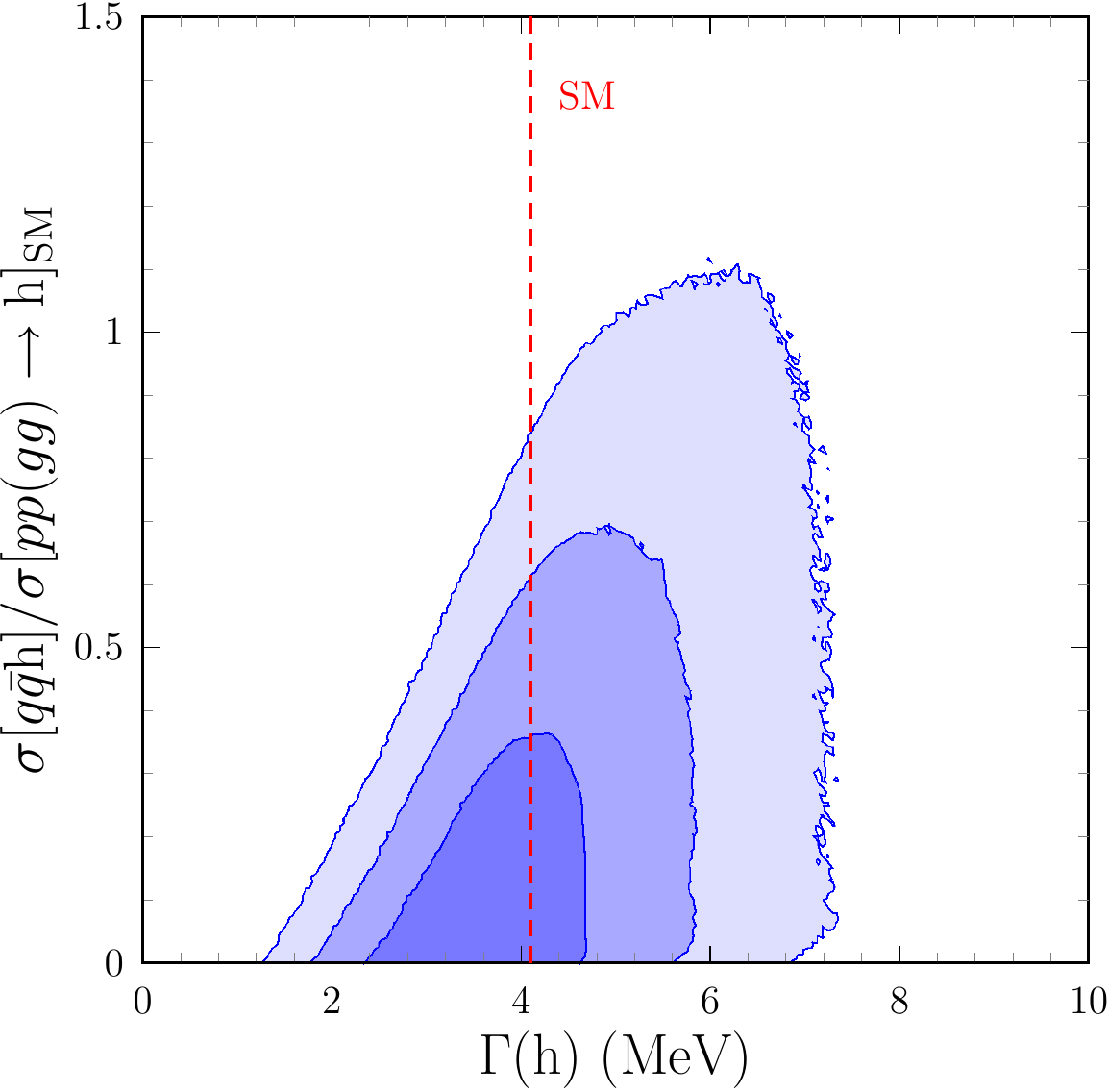}}\quad
\subfigure[${\sigma[q\bar q\nh]}$ vs. ${\sigma[pp(gg)\to\nh]}$, ${\sigma[q\bar q\nh]\not\subset\text{ ggF}}$\label{sFIG:qqH:ggF:gFC-1}]{\includegraphics[height=0.43\textwidth]{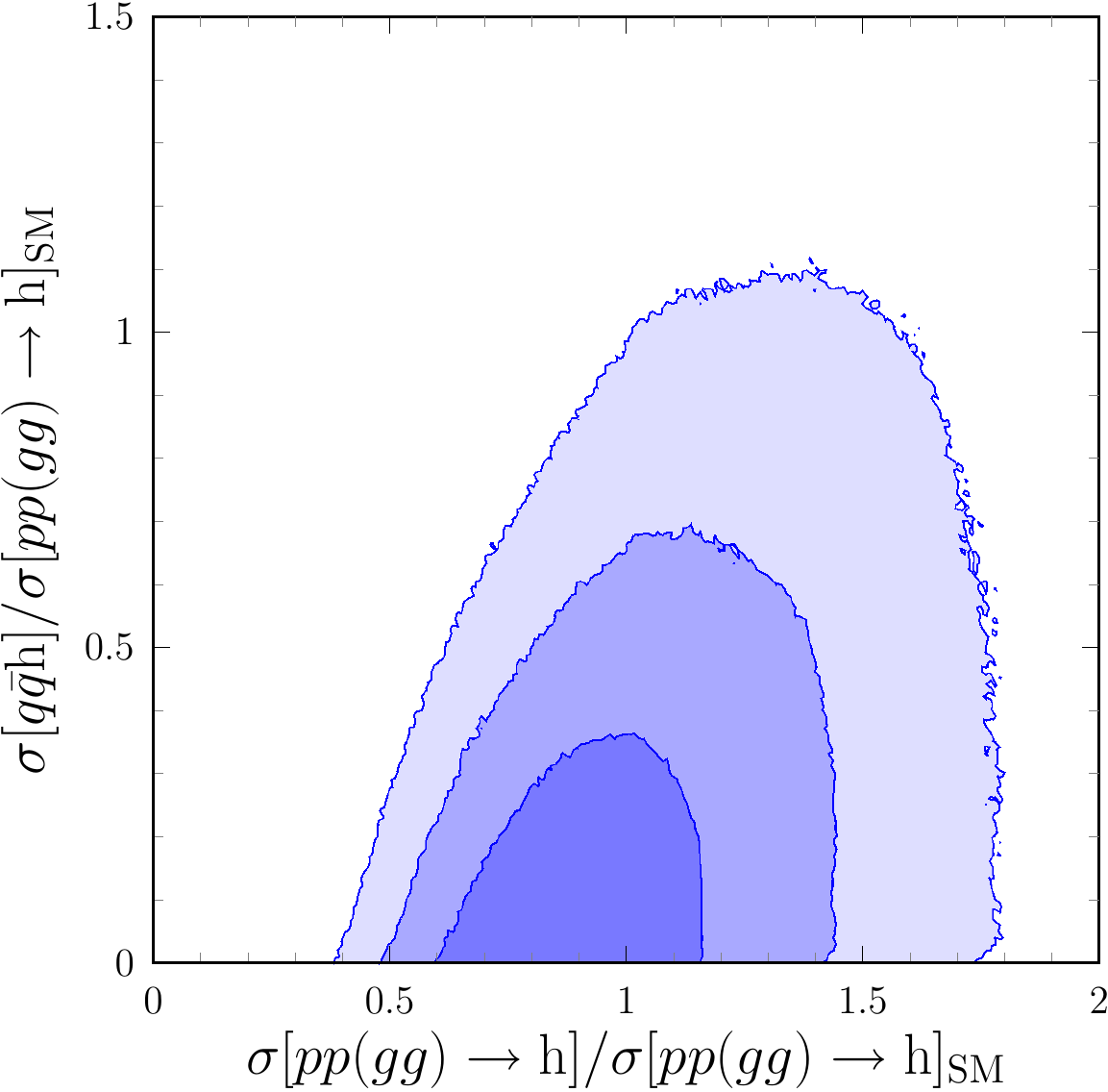}}\\
\caption{Effect of including $q\bar q\to\nh$ production in ggF in analyses of Higgs signal strengths; darker to lighter regions correspond to 68, 95 and 99\% CL.\label{FIG:qqH}}
\end{center}
\end{figure}

With the deviations with respect to the SM of the couplings of $\nh$ and their implications for decays and production mechanisms, one can impose the experimental constraints of section \ref{ssSEC:h:constraints} and explore the allowed values of $\cab$ and the gFC parameters $\nq{}{f}$ in \refEQ{eq:FC:Matrices:00}. For the results presented in the following we consider the most conservative situation, i.e. all parameters are free to vary simultaneously. Compared to restricted situations where not all parameters are considered simultaneously, this offers a safer interpretation of excluded regions (they are excluded whatever the values of the parameters not displayed) at the price, of course, of larger allowed regions.\\ 
Figure \ref{FIG:nf:cab} shows $\nq{}{f}$ vs. $\cab$ for all quarks and leptons. Some comments are in order.
\begin{itemize}
\item As expected, for $\cab\to 0$, the constraints on $\nq{}{f}$ disappear.
\item For $u$, $c$, $d$ and $s$ quarks, the allowed regions are almost identical, as one could anticipate from their irrelevant role, within the SM, in the available production $\times$ decay Higgs signal strengths. The corresponding $\nq{}{f}$'s appear to be effectively limited by the contributions to the Higgs width.
\item Surprisingly, the allowed size of $\abs{\nq{}{t}}$ appears to be independent of $\cab$: this will be discussed in connection with Fig. \ref{sFIG:bqt:aqt} below.
\item The $\nq{}{b}$ and $\nq{}{\tau}$ cases are also similar, with allowed regions differing from the $u$, $c$, $d$, $s$ cases for $\abs{\nq{}{q}}$'s below 10-15 GeV and not small $\cab$.
\item For $\nq{}{e}$ and $\nq{}{\mu}$, the allowed regions are much more constrained owing to the bounds set by dedicated $pp\to\nh\to$ $e^+e^-$, $\mu^+\mu^-$ analyses such as \cite{Khachatryan:2014aep,Aaboud:2017ojs}.
\end{itemize}
Although Fig. \ref{FIG:nf:cab} shows absolute bounds on $\abs{\nq{}{f}}$'s, it does not give information on $\arg(\nq{}{f})$'s and cannot be directly read in terms of the scalar and pseudoscalar couplings of $\nh$ in \refEQS{eq:HiggsScalarFactor:00}--\eqref{eq:HiggsPseudoScalarFactor:00}. Considering that, Figure \ref{FIG:bf:af} shows $\bar b_f$ vs $\bar a_f$ with
\begin{equation}
\bar a_f\equiv \sab m_f+\cab\re{\nq{}{f}},\quad \bar b_f=\cab\im{\nq{}{f}}\,.
\end{equation}
Furthermore, to maintain some information on $\cab$, allowed regions corresponding to $\abs{\cab}<0.01$, to $0.01<\abs{\cab}<0.1$ and to $0.1<\abs{\cab}$ are displayed. One can notice that
\begin{itemize}
\item for the first and second fermion generations, there is no dependence on $\arg(\nq{}{f})$, since only decays, with rates proportional to $\abs{\bar a_f}^2+\abs{\bar b_f}^2$, are relevant. For quarks, the allowed region for $\abs{\cab}<0.01$ is smaller: this is simply due to the perturbativity requirement in \refEQ{eq:perturbativity:00}.
\item For the top quark, two separate regions are allowed: this is also expected since independent sign changes in both $\bar a_t$ and $\bar b_t$ (together with sign changes in $\cab$, $\sab$) do not alter the predictions. For $\abs{\cab}<0.01$ the allowed regions are quite reduced and placed around $(\bar a_t,\bar b_t)=(\pm m_t,0)$; with $0.01<\abs{\cab} <0.1$ their size increases and only for $\abs{\cab}>0.1$ the interplay of (i) pseudoscalar contributions to $gg\to\nh$ and $\nh\to\gamma\gamma$, and (ii) $W$-top(scalar) interference in $\nh\to\gamma\gamma$ gives rise to larger regions.
\item For $b$ and $\tau$, the regions for not too small mixing, $\abs{\cab}>0.01$, are ring-shaped; $m_b$ and $m_\tau$ set the radii of such regions, as could be expected from the agreement of $\nh\to b\bar b$ and $\nh\to \tau\bar\tau$ signal strengths with SM expectations. For small mixing, $\abs{\cab}<0.01$, the perturbativity requirement on $\abs{\nq{}{b}}$, $\abs{\nq{}{\tau}}$ limits the allowed departure from $(\bar a_f,\bar b_f)=(\pm m_f,0)$, giving in fact, for the $b$ case, two disjoint patches.
\end{itemize}
%
\begin{figure}[h!tb]
\begin{center}
\subfigure[$u$\label{sFIG:bqu:aqu}]{\includegraphics[height=0.3\textwidth]{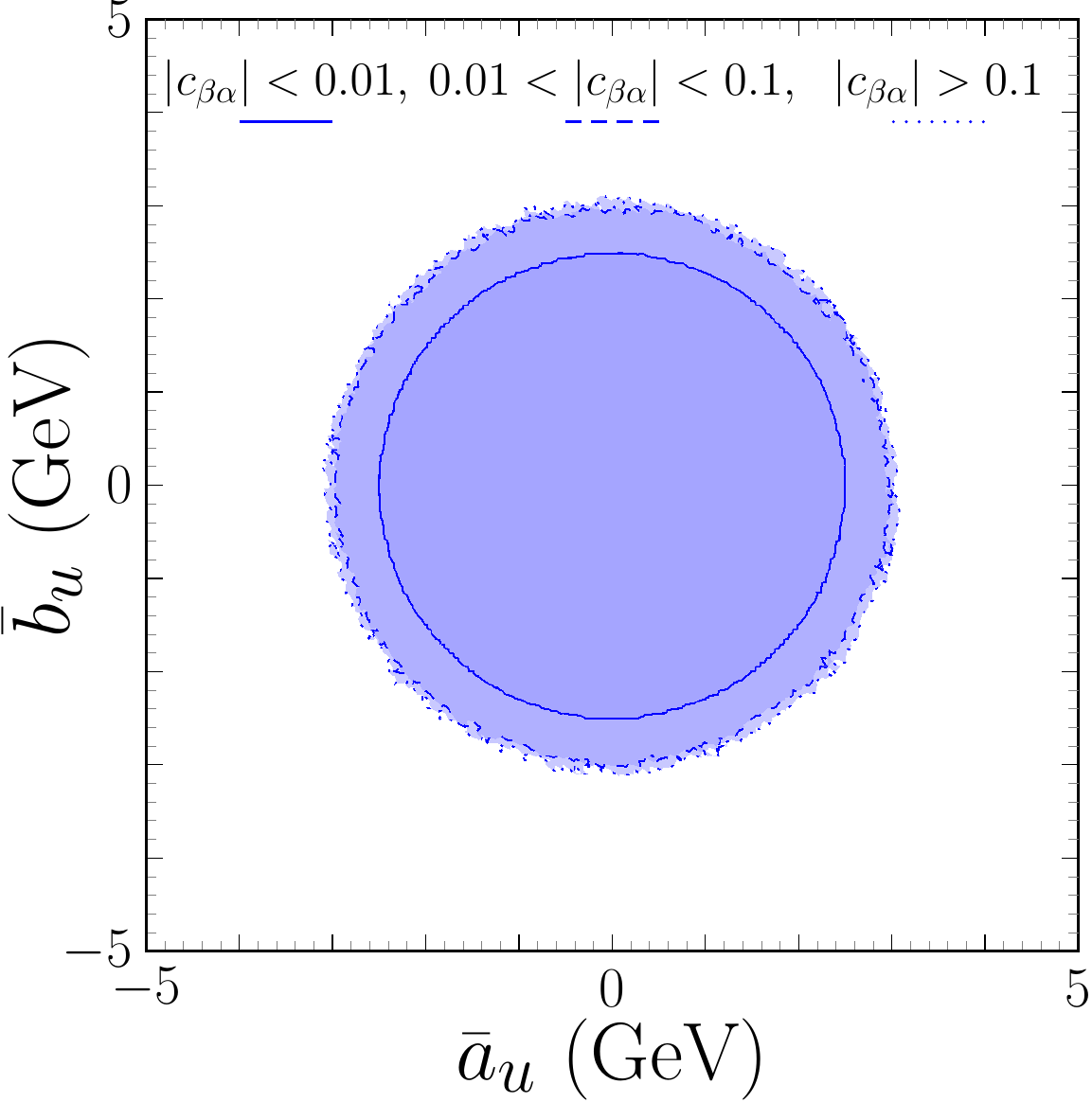}}\quad
\subfigure[$c$\label{sFIG:bqc:aqc}]{\includegraphics[height=0.3\textwidth]{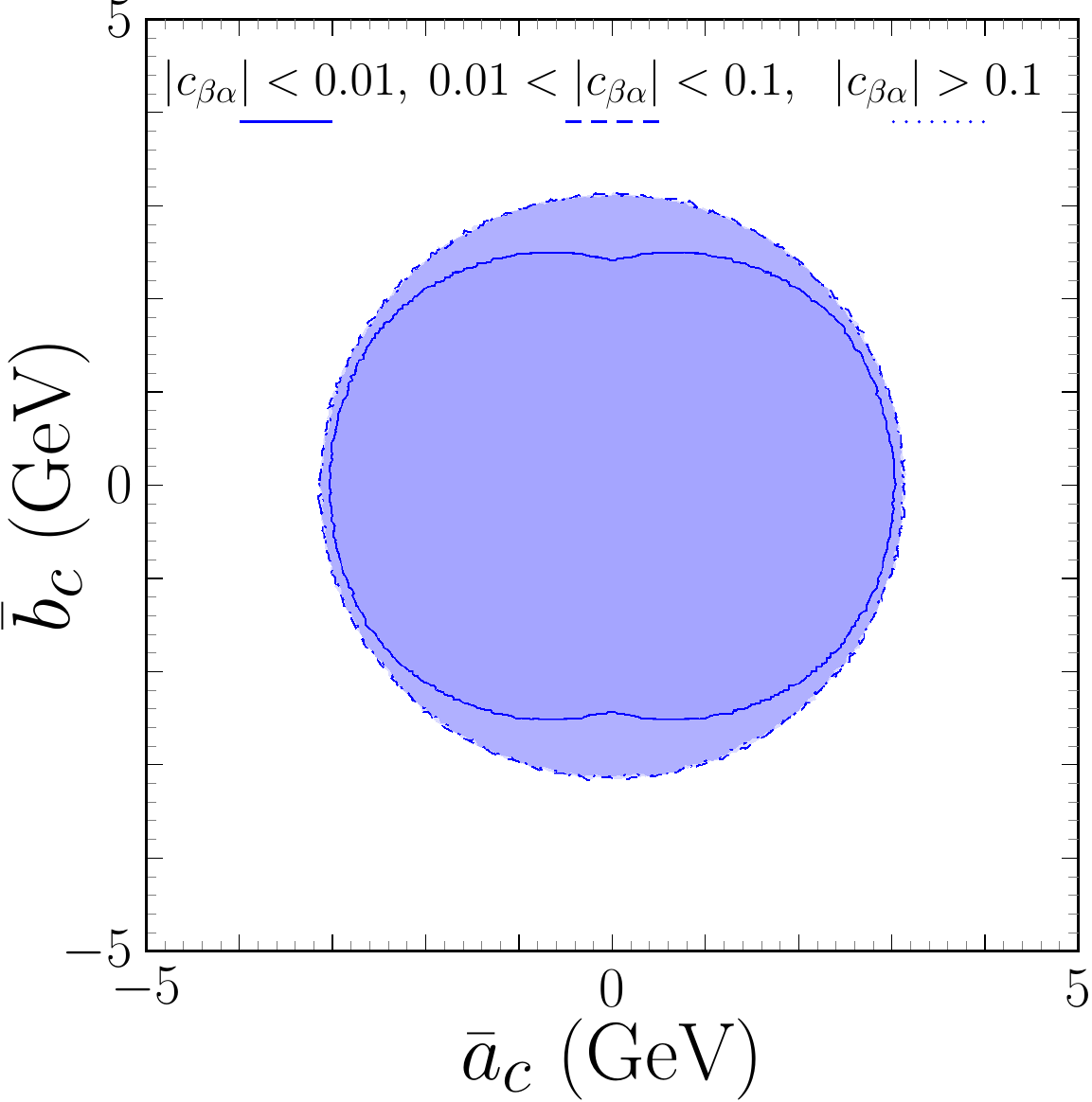}}\quad
\subfigure[$t$\label{sFIG:bqt:aqt}]{\includegraphics[height=0.3\textwidth]{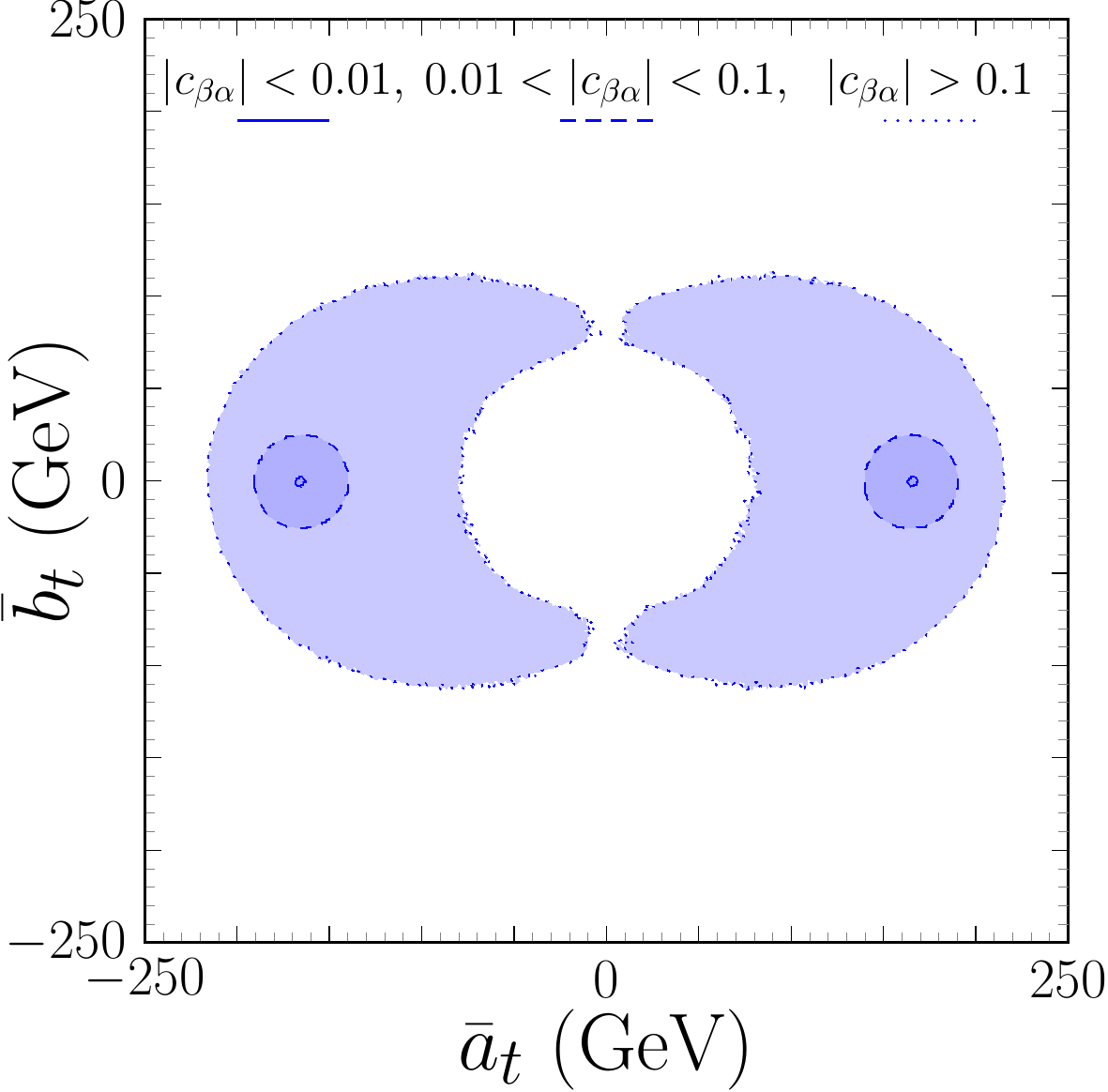}}\\
\subfigure[$d$\label{sFIG:bqd:aqd}]{\includegraphics[height=0.3\textwidth]{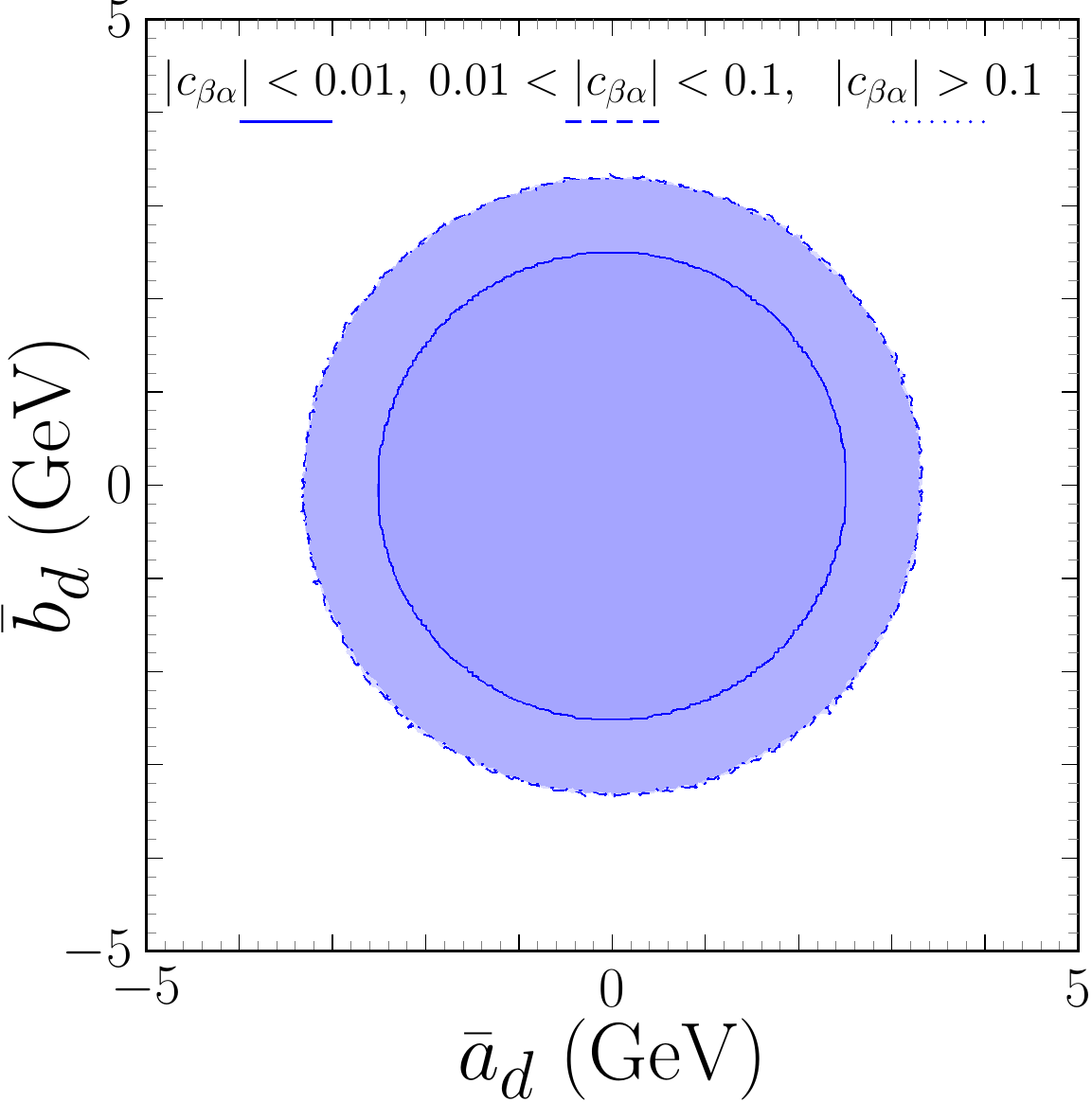}}\quad
\subfigure[$s$\label{sFIG:bqs:aqs}]{\includegraphics[height=0.3\textwidth]{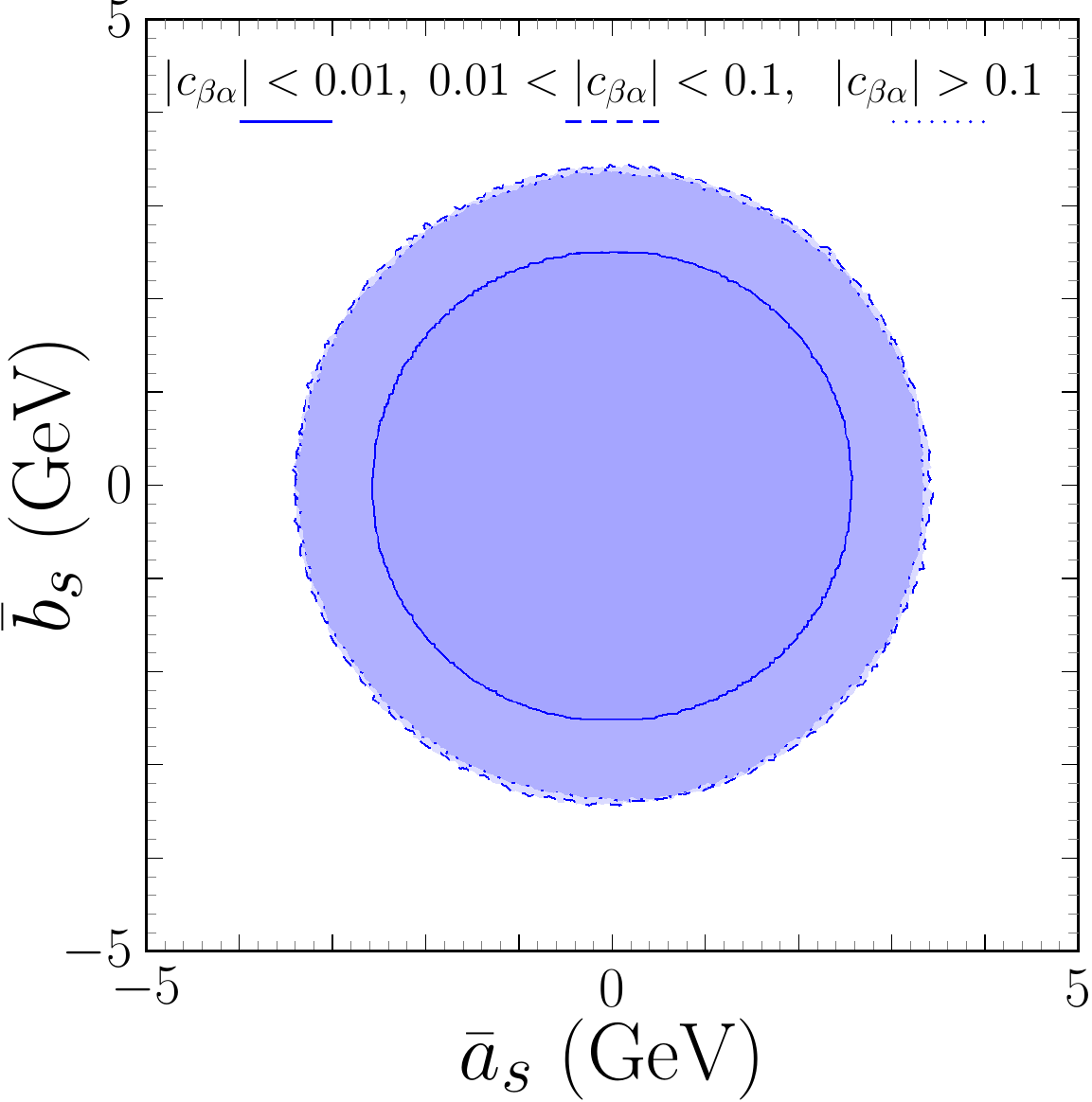}}\quad
\subfigure[$b$\label{sFIG:bqb:aqb}]{\includegraphics[height=0.3\textwidth]{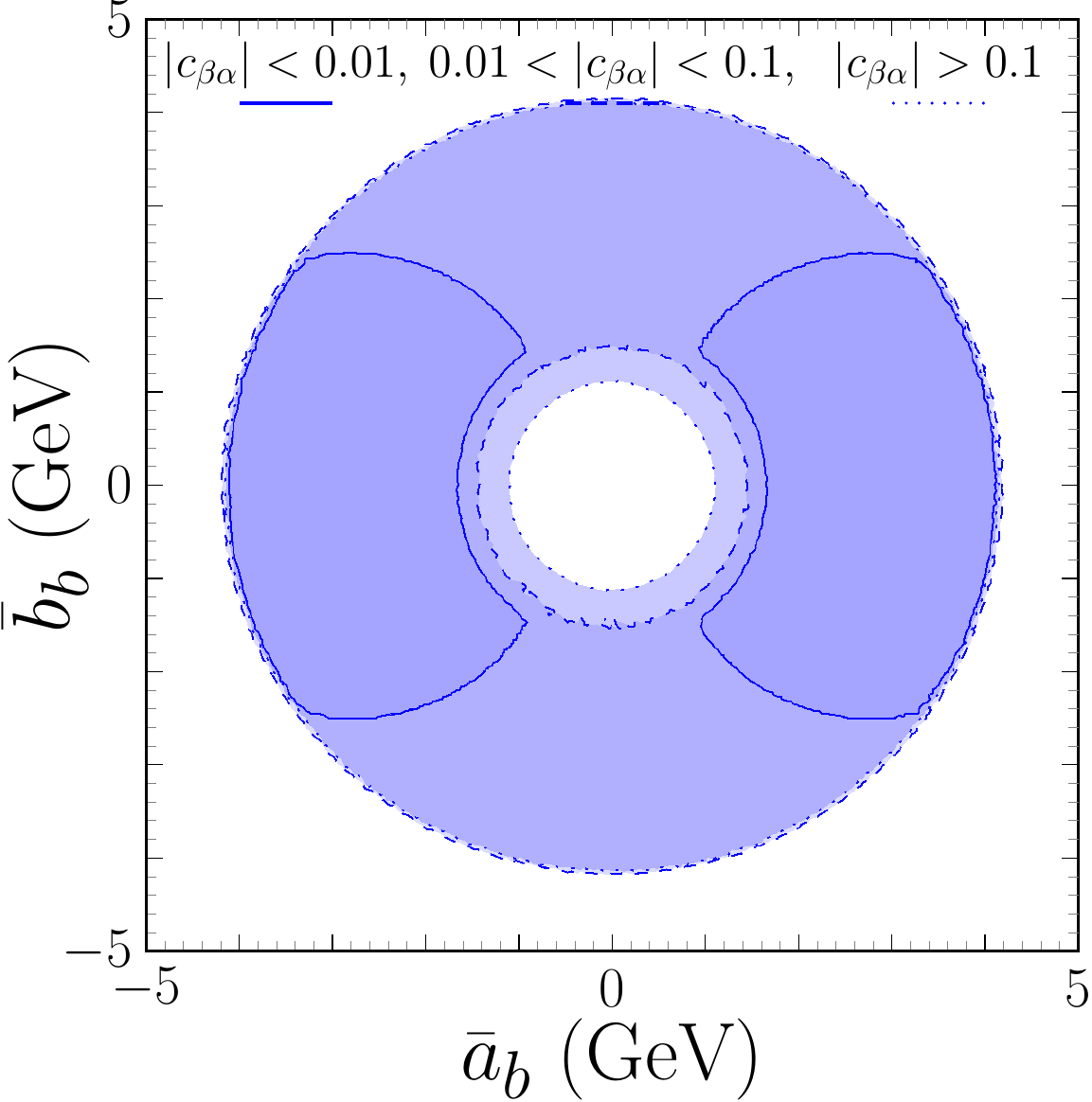}}\\
\subfigure[$e$\label{sFIG:ble:ale}]{\includegraphics[height=0.3\textwidth]{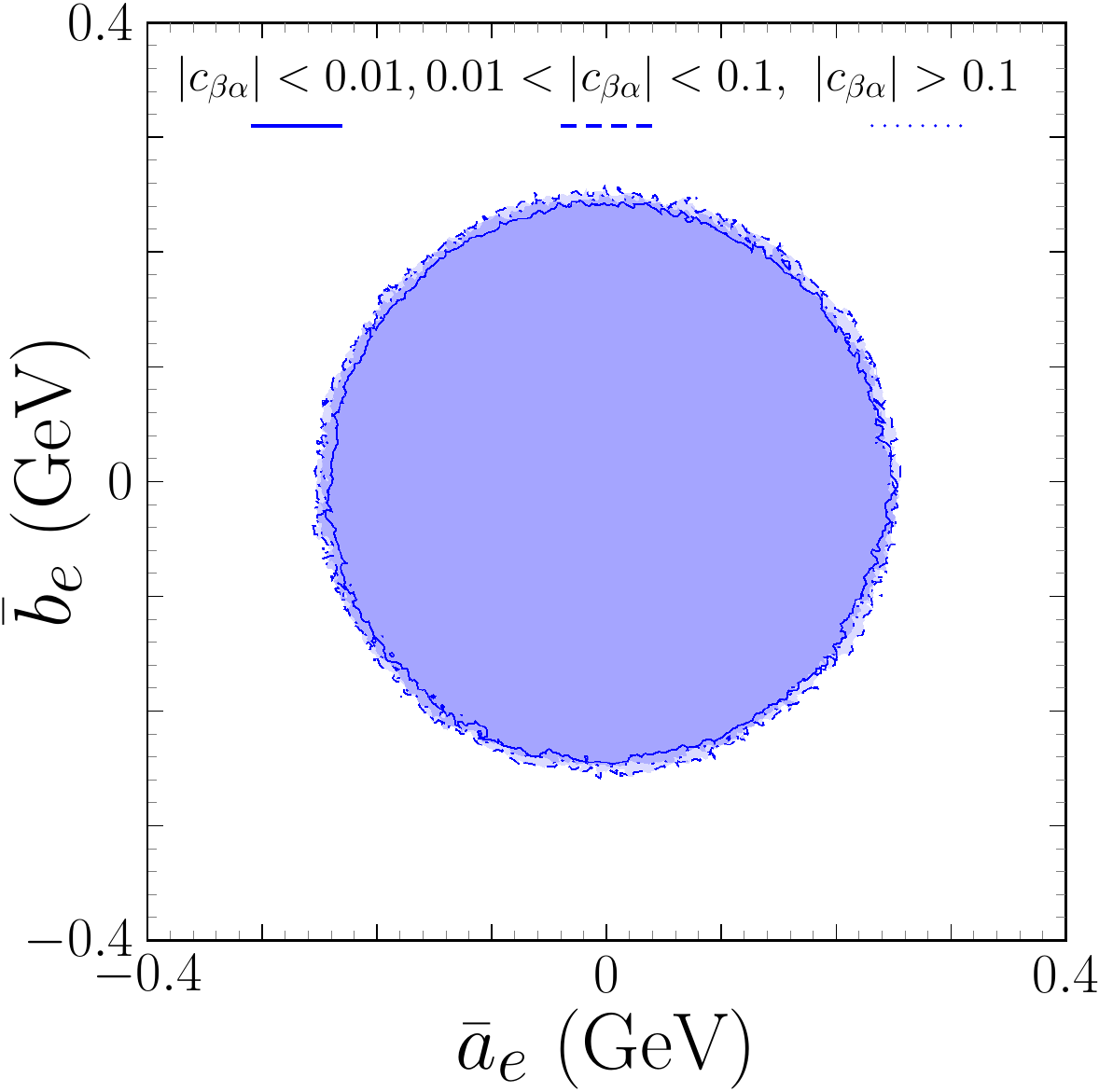}}\quad
\subfigure[$\mu$\label{sFIG:blm:alm}]{\includegraphics[height=0.3\textwidth]{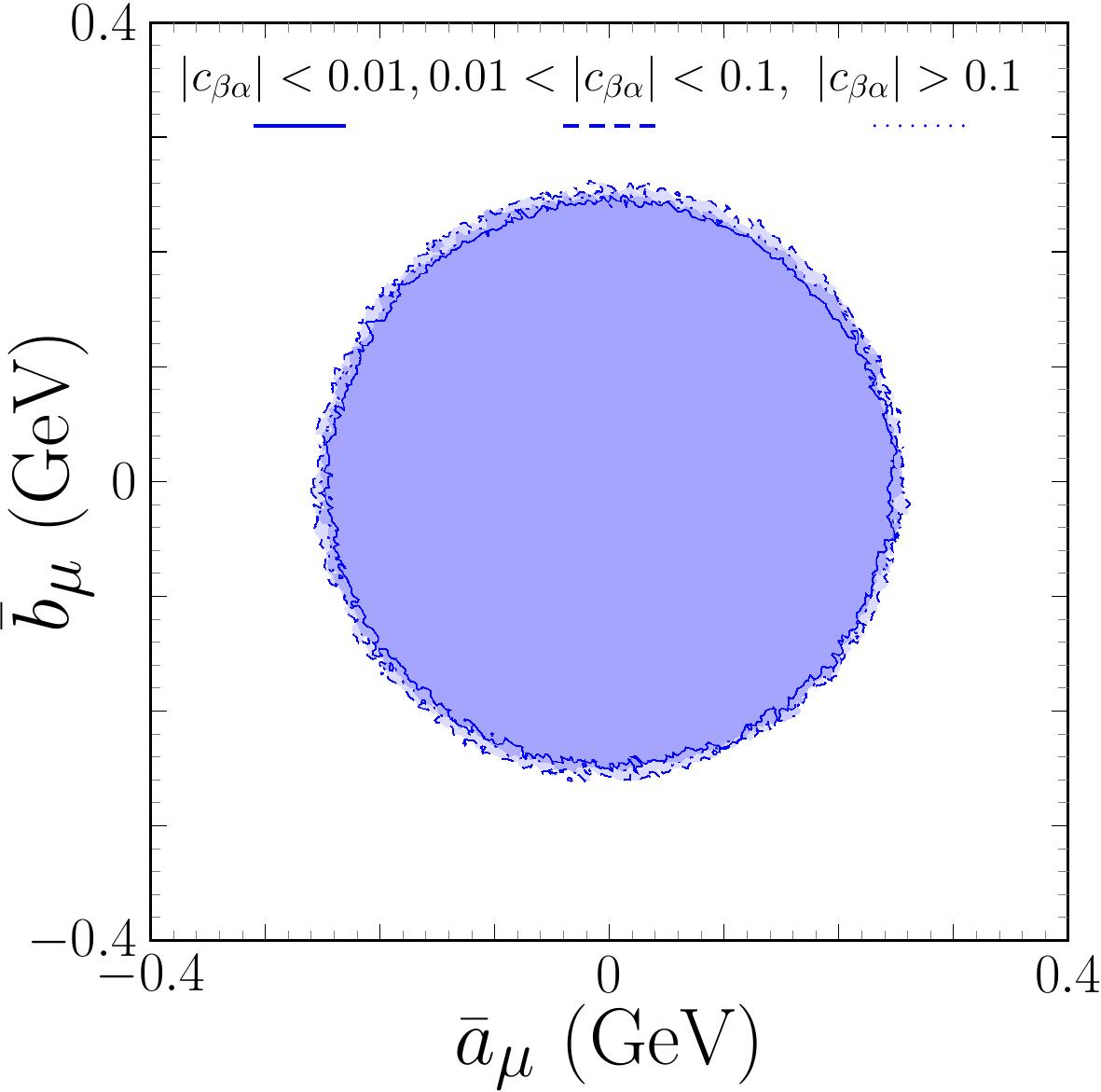}}\quad
\subfigure[$\tau$\label{sFIG:blt:alt}]{\includegraphics[height=0.3\textwidth]{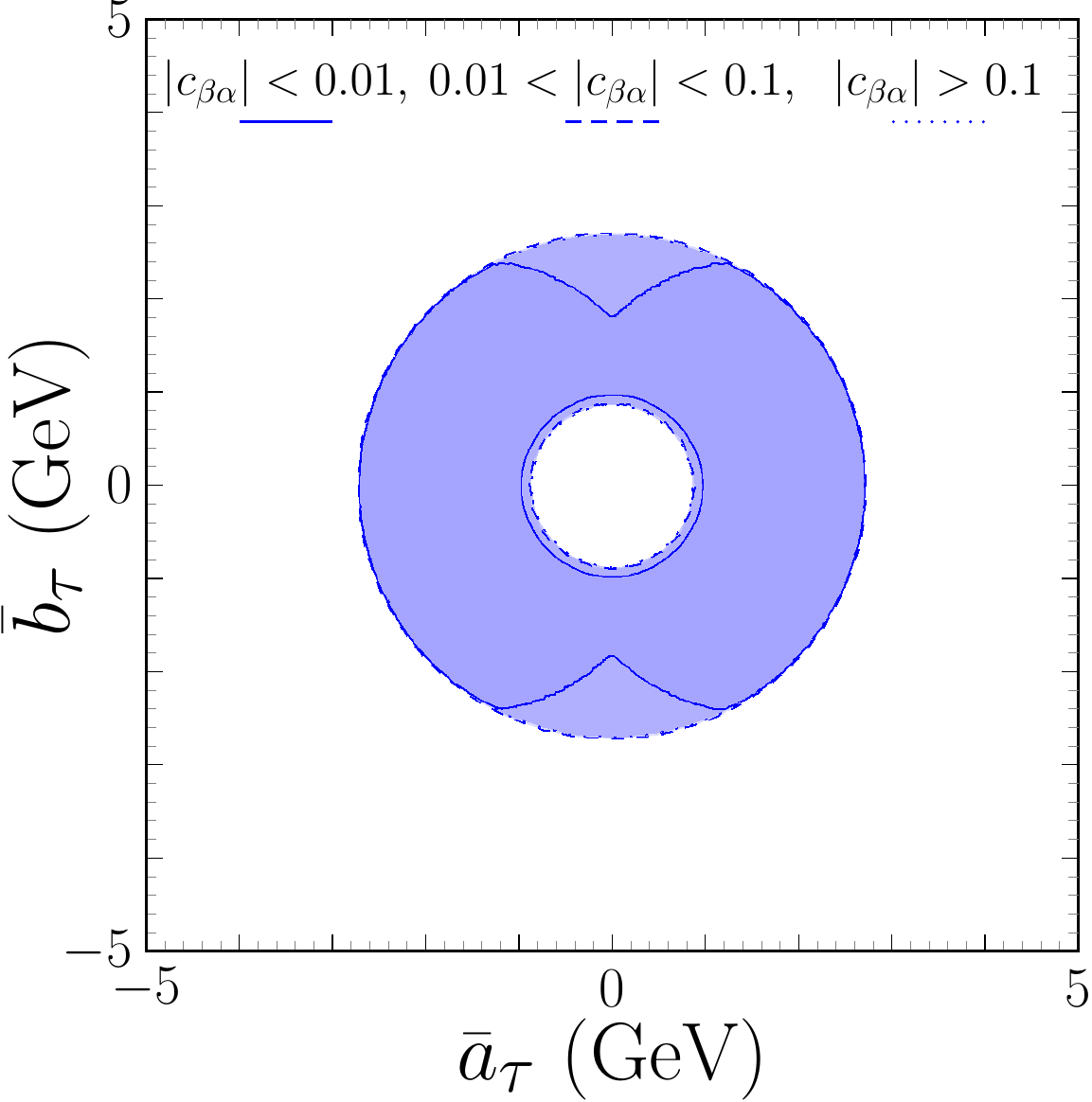}}
\caption{Allowed regions at 99\% CL for pseudoscalar  vs. scalar couplings for the different fermions $f$ with $\mathscr L_{\nh\bar ff}=-\frac{\nh}{v}\bar f(\bar a_f+i\bar b_f\gamma_5)f$.\label{FIG:bf:af}}
\end{center}
\end{figure}
To close this section we recall the discussion on $\bar qq\to\nh$ production in section \ref{ssSEC:h:production}: as commented there, values of $\nq{}{f}$ in agreement with the SM-like Higgs signal strengths could potentially give production cross sections not far from the dominating SM ones. Figure \ref{FIG:qqH} shows  
\begin{equation}
\frac{\sigma[q\bar q\nh]}{\sigma[pp(gg)\to\nh]_{\rm SM}}\equiv \left(\sum_{q=u,c,d,s}\sigma[pp(q\bar q)\to\nh]\right) {\Big /} \sigma[pp(gg)\to\nh]_{\rm SM},
\end{equation}
vs. the total Higgs width and vs. the gluon-gluon fusion production cross section in two different analyses: in Figures \ref{sFIG:qqH:Gh:gFC-0} and \ref{sFIG:qqH:ggF:gFC-0}, $\sigma[q\bar q\nh]$ is added to the ggF production cross section, while in Figures \ref{sFIG:qqH:Gh:gFC-1} and \ref{sFIG:qqH:ggF:gFC-1} it is not (and therefore, in the analysis, it does not affect directly observables constrained by experiment). Comparing \ref{sFIG:qqH:Gh:gFC-0}-\ref{sFIG:qqH:ggF:gFC-0} with \ref{sFIG:qqH:Gh:gFC-1}-\ref{sFIG:qqH:ggF:gFC-1}, one can notice that the constraints from Higgs signal strengths are able to bound the size of $\sigma[q\bar q\nh]$, even if there is room for an overall $q\bar q\to\nh$ cross section which is quite sizable, not far from the complete SM Higgs production cross section. Furthermore, when $\sigma[q\bar q\nh]$ is added to the ggF production cross section, the agreement with the observed Higgs signal strengths allows for a smaller amount of $q\bar q\to\nh$, and, for sizable $q\bar q\to\nh$, it is achieved at the cost of (i) reducing the ggF production cross section and (ii) increasing the total width $\Gamma(\nh)$, as the shape of the allowed regions in Figures \ref{sFIG:qqH:Gh:gFC-0} and \ref{sFIG:qqH:ggF:gFC-0} shows. For the results in Figures \ref{FIG:nf:cab} and \ref{FIG:bf:af}, the bounds on the the different $\bar a_f$, $\bar b_f$ do not differ in both analyses. It should be finally mentioned that, in connection with the previous comments and the size of $\sigma[pp(q\bar q)\to\nh]$, it might be interesting to analyse, for the remaining neutral scalars $\nH$ and $\nA$, the cross sections for $pp(q\bar q)\to\nH,\nA$ at the LHC.
%

%

%

\clearpage
\section*{Conclusions\label{SEC:Conclusions}}
In this paper we analyse the question of general Flavour Conservation in extensions of the SM with additional scalar doublets, in particular the 2HDM. 
The effect of the one loop Renomalization Group Evolution of the Yukawa coupling matrices on gFC scenarios is discussed in detail. In particular it is to be stressed that in the absence of Yukawa couplings with right-handed neutrinos, gFC in the lepton sector is stable. For the quark sector, some one loop RGE stable scenarios are discussed, including the case of a Cabibbo-like quark mixing matrix.
At a phenomenological level, we discuss the constraints that existing data on flavour conserving processes, in particular the ones related to the Higgs, impose on the parameters describing gFC in the different fermion sectors, including a detailed numerical analysis of that parameter space. Direct $q\bar q\to\nh$ production is also considered in detail: although it is completely negligible in the SM, that might not be the case in scenarios such as 2HDM, and it may even be relevant for the production of the additional non-SM neutral scalars.

\section*{Acknowledgments}
The authors thank Luca Fiorini for discussions. This work is partially supported by Spanish MINECO under grant FPA2015-68318-R, FPA2017-85140-C3-3-P and by the Severo Ochoa Excellence Center Project SEV-2014-0398, by Generalitat Valenciana under grant GVPROMETEOII 2014-049 and by Funda\c{c}\~ao para a Ci\^encia e a Tecnologia (FCT, Portugal) through the projects CERN/FIS-NUC/0010/2015 and CFTP-FCT Unit 777 (UID/FIS/00777/2013) which are partially funded through POCTI (FEDER), COMPETE, QREN and EU. MN acknowledges support from FCT through postdoctoral grant SFRH/BPD/112999/2015.

\appendix
\clearpage
%
%
\section{RGE details\label{APP:RGEdetails}}
The analysis of the RGE of the quark Yukawa couplings and the stability of the gFC scenario in \refEQ{eq:FC:Matrices:00} has been presented in detail for the set $\{\Yd{\alpha}\Ydd{\beta}\}$ in section \ref{sSEC:RGE:Yukawas} and \ref{sSEC:RGE:FC}. We reproduce in this appendix the equations relevant for $\{\Yd{\alpha}\Ydd{\beta}\}$ and also for $\{\Ydd{\alpha}\Yd{\beta}\}$, $\{\Yu{\alpha}\Yud{\beta}\}$ and $\{\Yud{\alpha}\Yu{\beta}\}$, omitted for conciseness in section \ref{SEC:RGE}. In correspondence with \refEQS{eq:RGEbilineardecomposition:00}, \eqref{eq:DerY:f:00} and \eqref{eq:DerY:g:00},
\begin{alignat}{2}
\der(\Yd{\alpha}\Ydd{\beta})&=\DYf{d_L}{\alpha\beta}(\Ydbase)+\DYg{d_L}{\alpha\beta}(\Ydbase,\Yubase)\,,\quad 
\der(\Yu{\alpha}\Yud{\beta})&=\DYf{u_L}{\alpha\beta}(\Yubase)+\DYg{u_L}{\alpha\beta}(\Ydbase,\Yubase)\,,\nonumber\\
\der(\Ydd{\alpha}\Yd{\beta})&=\DYf{d_R}{\alpha\beta}(\Ydbase)+\DYg{d_R}{\alpha\beta}(\Ydbase,\Yubase)\,,\quad 
\der(\Yud{\alpha}\Yu{\beta})&=\DYf{u_R}{\alpha\beta}(\Yubase)+\DYg{u_R}{\alpha\beta}(\Ydbase,\Yubase)\,,
\label{eq:RGEbilineardecomposition:All:00}
\end{alignat}
with
\begin{align}
&\DYf{d_L}{\alpha\beta}(\Ydbase)=
2a_d\Yd{\alpha}\Ydd{\beta}+\sum_{\rho=1}^{n=2}\left[T^d_{\alpha,\rho}\Yd{\rho}\Ydd{\beta}+T^{d\ast}_{\beta,\rho}\Yd{\alpha}\Ydd{\rho}\right]+2\Yd{\alpha}\YdH{R}\Ydd{\beta}+\frac{1}{2}\YdH{L}\Yd{\alpha}\Ydd{\beta}+\frac{1}{2}\Yd{\alpha}\Ydd{\beta}\YdH{L}\,,\nonumber\\
&\DYg{d_L}{\alpha\beta}(\Ydbase,\Yubase)=
\frac{1}{2}\YuH{L}\Yd{\alpha}\Ydd{\beta} +\frac{1}{2}\Yd{\alpha}\Ydd{\beta}\YuH{L} -2\sum_{\rho=1}^{n=2}\left[\Yu{\rho}\Yud{\alpha}\Yd{\rho}\Ydd{\beta}+\Yd{\alpha}\Ydd{\rho}\Yu{\beta}\Yud{\rho}\right],
\label{eq:DerYdYdd:fg:00}
\end{align}
\begin{align}
&\DYf{d_R}{\alpha\beta}(\Ydbase)=
2a_d\Ydd{\alpha}\Yd{\beta}+\sum_{\rho=1}^{n=2}\left[T^{d\ast}_{\alpha,\rho}\Ydd{\rho}\Yd{\beta}+T^d_{\beta,\rho}\Ydd{\alpha}\Yd{\rho}\right]+\Ydd{\alpha}\YdH{L}\Yd{\beta}
+\YdH{R}\Ydd{\alpha}\Yd{\beta}+\Ydd{\alpha}\Yd{\beta}\YdH{R}\,,\nonumber\\
&\DYg{d_R}{\alpha\beta}(\Ydbase,\Yubase)=
\Ydd{\alpha}\YuH{L}\Yd{\beta}-2\sum_{\rho=1}^{n=2}\left[\Ydd{\rho}\Yu{\alpha}\Yud{\rho}\Yd{\beta}+\Ydd{\alpha}\Yu{\rho}\Yud{\beta}\Yd{\rho}\right],
\label{eq:DerYddYd:fg:00}
\end{align}
\begin{align}
&\DYf{u_L}{\alpha\beta}(\Yubase)=
2a_u\Yu{\alpha}\Yud{\beta}+\sum_{\rho=1}^{n=2}\left[T^u_{\alpha,\rho}\Yu{\rho}\Yud{\beta}+T^{d\ast}_{\beta,\rho}\Yu{\alpha}\Yud{\rho}\right]+2\Yu{\alpha}\YuH{R}\Yud{\beta}
+\frac{1}{2}\YuH{L}\Yu{\alpha}\Yud{\beta}+\frac{1}{2}\Yu{\alpha}\Yud{\beta}\YuH{L}\,,\nonumber\\
&\DYg{u_L}{\alpha\beta}(\Ydbase,\Yubase)=
\frac{1}{2}\YdH{L}\Yu{\alpha}\Yud{\beta} +\frac{1}{2}\Yu{\alpha}\Yud{\beta}\YdH{L} -2\sum_{\rho=1}^{n=2}\left[\Yd{\rho}\Ydd{\alpha}\Yu{\rho}\Yud{\beta}+\Yu{\alpha}\Yud{\rho}\Yd{\beta}\Ydd{\rho}\right],
\label{eq:DerYuYud:fg:00}
\end{align}
and
\begin{align}
&\DYf{u_R}{\alpha\beta}(\Ydbase)=
2a_u\Yud{\alpha}\Yu{\beta}+\sum_{\rho=1}^{n=2}\left[T^{u\ast}_{\alpha,\rho}\Yud{\rho}\Yu{\beta}+T^u_{\beta,\rho}\Yud{\alpha}\Yu{\rho}\right]+\Yud{\alpha}\YuH{L}\Yu{\beta}
+\YuH{R}\Yud{\alpha}\Yu{\beta}+\Yud{\alpha}\Yu{\beta}\YuH{R}\,,\nonumber\\
&\DYg{u_R}{\alpha\beta}(\Ydbase,\Yubase)=
\Yud{\alpha}\YdH{L}\Yu{\beta}-2\sum_{\rho=1}^{n=2}\left[\Yud{\rho}\Yd{\alpha}\Ydd{\rho}\Yu{\beta}+\Yud{\alpha}\Yd{\rho}\Ydd{\beta}\Yu{\rho}\right].
\label{eq:DerYudYu:fg:00}
\end{align}
The RGE of the commutation relations of \refEQ{eq:abelian:01b} reads
\begin{align}
\der\COMM{\Yd{\alpha}\Ydd{\beta}}{\Yd{\gamma}\Ydd{\delta}}&=\COMM{\DYg{d_L}{\alpha\beta}(\Ydbase,\Yubase)}{\Yd{\gamma}\Ydd{\delta}}+\COMM{\Yd{\alpha}\Ydd{\beta}}{\DYg{d_L}{\gamma\delta}(\Ydbase,\Yubase)}\,,\nonumber\\
\der\COMM{\Ydd{\alpha}\Yd{\beta}}{\Ydd{\gamma}\Yd{\delta}}&=\COMM{\DYg{d_R}{\alpha\beta}(\Ydbase,\Yubase)}{\Ydd{\gamma}\Yd{\delta}}+\COMM{\Ydd{\alpha}\Yd{\beta}}{\DYg{d_R}{\gamma\delta}(\Ydbase,\Yubase)}\,,\nonumber\\
\der\COMM{\Yu{\alpha}\Yud{\beta}}{\Yu{\gamma}\Yud{\delta}}&=\COMM{\DYg{u_L}{\alpha\beta}(\Ydbase,\Yubase)}{\Yu{\gamma}\Yud{\delta}}+\COMM{\Yu{\alpha}\Yud{\beta}}{\DYg{u_L}{\gamma\delta}(\Ydbase,\Yubase)}\,,\nonumber\\
\der\COMM{\Yud{\alpha}\Yu{\beta}}{\Yud{\gamma}\Yu{\delta}}&=\COMM{\DYg{u_R}{\alpha\beta}(\Ydbase,\Yubase)}{\Yud{\gamma}\Yd{\delta}}+\COMM{\Yud{\alpha}\Yu{\beta}}{\DYg{u_R}{\gamma\delta}(\Ydbase,\Yubase)}\,,
\label{eq:DerComm:All:00}
\end{align}
which, following the discussion in section \ref{sSEC:RGE:FC}, lead to (summation over $h=1,2$ understood)
\begin{align}
\{d_L\}\equiv\frac{v^2}{2}&\UdLd\left(\der\COMM{\wYDd{i}\wYDdd{j}}{\wYDd{k}\wYDdd{l}}\right)\UdL=\label{eq:der:YdYdd:00}\\
&\mYDd{i}\mYDdd{j}\,\CKMd\mYDu{h}\mYDud{h}\CKM\,\mYDd{k}\mYDdd{l}-\mYDd{k}\mYDdd{l}\,\CKMd\mYDu{h}\mYDud{h}\CKM\,\mYDd{i}\mYDdd{j}\nonumber\\
&-2\COMM{\CKMd\mYDu{h}\mYDud{i}\CKM}{\mYDd{k}\mYDdd{l}}\mYDd{h}\mYDdd{j}-2\mYDd{i}\mYDdd{h}\COMM{\CKMd\mYDu{j}\mYDud{h}\CKM}{\mYDd{k}\mYDdd{l}}\nonumber\\
&+2\COMM{\CKMd\mYDu{h}\mYDud{k}\CKM}{\mYDd{i}\mYDdd{j}}\mYDd{h}\mYDdd{l}+2\mYDd{k}\mYDdd{h}\COMM{\CKMd\mYDu{l}\mYDud{h}\CKM}{\mYDd{i}\mYDdd{j}},
\nonumber
\end{align}
\begin{align}
\{d_R\}\equiv\frac{v^2}{2}&\UdRd\left(\der\COMM{\wYDdd{i}\wYDd{j}}{\wYDdd{k}\wYDd{l}}\right)\UdR=\label{eq:der:YddYd:00}\\
&\COMM{\mYDdd{i}\,\CKMd\mYDu{h}\mYDud{h}\CKM\,\mYDd{j}}{\mYDdd{k}\mYDd{l}}-\COMM{\mYDdd{k}\,\CKMd\mYDu{h}\mYDud{h}\CKM\,\mYDd{l}}{\mYDdd{i}\mYDd{j}}\nonumber\\
&-2\COMM{\mYDdd{h}\,\CKMd\mYDu{i}\mYDud{h}\CKM\,\mYDd{j}}{\mYDdd{k}\mYDd{l}}
 -2\COMM{\mYDdd{i}\,\CKMd\mYDu{h}\mYDud{j}\CKM\,\mYDd{h}}{\mYDdd{k}\mYDd{l}}\nonumber\\
&+2\COMM{\mYDdd{h}\,\CKMd\mYDu{k}\mYDud{h}\CKM\,\mYDd{l}}{\mYDdd{i}\mYDd{j}}
 +2\COMM{\mYDdd{k}\,\CKMd\mYDu{h}\mYDud{l}\CKM\,\mYDd{h}}{\mYDdd{i}\mYDd{j}},
\nonumber
\end{align}
\begin{align}
\{u_L\}\equiv\frac{v^2}{2}&\UuLd\left(\der\COMM{\wYDu{i}\wYDud{j}}{\wYDu{k}\wYDud{l}}\right)\UuL=\label{eq:der:YuYud:00}\\
&\mYDu{i}\mYDud{j}\,\CKM\mYDd{h}\mYDdd{h}\CKMd\,\mYDu{k}\mYDud{l}-\mYDu{k}\mYDud{l}\,\CKM\mYDd{h}\mYDdd{h}\CKMd\,\mYDu{i}\mYDud{j}\nonumber\\
&-2\COMM{\CKM\mYDd{h}\mYDdd{i}\CKMd}{\mYDu{k}\mYDud{l}}\mYDu{h}\mYDud{j}-2\mYDu{i}\mYDud{h}\COMM{\CKM\mYDd{j}\mYDdd{h}\CKMd}{\mYDu{k}\mYDud{l}}\nonumber\\
&+2\COMM{\CKM\mYDd{h}\mYDdd{k}\CKMd}{\mYDu{i}\mYDud{j}}\mYDu{h}\mYDud{l}+2\mYDu{k}\mYDud{h}\COMM{\CKM\mYDd{l}\mYDdd{h}\CKMd}{\mYDu{i}\mYDud{j}},
\nonumber
\end{align}
and
\begin{align}
\{u_R\}\equiv\frac{v^2}{2}&\UuRd\left(\der\COMM{\wYDud{i}\wYDu{j}}{\wYDud{k}\wYDu{l}}\right)\UuR=\label{eq:der:YudYu:00}\\
&\COMM{\mYDud{i}\,\CKM\mYDd{h}\mYDdd{h}\CKMd\,\mYDu{j}}{\mYDud{k}\mYDu{l}}-\COMM{\mYDud{k}\,\CKM\mYDd{h}\mYDdd{h}\CKMd\,\mYDu{l}}{\mYDud{i}\mYDu{j}}\nonumber\\
&-2\COMM{\mYDud{h}\,\CKM\mYDd{i}\mYDdd{h}\CKMd\,\mYDu{j}}{\mYDud{k}\mYDu{l}}
 -2\COMM{\mYDud{i}\,\CKM\mYDd{h}\mYDdd{j}\CKMd\,\mYDu{h}}{\mYDud{k}\mYDu{l}}\nonumber\\
&+2\COMM{\mYDud{h}\,\CKM\mYDd{k}\mYDdd{h}\CKMd\,\mYDu{l}}{\mYDud{i}\mYDu{j}}
 +2\COMM{\mYDud{k}\,\CKM\mYDd{h}\mYDdd{l}\CKMd\,\mYDu{h}}{\mYDud{i}\mYDu{j}}.
\nonumber
\end{align}
In order to compute the matrix elements of \refEQS{eq:der:YdYdd:00}--\eqref{eq:der:YudYu:00}, we notice that
\begin{equation}
\left(\COMM{\CKMd\mYDu{i}\mYDud{j}\CKM}{\mYDdd{k}\mYDd{l}}\right)_{ab}=
\sum_{q=1}^3\Vc{qa}\V{qb}\myDu{i}{q}\myDuc{j}{q}(\myDdc{k}{b}\myDd{l}{b}-\myDdc{k}{a}\myDd{l}{a}),
\end{equation}
and
\begin{equation}
\left(\COMM{\CKM\mYDd{i}\mYDdd{j}\CKMd}{\mYDu{k}\mYDud{l}}\right)_{ab}=
\sum_{q=1}^3\V{aq}\Vc{bq}\myDd{i}{q}\myDdc{j}{q}(\myDu{k}{b}\myDuc{l}{b}-\myDu{k}{a}\myDuc{l}{a}).
\end{equation}
Then, with the parameters in \refEQ{eq:DiagY:01}, the matrix elements $(a,b)$ of \refEQS{eq:der:YdYdd:00}--\eqref{eq:der:YudYu:00} read
\begin{align}
\{d_L\}_{ab}=\sum_{q=1}^3\sum_{{\rm h}=1}^{n=2}\Vc{qa}\V{qb}\Big\{&
\abs{\myDu{h}{q}}^2 \left( \myDd{i}{a}\myDdc{j}{a}\myDd{k}{b}\myDdc{l}{b} - \myDd{i}{b}\myDdc{j}{b}\myDd{k}{a}\myDdc{l}{a} \right)\label{eq:RGEStab:Matrix:YdYdd:00}\\
&-2\left(\myDu{h}{q}\myDuc{i}{q}\myDd{h}{b}\myDdc{j}{b}+\myDu{j}{q}\myDuc{h}{q}\myDd{i}{a}\myDdc{h}{a}\right)\left(\myDd{k}{b}\myDdc{l}{b}-\myDd{k}{a}\myDdc{l}{a}\right)\nonumber\\
&+2\left(\myDu{h}{q}\myDuc{k}{q}\myDd{h}{b}\myDdc{l}{b}+\myDu{l}{q}\myDuc{h}{q}\myDd{k}{a}\myDdc{h}{a}\right)\left(\myDd{i}{b}\myDdc{j}{b}-\myDd{i}{a}\myDdc{j}{a}\right)
\Big\}.
\nonumber
\end{align}
\begin{align}
&\{d_R\}_{ab}=\label{eq:RGEStab:Matrix:YddYd:00}\\
&\sum_{q=1}^3\sum_{{\rm h}=1}^{n=2}\Vc{qa}\V{qb}\Big\{\left(\myDdc{k}{b}\myDd{l}{b}-\myDdc{k}{a}\myDd{l}{a}\right)
\left(\abs{\myDu{h}{q}}^2\myDdc{i}{a}\myDd{j}{b}-2\myDu{i}{q}\myDuc{h}{q}\myDdc{h}{a}\myDd{j}{b}-2\myDu{h}{q}\myDuc{j}{q}\myDdc{i}{a}\myDd{h}{b}\right)\nonumber\\
&\phantom{xxxxxxxxx}-\left(\myDdc{i}{b}\myDd{j}{b}-\myDdc{i}{a}\myDd{j}{a}\right)
\left(\abs{\myDu{h}{q}}^2\myDdc{k}{a}\myDd{l}{b}-2\myDu{k}{q}\myDuc{h}{q}\myDdc{h}{a}\myDd{l}{b}-2\myDu{h}{q}\myDuc{l}{q}\myDdc{k}{a}\myDd{h}{b}\right)\Big\},
\nonumber
\end{align}
\begin{align}
\{u_L\}_{ab}=\sum_{q=1}^3\sum_{{\rm h}=1}^{n=2}\V{aq}\Vc{bq}\Big\{&
\abs{\myDd{h}{q}}^2 \left( \myDu{i}{a}\myDuc{j}{a}\myDu{k}{b}\myDuc{l}{b} - \myDu{i}{b}\myDuc{j}{b}\myDu{k}{a}\myDuc{l}{a} \right)\label{eq:RGEStab:Matrix:YuYud:00}\\
&-2\left(\myDd{h}{q}\myDdc{i}{q}\myDu{h}{b}\myDuc{j}{b}+\myDd{j}{q}\myDdc{h}{q}\myDu{i}{a}\myDuc{h}{a}\right)\left(\myDu{k}{b}\myDuc{l}{b}-\myDu{k}{a}\myDuc{l}{a}\right)\nonumber\\
&+2\left(\myDd{h}{q}\myDdc{k}{q}\myDu{h}{b}\myDuc{l}{b}+\myDd{l}{q}\myDdc{h}{q}\myDu{k}{a}\myDuc{h}{a}\right)\left(\myDu{i}{b}\myDuc{j}{b}-\myDu{i}{a}\myDuc{j}{a}\right)
\Big\},
\nonumber
\end{align}
and
\begin{align}
&\{u_R\}_{ab}=\label{eq:RGEStab:Matrix:YudYu:00}\\
&\sum_{q=1}^3\sum_{{\rm h}=1}^{n=2}\V{aq}\Vc{bq}\Big\{\left(\myDuc{k}{b}\myDu{l}{b}-\myDuc{k}{a}\myDu{l}{a}\right)
\left(\abs{\myDd{h}{q}}^2\myDuc{i}{a}\myDu{j}{b}-2\myDd{i}{q}\myDdc{h}{q}\myDuc{h}{a}\myDu{j}{b}-2\myDd{h}{q}\myDdc{j}{q}\myDuc{i}{a}\myDu{h}{b}\right)\nonumber\\
&\phantom{xxxxxxxxx}-\left(\myDuc{i}{b}\myDu{j}{b}-\myDuc{i}{a}\myDu{j}{a}\right)
\left(\abs{\myDd{h}{q}}^2\myDuc{k}{a}\myDu{l}{b}-2\myDd{k}{q}\myDdc{h}{q}\myDuc{h}{a}\myDu{l}{b}-2\myDd{h}{q}\myDdc{l}{q}\myDuc{k}{a}\myDu{h}{b}\right)\Big\}.
\nonumber
\end{align}
For diagonal elements, $a=b$, the right-hand sides of \refEQS{eq:RGEStab:Matrix:YdYdd:00}--\eqref{eq:RGEStab:Matrix:YudYu:00} are identically zero. 
For  $i=j$ and $k=l$, by construction, we have in addition $\{q_X\}_{ba}=-\{q_X\}_{ab}^\ast$ ($q=u,d$, $X=L,R$). For illustration, we show in the following \refEQS{eq:RGEStab:Matrix:YdYdd:00}--\eqref{eq:RGEStab:Matrix:YudYu:00} for 2HDM and $i=j=1$, $k=l=2$.
\begin{align}
&\{d_L\}_{d_ad_b}=\sum_{u_q=1}^3\Vc{u_qd_a}\V{u_qd_b}\Big\{
(m_{u_q}^2+\abs{\nq{}{u_q}}^2) \left( m_{d_a}^2\abs{\nq{}{d_b}}^2 - m_{d_b}^2\abs{\nq{}{d_a}}^2 \right)\label{eq:RGEStab:Matrix:YdYdd:01}\\
&-2\left(\abs{\nq{}{d_b}}^2-\abs{\nq{}{d_a}}^2\right)\left(m_{u_q}m_{d_b}(m_{u_q}m_{d_b}+\nq{}{u_q}\nq{}{d_b})+m_{u_q}m_{d_a}(m_{u_q}m_{d_a}+\nqc{}{u_q}\nqc{}{d_a})\right)\nonumber\\
&+2\left(m_{d_b}^2-m_{d_a}^2\right)\left(\nqc{}{u_q}\nqc{}{d_b}(m_{u_q}m_{d_b}+\nq{}{u_q}\nq{}{d_b})+\nq{}{u_q}\nq{}{d_a}(m_{u_q}m_{d_a}+\nqc{}{u_q}\nqc{}{d_a})\right)
\Big\},
\nonumber
\end{align}
\begin{align}
&\{d_R\}_{d_ad_b}=\sum_{u_q=1}^3\Vc{u_qd_a}\V{u_qd_b}\Big\{\label{eq:RGEStab:Matrix:YddYd:01}\\
&\left(\abs{\nq{}{d_b}}^2-\abs{\nq{}{d_a}}^2\right)(m_{u_q}^2+\abs{\nq{}{u_q}}^2)m_{d_a}m_{d_b}\nonumber\\
&-2\left(\abs{\nq{}{d_b}}^2-\abs{\nq{}{d_a}}^2\right)\left(m_{u_q}m_{d_b}(m_{u_q}m_{d_a}+\nqc{}{u_q}\nqc{}{d_a})+m_{u_q}m_{d_a}(m_{u_q}m_{d_b}+\nq{}{u_q}\nq{}{d_b})\right)\nonumber\\
&-\left(m_{d_b}^2-m_{d_a}^2\right)(m_{u_q}^2+\abs{\nq{}{u_q}}^2)\nqc{}{d_a}\nq{}{d_b}\nonumber\\
&+2\left(m_{d_b}^2-m_{d_a}^2\right)\left(\nq{}{u_q}\nq{}{d_b}(m_{u_q}m_{d_a}+\nqc{}{u_q}\nqc{}{d_a})+\nqc{}{u_q}\nqc{}{d_a}(m_{u_q}m_{d_b}+\nq{}{u_q}\nq{}{d_b})\right)\Big\},
\nonumber
\end{align}
\begin{align}
&\{u_L\}_{u_au_b}=\sum_{d_q=1}^3\V{u_ad_q}\Vc{u_bd_q}\Big\{
(m_{d_q}^2+\abs{\nq{}{d_q}}^2) \left( m_{u_a}^2\abs{\nq{}{u_b}}^2 - m_{u_b}^2\abs{\nq{}{u_a}}^2 \right)\label{eq:RGEStab:Matrix:YuYud:01}\\
&-2\left(\abs{\nq{}{u_b}}^2-\abs{\nq{}{u_a}}^2\right)\left(m_{d_q}m_{u_b}(m_{d_q}m_{u_b}+\nq{}{d_q}\nq{}{u_b})+m_{d_q}m_{u_a}(m_{d_q}m_{u_a}+\nqc{}{d_q}\nqc{}{u_a})\right)\nonumber\\
&+2\left(m_{u_b}^2-m_{u_a}^2\right)\left(\nqc{}{d_q}\nqc{}{u_b}(m_{d_q}m_{u_b}+\nq{}{d_q}\nq{}{u_b})+\nq{}{d_q}\nq{}{u_a}(m_{d_q}m_{u_a}+\nqc{}{d_q}\nqc{}{u_a})\right)
\Big\},
\nonumber
\end{align}
and
\begin{align}
&\{u_R\}_{u_au_b}=\sum_{q=1}^3\V{u_ad_q}\Vc{u_bd_q}\Big\{\label{eq:RGEStab:Matrix:YudYu:01}\\
&\left(\abs{\nq{}{u_b}}^2-\abs{\nq{}{u_a}}^2\right)(m_{d_q}^2+\abs{\nq{}{d_q}}^2)m_{u_a}m_{u_b}\nonumber\\
&-2\left(\abs{\nq{}{u_b}}^2-\abs{\nq{}{u_a}}^2\right)
\left(m_{d_q}m_{u_b}(m_{d_q}m_{u_a}+\nqc{}{d_q}\nqc{}{u_a})+m_{d_q}m_{u_a}(m_{d_q}m_{u_b}+\nq{}{d_q}\nq{}{u_b})\right)\nonumber\\
&-\left(m_{u_b}^2-m_{u_a}^2\right)(m_{d_q}^2+\abs{\nq{}{d_q}}^2)\nqc{}{u_a}\nq{}{u_b}\nonumber\\
&+2\left(m_{u_b}^2-m_{u_a}^2\right)
\left(\nq{}{d_q}\nq{}{u_b}(m_{d_q}m_{u_a}+\nqc{}{d_q}\nqc{}{u_a})+\nqc{}{d_q}\nqc{}{u_a}(m_{d_q}m_{u_b}+\nq{}{d_q}\nq{}{u_b})\right)
\Big\}.
\nonumber
\end{align}
The formal generalisation of the conditions in this appendix and in section \ref{SEC:RGE} to the case of models with $n$ Higgs doublets instead of 2 is almost straightforward.

\clearpage
\providecommand{\href}[2]{#2}\begingroup\raggedright\endgroup

%


\begin{thebibliography}{10%
0}

\bibitem{Lee:1973iz}
T.~Lee, {\it {A Theory of Spontaneous T Violation}},  {\em Phys.Rev.} {\bf D8}
  (1973) 1226--1239.

\bibitem{Branco:2011iw}
G.~Branco, P.~Ferreira, L.~Lavoura, M.~Rebelo, M.~Sher, {\em et~al.}, {\it
  {Theory and phenomenology of two-Higgs-doublet models}},  {\em Phys.Rept.}
  {\bf 516} (2012) 1--102, [\href{http://xxx.lanl.gov/abs/1106.0034}{{\tt
  1106.0034}}].

\bibitem{Ivanov:2017dad}
I.~P. Ivanov, {\it {Building and testing models with extended Higgs sectors}},
  {\em Prog. Part. Nucl. Phys.} {\bf 95} (2017) 160--208,
  [\href{http://xxx.lanl.gov/abs/1702.03776}{{\tt 1702.03776}}].

\bibitem{Aad:2012tfa}
{\bf ATLAS} Collaboration, G.~Aad {\em et~al.}, {\it {Observation of a new
  particle in the search for the Standard Model Higgs boson with the ATLAS
  detector at the LHC}},  {\em Phys. Lett.} {\bf B716} (2012) 1--29,
  [\href{http://xxx.lanl.gov/abs/1207.7214}{{\tt 1207.7214}}].

\bibitem{Chatrchyan:2012xdj}
{\bf CMS} Collaboration, S.~Chatrchyan {\em et~al.}, {\it {Observation of a new
  boson at a mass of 125 GeV with the CMS experiment at the LHC}},  {\em Phys.
  Lett.} {\bf B716} (2012) 30--61,
  [\href{http://xxx.lanl.gov/abs/1207.7235}{{\tt 1207.7235}}].

\bibitem{Haber:1978jt}
H.~E. Haber, G.~L. Kane, and T.~Sterling, {\it {The Fermion Mass Scale and
  Possible Effects of Higgs Bosons on Experimental Observables}},  {\em Nucl.
  Phys.} {\bf B161} (1979) 493--532.

\bibitem{Donoghue:1978cj}
J.~F. Donoghue and L.~F. Li, {\it {Properties of Charged Higgs Bosons}},  {\em
  Phys. Rev.} {\bf D19} (1979) 945.

\bibitem{Abbott:1979dt}
L.~F. Abbott, P.~Sikivie, and M.~B. Wise, {\it {Constraints on Charged Higgs
  Couplings}},  {\em Phys. Rev.} {\bf D21} (1980) 1393.

\bibitem{Hall:1981bc}
L.~J. Hall and M.~B. Wise, {\it {Flavor changing Higgs - boson couplings}},
  {\em Nucl. Phys.} {\bf B187} (1981) 397.

\bibitem{Barger:1989fj}
V.~D. Barger, J.~L. Hewett, and R.~J.~N. Phillips, {\it {New Constraints on the
  Charged Higgs Sector in Two Higgs Doublet Models}},  {\em Phys. Rev.} {\bf
  D41} (1990) 3421--3441.

\bibitem{Atwood:1996vj}
D.~Atwood, L.~Reina, and A.~Soni, {\it {Phenomenology of two Higgs doublet
  models with flavor changing neutral currents}},  {\em Phys. Rev.} {\bf D55}
  (1997) 3156--3176, [\href{http://xxx.lanl.gov/abs/hep-ph/9609279}{{\tt
  hep-ph/9609279}}].

\bibitem{WahabElKaffas:2007xd}
A.~Wahab El~Kaffas, P.~Osland, and O.~M. Ogreid, {\it {Constraining the
  Two-Higgs-Doublet-Model parameter space}},  {\em Phys. Rev.} {\bf D76} (2007)
  095001, [\href{http://xxx.lanl.gov/abs/0706.2997}{{\tt 0706.2997}}].

\bibitem{Aoki:2009ha}
M.~Aoki, S.~Kanemura, K.~Tsumura, and K.~Yagyu, {\it {Models of Yukawa
  interaction in the two Higgs doublet model, and their collider
  phenomenology}},  {\em Phys. Rev.} {\bf D80} (2009) 015017,
  [\href{http://xxx.lanl.gov/abs/0902.4665}{{\tt 0902.4665}}].

\bibitem{Mahmoudi:2009zx}
F.~Mahmoudi and O.~Stal, {\it {Flavor constraints on the two-Higgs-doublet
  model with general Yukawa couplings}},  {\em Phys. Rev.} {\bf D81} (2010)
  035016, [\href{http://xxx.lanl.gov/abs/0907.1791}{{\tt 0907.1791}}].

\bibitem{Deschamps:2009rh}
O.~Deschamps, S.~Descotes-Genon, S.~Monteil, V.~Niess, S.~T'Jampens, and
  V.~Tisserand, {\it {The Two Higgs Doublet of Type II facing flavour physics
  data}},  {\em Phys. Rev.} {\bf D82} (2010) 073012,
  [\href{http://xxx.lanl.gov/abs/0907.5135}{{\tt 0907.5135}}].

\bibitem{Crivellin:2013wna}
A.~Crivellin, A.~Kokulu, and C.~Greub, {\it {Flavor-phenomenology of
  two-Higgs-doublet models with generic Yukawa structure}},  {\em Phys.Rev.}
  {\bf D87} (2013), no.~9 094031,
  [\href{http://xxx.lanl.gov/abs/1303.5877}{{\tt 1303.5877}}].

\bibitem{Broggio:2014mna}
A.~Broggio, E.~J. Chun, M.~Passera, K.~M. Patel, and S.~K. Vempati, {\it
  {Limiting two-Higgs-doublet models}},  {\em JHEP} {\bf 11} (2014) 058,
  [\href{http://xxx.lanl.gov/abs/1409.3199}{{\tt 1409.3199}}].

\bibitem{Das:2015qva}
D.~Das, {\it {New limits on tan $\beta$ for 2HDMs with Z$_2$ symmetry}},  {\em
  Int. J. Mod. Phys.} {\bf A30} (2015), no.~26 1550158,
  [\href{http://xxx.lanl.gov/abs/1501.02610}{{\tt 1501.02610}}].

\bibitem{Gaitan:2015hga}
R.~Gaitán, R.~Martinez, and J.~H.~M. de~Oca, {\it {Rare top decay $t
  \rightarrow c \gamma$ with flavor changing neutral scalar interactions in two
  Higgs doublet model}},  {\em Phys. Rev.} {\bf D94} (2016), no.~9 094038,
  [\href{http://xxx.lanl.gov/abs/1503.04391}{{\tt 1503.04391}}].

\bibitem{Altunkaynak:2015twa}
B.~Altunkaynak, W.-S. Hou, C.~Kao, M.~Kohda, and B.~McCoy, {\it {Flavor
  Changing Heavy Higgs Interactions at the LHC}},  {\em Phys. Lett.} {\bf B751}
  (2015) 135--142, [\href{http://xxx.lanl.gov/abs/1506.00651}{{\tt
  1506.00651}}].

\bibitem{Arhrib:2015maa}
A.~Arhrib, R.~Benbrik, C.-H. Chen, M.~Gomez-Bock, and S.~Semlali, {\it
  {Two-Higgs-doublet type-II and -III models and $t\rightarrow c h$ at the
  LHC}},  {\em Eur. Phys. J.} {\bf C76} (2016), no.~6 328,
  [\href{http://xxx.lanl.gov/abs/1508.06490}{{\tt 1508.06490}}].

\bibitem{Kim:2015zla}
C.~S. Kim, Y.~W. Yoon, and X.-B. Yuan, {\it {Exploring top quark FCNC within
  2HDM type III in association with flavor physics}},  {\em JHEP} {\bf 12}
  (2015) 038, [\href{http://xxx.lanl.gov/abs/1509.00491}{{\tt 1509.00491}}].

\bibitem{Enomoto:2015wbn}
T.~Enomoto and R.~Watanabe, {\it {Flavor constraints on the Two Higgs Doublet
  Models of Z$_{2}$ symmetric and aligned types}},  {\em JHEP} {\bf 05} (2016)
  002, [\href{http://xxx.lanl.gov/abs/1511.05066}{{\tt 1511.05066}}].

\bibitem{Benbrik:2015evd}
R.~Benbrik, C.-H. Chen, and T.~Nomura, {\it {$h,Z\to \ell_i \bar\ell_j$,
  $\Delta a_{\mu}$, $\tau\to (3\mu,\mu \gamma)$ in generic two-Higgs-doublet
  models}},  {\em Phys. Rev.} {\bf D93} (2016), no.~9 095004,
  [\href{http://xxx.lanl.gov/abs/1511.08544}{{\tt 1511.08544}}].

\bibitem{Cline:2015lqp}
J.~M. Cline, {\it {Scalar doublet models confront $\tau$ and b anomalies}},
  {\em Phys. Rev.} {\bf D93} (2016), no.~7 075017,
  [\href{http://xxx.lanl.gov/abs/1512.02210}{{\tt 1512.02210}}].

\bibitem{Han:2017pfo}
L.~Wang, F.~Zhang, and X.-F. Han, {\it {Two-Higgs-doublet model of type-II
  confronted with the LHC run-I and run-II data}},  {\em Phys. Rev.} {\bf D95}
  (2017), no.~11 115014, [\href{http://xxx.lanl.gov/abs/1701.02678}{{\tt
  1701.02678}}].

\bibitem{Gori:2017qwg}
S.~Gori, H.~E. Haber, and E.~Santos, {\it {High scale flavor alignment in
  two-Higgs doublet models and its phenomenology}},  {\em JHEP} {\bf 06} (2017)
  110, [\href{http://xxx.lanl.gov/abs/1703.05873}{{\tt 1703.05873}}].

\bibitem{Arbey:2017gmh}
A.~Arbey, F.~Mahmoudi, O.~Stal, and T.~Stefaniak, {\it {Status of the Charged
  Higgs Boson in Two Higgs Doublet Models}},  {\em Eur. Phys. J.} {\bf C78}
  (2018), no.~3 182, [\href{http://xxx.lanl.gov/abs/1706.07414}{{\tt
  1706.07414}}].

\bibitem{Deshpande:1977rw}
N.~G. Deshpande and E.~Ma, {\it {Pattern of Symmetry Breaking with Two Higgs
  Doublets}},  {\em Phys. Rev.} {\bf D18} (1978) 2574.

\bibitem{LopezHonorez:2006gr}
L.~Lopez~Honorez, E.~Nezri, J.~F. Oliver, and M.~H.~G. Tytgat, {\it {The Inert
  Doublet Model: An Archetype for Dark Matter}},  {\em JCAP} {\bf 0702} (2007)
  028, [\href{http://xxx.lanl.gov/abs/hep-ph/0612275}{{\tt hep-ph/0612275}}].

\bibitem{Bona:2005vz}
{\bf UTfit} Collaboration, M.~Bona {\em et~al.}, {\it {The 2004 UTfit
  collaboration report on the status of the unitarity triangle in the standard
  model}},  {\em JHEP} {\bf 07} (2005) 028,
  [\href{http://xxx.lanl.gov/abs/hep-ph/0501199}{{\tt hep-ph/0501199}}].

\bibitem{Charles:2004jd}
{\bf CKMfitter Group} Collaboration, J.~Charles, A.~Hocker, H.~Lacker,
  S.~Laplace, F.~R. Le~Diberder, J.~Malcles, J.~Ocariz, M.~Pivk, and L.~Roos,
  {\it {CP violation and the CKM matrix: Assessing the impact of the asymmetric
  $B$ factories}},  {\em Eur. Phys. J.} {\bf C41} (2005) 1--131,
  [\href{http://xxx.lanl.gov/abs/hep-ph/0406184}{{\tt hep-ph/0406184}}].

\bibitem{Botella:2005fc}
F.~Botella, G.~Branco, M.~Nebot, and M.~Rebelo, {\it {New physics and evidence
  for a complex CKM}},  {\em Nucl.Phys.} {\bf B725} (2005) 155--172,
  [\href{http://xxx.lanl.gov/abs/hep-ph/0502133}{{\tt hep-ph/0502133}}].

\bibitem{Turok:1990zg}
N.~Turok and J.~Zadrozny, {\it {Electroweak baryogenesis in the two doublet
  model}},  {\em Nucl. Phys.} {\bf B358} (1991) 471--493.

\bibitem{Guo:2016ixx}
H.-K. Guo, Y.-Y. Li, T.~Liu, M.~Ramsey-Musolf, and J.~Shu, {\it
  {Lepton-Flavored Electroweak Baryogenesis}},  {\em Phys. Rev.} {\bf D96}
  (2017), no.~11 115034, [\href{http://xxx.lanl.gov/abs/1609.09849}{{\tt
  1609.09849}}].

\bibitem{Fuyuto:2017ewj}
K.~Fuyuto, W.-S. Hou, and E.~Senaha, {\it {Electroweak baryogenesis driven by
  extra top Yukawa couplings}},  {\em Phys. Lett.} {\bf B776} (2018) 402--406,
  [\href{http://xxx.lanl.gov/abs/1705.05034}{{\tt 1705.05034}}].

\bibitem{Georgi:1978ri}
H.~Georgi and D.~V. Nanopoulos, {\it {Suppression of Flavor Changing Effects
  From Neutral Spinless Meson Exchange in Gauge Theories}},  {\em Phys. Lett.}
  {\bf B82} (1979) 95.

\bibitem{Glashow:1976nt}
S.~L. Glashow and S.~Weinberg, {\it {Natural Conservation Laws for Neutral
  Currents}},  {\em Phys.Rev.} {\bf D15} (1977) 1958.

\bibitem{Paschos:1976ay}
E.~Paschos, {\it {Diagonal Neutral Currents}},  {\em Phys.Rev.} {\bf D15}
  (1977) 1966.

\bibitem{Ko:2012hd}
P.~Ko, Y.~Omura, and C.~Yu, {\it {A Resolution of the Flavor Problem of Two
  Higgs Doublet Models with an Extra $U(1)_H$ Symmetry for Higgs Flavor}},
  {\em Phys. Lett.} {\bf B717} (2012) 202--206,
  [\href{http://xxx.lanl.gov/abs/1204.4588}{{\tt 1204.4588}}].

\bibitem{Campos:2017dgc}
M.~D. Campos, D.~Cogollo, M.~Lindner, T.~Melo, F.~S. Queiroz, and
  W.~Rodejohann, {\it {Neutrino Masses and Absence of Flavor Changing
  Interactions in the 2HDM from Gauge Principles}},  {\em JHEP} {\bf 08} (2017)
  092, [\href{http://xxx.lanl.gov/abs/1705.05388}{{\tt 1705.05388}}].

\bibitem{Gatto:1978dy}
R.~Gatto, G.~Morchio, and F.~Strocchi, {\it {Natural Flavor Conservation in the
  Neutral Currents and the Determination of the Cabibbo Angle}},  {\em Phys.
  Lett.} {\bf 80B} (1979) 265--268.

\bibitem{Gatto:1979mr}
R.~Gatto, G.~Morchio, G.~Sartori, and F.~Strocchi, {\it {Natural Flavor
  Conservation in Higgs Induced Neutral Currents and the Quark Mixing Angles}},
   {\em Nucl. Phys.} {\bf B163} (1980) 221--253.

\bibitem{Sartori:1979gt}
G.~Sartori, {\it {Discrete Symmetries, Natural Flavor Conservation and Weak
  Mixing Angles}},  {\em Phys. Lett.} {\bf 82B} (1979) 255--259.

\bibitem{Grimus:1986mh}
W.~Grimus and G.~Ecker, {\it {On the Simultaneous Diagonalizability of
  Matrices}},  {\em J. Phys.} {\bf A19} (1986) 3917.

\bibitem{Barbieri:1978qh}
R.~Barbieri, R.~Gatto, and F.~Strocchi, {\it {Quark Mass Matrix and Discrete
  Symmetries in the SU(2) X U(1) Model}},  {\em Phys. Lett.} {\bf 74B} (1978)
  344--346.

\bibitem{Sartori:1979ms}
G.~Sartori, {\it {The Concrete Realization of the Symmetries Responsible for
  Natural Flavor Conservation Laws}},  {\em Nuovo Cim.} {\bf A55} (1980) 377.

\bibitem{Gatto:1979sh}
R.~Gatto, G.~Morchio, and F.~Strocchi, {\it {Symmetrie Leading to Flavor
  Conservation in Higgs Induced Neutral Currents and Implications on the
  Cabibbo Angle}},  {\em Phys. Lett.} {\bf 83B} (1979) 348--350.

\bibitem{Segre:1979gs}
G.~Segre and H.~A. Weldon, {\it {Natural Flavor Conservation and the Absence of
  Radiatively Induced Cabibbo Angles}},  {\em Phys. Lett.} {\bf 86B} (1979)
  291--293.

\bibitem{Segre:1979rt}
G.~Segre and H.~A. Weldon, {\it {The Conflict Between Natural Flavor
  Conservation of Higgs Couplings and Cabibbo Mixing in $SU(2)_L \times
  U(1)$}},  {\em Annals Phys.} {\bf 124} (1980) 37.

\bibitem{Rothman:1980ev}
A.~C. Rothman and K.~Kang, {\it {Natural Flavor Conservation}},  {\em Phys.
  Rev.} {\bf D23} (1981) 2657.

\bibitem{Kang:1980yg}
K.~Kang and A.~C. Rothman, {\it {Generalized Mixing Angles in Gauge Theories
  With Natural Flavor Conservation}},  {\em Phys. Rev.} {\bf D24} (1981) 167.

\bibitem{Leurer:1992wg}
M.~Leurer, Y.~Nir, and N.~Seiberg, {\it {Mass matrix models}},  {\em Nucl.
  Phys.} {\bf B398} (1993) 319--342,
  [\href{http://xxx.lanl.gov/abs/hep-ph/9212278}{{\tt hep-ph/9212278}}].

\bibitem{Cheng:1987rs}
T.~P. Cheng and M.~Sher, {\it {Mass Matrix Ansatz and Flavor Nonconservation in
  Models with Multiple Higgs Doublets}},  {\em Phys. Rev.} {\bf D35} (1987)
  3484.

\bibitem{Antaramian:1992ya}
A.~Antaramian, L.~J. Hall, and A.~Rasin, {\it {Flavor changing interactions
  mediated by scalars at the weak scale}},  {\em Phys. Rev. Lett.} {\bf 69}
  (1992) 1871--1873, [\href{http://xxx.lanl.gov/abs/hep-ph/9206205}{{\tt
  hep-ph/9206205}}].

\bibitem{Hall:1993ca}
L.~J. Hall and S.~Weinberg, {\it {Flavor changing scalar interactions}},  {\em
  Phys. Rev.} {\bf D48} (1993) R979--R983,
  [\href{http://xxx.lanl.gov/abs/hep-ph/9303241}{{\tt hep-ph/9303241}}].

\bibitem{Wu:1994ja}
Y.~L. Wu and L.~Wolfenstein, {\it {Sources of CP violation in the two Higgs
  doublet model}},  {\em Phys. Rev. Lett.} {\bf 73} (1994) 1762--1764,
  [\href{http://xxx.lanl.gov/abs/hep-ph/9409421}{{\tt hep-ph/9409421}}].

\bibitem{Datta:2008qn}
A.~Datta, {\it {Suppressed FCNC in New Physics with Shared Flavor Symmetry}},
  {\em Phys. Rev.} {\bf D78} (2008) 095004,
  [\href{http://xxx.lanl.gov/abs/0807.0795}{{\tt 0807.0795}}].

\bibitem{Joshipura:1990xm}
A.~S. Joshipura, {\it {Neutral Higgs and CP violation}},  {\em Mod. Phys.
  Lett.} {\bf A6} (1991) 1693--1700.

\bibitem{Joshipura:1990pi}
A.~S. Joshipura and S.~D. Rindani, {\it {Naturally suppressed flavor violations
  in two Higgs doublet models}},  {\em Phys.Lett.} {\bf B260} (1991) 149--153.

\bibitem{Joshipura:2007cs}
A.~S. Joshipura and B.~P. Kodrani, {\it {Minimal flavour violations and tree
  level FCNC}},  {\em Phys. Rev.} {\bf D77} (2008) 096003,
  [\href{http://xxx.lanl.gov/abs/0710.3020}{{\tt 0710.3020}}].

\bibitem{Joshipura:2010tz}
A.~S. Joshipura and B.~P. Kodrani, {\it {Fermion number conservation and two
  Higgs doublet models without tree level flavour changing neutral currents}},
  {\em Phys. Rev.} {\bf D82} (2010) 115013,
  [\href{http://xxx.lanl.gov/abs/1004.3637}{{\tt 1004.3637}}].

\bibitem{Lavoura:1994ty}
L.~Lavoura, {\it {Models of CP violation exclusively via neutral scalar
  exchange}},  {\em Int.J.Mod.Phys.} {\bf A9} (1994) 1873--1888.

\bibitem{Branco:1996bq}
G.~Branco, W.~Grimus, and L.~Lavoura, {\it {Relating the scalar flavor changing
  neutral couplings to the CKM matrix}},  {\em Phys.Lett.} {\bf B380} (1996)
  119--126, [\href{http://xxx.lanl.gov/abs/hep-ph/9601383}{{\tt
  hep-ph/9601383}}].

\bibitem{Botella:2009pq}
F.~Botella, G.~Branco, and M.~Rebelo, {\it {Minimal Flavour Violation and
  Multi-Higgs Models}},  {\em Phys.Lett.} {\bf B687} (2010) 194--200,
  [\href{http://xxx.lanl.gov/abs/0911.1753}{{\tt 0911.1753}}].

\bibitem{Botella:2011ne}
F.~Botella, G.~Branco, M.~Nebot, and M.~Rebelo, {\it {Two-Higgs Leptonic
  Minimal Flavour Violation}},  {\em JHEP} {\bf 1110} (2011) 037,
  [\href{http://xxx.lanl.gov/abs/1102.0520}{{\tt 1102.0520}}].

\bibitem{Botella:2015hoa}
F.~J. Botella, G.~C. Branco, M.~Nebot, and M.~N. Rebelo, {\it {Flavour Changing
  Higgs Couplings in a Class of Two Higgs Doublet Models}},  {\em Eur. Phys.
  J.} {\bf C76} (2016), no.~3 161,
  [\href{http://xxx.lanl.gov/abs/1508.05101}{{\tt 1508.05101}}].

\bibitem{Alves:2017xmk}
J.~M. Alves, F.~J. Botella, G.~C. Branco, F.~Cornet-Gomez, and M.~Nebot, {\it
  {Controlled Flavour Changing Neutral Couplings in Two Higgs Doublet Models}},
   {\em Eur. Phys. J.} {\bf C77} (2017), no.~9 585,
  [\href{http://xxx.lanl.gov/abs/1703.03796}{{\tt 1703.03796}}].

\bibitem{Pich:2009sp}
A.~Pich and P.~Tuzon, {\it {Yukawa Alignment in the Two-Higgs-Doublet Model}},
  {\em Phys.Rev.} {\bf D80} (2009) 091702,
  [\href{http://xxx.lanl.gov/abs/0908.1554}{{\tt 0908.1554}}].

\bibitem{Ecker:1989ay}
G.~Ecker, W.~Grimus, and H.~Neufeld, {\it {Spontaneous CP Violation and Neutral
  Flavor Conservation}},  {\em Phys. Lett.} {\bf B228} (1989) 401--405.

\bibitem{Serodio:2011hg}
H.~Serodio, {\it {Yukawa Alignment in a Multi Higgs Doublet Model: An effective
  approach}},  {\em Phys. Lett.} {\bf B700} (2011) 133--138,
  [\href{http://xxx.lanl.gov/abs/1104.2545}{{\tt 1104.2545}}].

\bibitem{Varzielas:2011jr}
I.~de~Medeiros~Varzielas, {\it {Family symmetries and alignment in multi-Higgs
  doublet models}},  {\em Phys. Lett.} {\bf B701} (2011) 597--600,
  [\href{http://xxx.lanl.gov/abs/1104.2601}{{\tt 1104.2601}}].

\bibitem{Wise:1980ux}
M.~B. Wise, {\it {Radiatively induced Flavor Changing Neutral Higgs boson
  couplings}},  {\em Phys. Lett.} {\bf B103} (1981) 121--123.

\bibitem{Frere:1985bu}
J.~M. Frere and Y.-P. Yao, {\it {Naturalness for multiscalar models and
  radiative stability}},  {\em Phys. Rev. Lett.} {\bf 55} (1985) 2386.

\bibitem{Cvetic:1997zd}
G.~Cvetic, S.~S. Hwang, and C.~S. Kim, {\it {One loop renormalization group
  equations of the general framework with two Higgs doublets}},  {\em Int. J.
  Mod. Phys.} {\bf A14} (1999) 769--798,
  [\href{http://xxx.lanl.gov/abs/hep-ph/9706323}{{\tt hep-ph/9706323}}].

\bibitem{Cvetic:1998uw}
G.~Cvetic, C.~S. Kim, and S.~S. Hwang, {\it {Higgs mediated flavor changing
  neutral currents in the general framework with two Higgs doublets: An RGE
  analysis}},  {\em Phys. Rev.} {\bf D58} (1998) 116003,
  [\href{http://xxx.lanl.gov/abs/hep-ph/9806282}{{\tt hep-ph/9806282}}].

\bibitem{Ferreira:2010xe}
P.~Ferreira, L.~Lavoura, and J.~P. Silva, {\it {Renormalization-group
  constraints on Yukawa alignment in multi-Higgs-doublet models}},  {\em
  Phys.Lett.} {\bf B688} (2010) 341--344,
  [\href{http://xxx.lanl.gov/abs/1001.2561}{{\tt 1001.2561}}].

\bibitem{Bijnens:2011gd}
J.~Bijnens, J.~Lu, and J.~Rathsman, {\it {Constraining General Two Higgs
  Doublet Models by the Evolution of Yukawa Couplings}},  {\em JHEP} {\bf 05}
  (2012) 118, [\href{http://xxx.lanl.gov/abs/1111.5760}{{\tt 1111.5760}}].

\bibitem{Botella:2015yfa}
F.~J. Botella, G.~C. Branco, A.~M. Coutinho, M.~N. Rebelo, and J.~I.
  Silva-Marcos, {\it {Natural Quasi-Alignment with two Higgs Doublets and RGE
  Stability}},  {\em Eur. Phys. J.} {\bf C75} (2015) 286,
  [\href{http://xxx.lanl.gov/abs/1501.07435}{{\tt 1501.07435}}].

\bibitem{Penuelas:2017ikk}
A.~Pe\~nuelas and A.~Pich, {\it {Flavour alignment in multi-Higgs-doublet
  models}},  {\em JHEP} {\bf 12} (2017) 084,
  [\href{http://xxx.lanl.gov/abs/1710.02040}{{\tt 1710.02040}}].

\bibitem{Ahn:2010zza}
Y.~H. Ahn and C.-H. Chen, {\it {New charged Higgs effects on
  $\Gamma_{K_{e2}}/\Gamma_{K_{2}}$, $f_{D_s}$ and ${\cal B}(B^+\to \tau^+ \nu)$
  in the Two-Higgs-Doublet model}},  {\em Phys. Lett.} {\bf B690} (2010)
  57--61, [\href{http://xxx.lanl.gov/abs/1002.4216}{{\tt 1002.4216}}].

\bibitem{Braeuninger:2010td}
C.~B. Braeuninger, A.~Ibarra, and C.~Simonetto, {\it {Radiatively induced
  flavour violation in the general two-Higgs doublet model with Yukawa
  alignment}},  {\em Phys. Lett.} {\bf B692} (2010) 189--195,
  [\href{http://xxx.lanl.gov/abs/1005.5706}{{\tt 1005.5706}}].

\bibitem{Botella:1994cs}
F.~J. Botella and J.~P. Silva, {\it {Jarlskog - like invariants for theories
  with scalars and fermions}},  {\em Phys. Rev.} {\bf D51} (1995) 3870--3875,
  [\href{http://xxx.lanl.gov/abs/hep-ph/9411288}{{\tt hep-ph/9411288}}].

\bibitem{Pakvasa:1977in}
S.~Pakvasa and H.~Sugawara, {\it {Discrete Symmetry and Cabibbo Angle}},  {\em
  Phys. Lett.} {\bf 73B} (1978) 61--64.

\bibitem{Wyler:1978fj}
D.~Wyler, {\it {The Cabibbo Angle in the SU(2)$_L \times$ U(1) Gauge
  Theories}},  {\em Phys. Rev.} {\bf D19} (1979) 330.

\bibitem{Nebot:2015wsa}
M.~Nebot and J.~P. Silva, {\it {Self-cancellation of a scalar in neutral meson
  mixing and implications for the LHC}},  {\em Phys. Rev.} {\bf D92} (2015),
  no.~8 085010, [\href{http://xxx.lanl.gov/abs/1507.07941}{{\tt 1507.07941}}].

\bibitem{Grimus:2004yh}
W.~Grimus and L.~Lavoura, {\it {Renormalization of the neutrino mass operators
  in the multi-Higgs-doublet standard model}},  {\em Eur. Phys. J.} {\bf C39}
  (2005) 219--227, [\href{http://xxx.lanl.gov/abs/hep-ph/0409231}{{\tt
  hep-ph/0409231}}].

\bibitem{Cheng:1973nv}
T.~P. Cheng, E.~Eichten, and L.-F. Li, {\it {Higgs Phenomena in Asymptotically
  Free Gauge Theories}},  {\em Phys. Rev.} {\bf D9} (1974) 2259.

\bibitem{Jung:2010ik}
M.~Jung, A.~Pich, and P.~Tuzon, {\it {Charged-Higgs phenomenology in the
  Aligned two-Higgs-doublet model}},  {\em JHEP} {\bf 1011} (2010) 003,
  [\href{http://xxx.lanl.gov/abs/1006.0470}{{\tt 1006.0470}}].

\bibitem{Gunion:2002zf}
J.~F. Gunion and H.~E. Haber, {\it {The CP conserving two Higgs doublet model:
  The Approach to the decoupling limit}},  {\em Phys. Rev.} {\bf D67} (2003)
  075019, [\href{http://xxx.lanl.gov/abs/hep-ph/0207010}{{\tt
  hep-ph/0207010}}].

\bibitem{Khachatryan:2016vau}
{\bf ATLAS, CMS} Collaboration, G.~Aad {\em et~al.}, {\it {Measurements of the
  Higgs boson production and decay rates and constraints on its couplings from
  a combined ATLAS and CMS analysis of the LHC pp collision data at $
  \sqrt{s}=7 $ and 8 TeV}},  {\em JHEP} {\bf 08} (2016) 045,
  [\href{http://xxx.lanl.gov/abs/1606.02266}{{\tt 1606.02266}}].

\bibitem{Gunion:1989we}
J.~F. Gunion, H.~E. Haber, G.~L. Kane, and S.~Dawson, {\it {The Higgs Hunter's
  Guide}},  {\em Front. Phys.} {\bf 80} (2000) 1--448.

\bibitem{Spira:1995rr}
M.~Spira, A.~Djouadi, D.~Graudenz, and P.~M. Zerwas, {\it {Higgs boson
  production at the LHC}},  {\em Nucl. Phys.} {\bf B453} (1995) 17--82,
  [\href{http://xxx.lanl.gov/abs/hep-ph/9504378}{{\tt hep-ph/9504378}}].

\bibitem{Dittmaier:2011ti}
{\bf LHC Higgs Cross Section Working Group} Collaboration, S.~Dittmaier {\em
  et~al.}, {\it {Handbook of LHC Higgs Cross Sections: 1. Inclusive
  Observables}},  \href{http://xxx.lanl.gov/abs/1101.0593}{{\tt 1101.0593}}.

\bibitem{Dittmaier:2012vm}
{\bf LHC Higgs Cross Section Working Group} Collaboration, S.~Dittmaier {\em
  et~al.}, {\it {Handbook of LHC Higgs Cross Sections: 2. Differential
  Distributions}},  \href{http://xxx.lanl.gov/abs/1201.3084}{{\tt 1201.3084}}.

\bibitem{Heinemeyer:2013tqa}
{\bf LHC Higgs Cross Section Working Group} Collaboration, S.~Heinemeyer {\em
  et~al.}, {\it {Handbook of LHC Higgs Cross Sections: 3. Higgs Properties}},
  \href{http://xxx.lanl.gov/abs/1307.1347}{{\tt 1307.1347}}.

\bibitem{deFlorian:2016spz}
{\bf LHC Higgs Cross Section Working Group} Collaboration, D.~de~Florian {\em
  et~al.}, {\it {Handbook of LHC Higgs Cross Sections: 4. Deciphering the
  Nature of the Higgs Sector}},  \href{http://xxx.lanl.gov/abs/1610.07922}{{\tt
  1610.07922}}.

\bibitem{Xing:2011aa}
Z.-z. Xing, H.~Zhang, and S.~Zhou, {\it {Impacts of the Higgs mass on vacuum
  stability, running fermion masses and two-body Higgs decays}},  {\em Phys.
  Rev.} {\bf D86} (2012) 013013, [\href{http://xxx.lanl.gov/abs/1112.3112}{{\tt
  1112.3112}}].

\bibitem{Georgi:1977gs}
H.~Georgi, S.~Glashow, M.~Machacek, and D.~V. Nanopoulos, {\it {Higgs Bosons
  from Two Gluon Annihilation in Proton Proton Collisions}},  {\em
  Phys.Rev.Lett.} {\bf 40} (1978) 692.

\bibitem{Zhou:2015wra}
Y.~Zhou, {\it {Constraining the Higgs boson coupling to light quarks in the
  $H\to ZZ$ final states}},  {\em Phys. Rev.} {\bf D93} (2016), no.~1 013019,
  [\href{http://xxx.lanl.gov/abs/1505.06369}{{\tt 1505.06369}}].

\bibitem{Perez:2015lra}
G.~Perez, Y.~Soreq, E.~Stamou, and K.~Tobioka, {\it {Prospects for measuring
  the Higgs boson coupling to light quarks}},  {\em Phys. Rev.} {\bf D93}
  (2016), no.~1 013001, [\href{http://xxx.lanl.gov/abs/1505.06689}{{\tt
  1505.06689}}].

\bibitem{Yu:2016rvv}
F.~Yu, {\it {Phenomenology of Enhanced Light Quark Yukawa Couplings and the
  $W^\pm h$ Charge Asymmetry}},  {\em JHEP} {\bf 02} (2017) 083,
  [\href{http://xxx.lanl.gov/abs/1609.06592}{{\tt 1609.06592}}].

\bibitem{Cohen:2017rsk}
J.~Cohen, S.~Bar-Shalom, G.~Eilam, and A.~Soni, {\it {Light-quarks Yukawa and
  new physics in exclusive high-$p_T$ Higgs + jet(b-jet) events}},
  \href{http://xxx.lanl.gov/abs/1705.09295}{{\tt 1705.09295}}.

\bibitem{Martin:2009iq}
A.~D. Martin, W.~J. Stirling, R.~S. Thorne, and G.~Watt, {\it {Parton
  distributions for the LHC}},  {\em Eur. Phys. J.} {\bf C63} (2009) 189--285,
  [\href{http://xxx.lanl.gov/abs/0901.0002}{{\tt 0901.0002}}].

\bibitem{Aaboud:2017xsd}
{\bf ATLAS} Collaboration, M.~Aaboud {\em et~al.}, {\it {Evidence for the $
  H\to b\overline{b} $ decay with the ATLAS detector}},  {\em JHEP} {\bf 12}
  (2017) 024, [\href{http://xxx.lanl.gov/abs/1708.03299}{{\tt 1708.03299}}].

\bibitem{Sirunyan:2017elk}
{\bf CMS} Collaboration, A.~M. Sirunyan {\em et~al.}, {\it {Evidence for the
  Higgs boson decay to a bottom quark-antiquark pair}},
  \href{http://xxx.lanl.gov/abs/1709.07497}{{\tt 1709.07497}}.

\bibitem{Sirunyan:2017khh}
{\bf CMS} Collaboration, A.~M. Sirunyan {\em et~al.}, {\it {Observation of the
  Higgs boson decay to a pair of $\tau$ leptons with the CMS detector}},  {\em
  Phys. Lett.} {\bf B779} (2018) 283--316,
  [\href{http://xxx.lanl.gov/abs/1708.00373}{{\tt 1708.00373}}].

\bibitem{Kauer:2012hd}
N.~Kauer and G.~Passarino, {\it {Inadequacy of zero-width approximation for a
  light Higgs boson signal}},  {\em JHEP} {\bf 08} (2012) 116,
  [\href{http://xxx.lanl.gov/abs/1206.4803}{{\tt 1206.4803}}].

\bibitem{Aad:2015xua}
{\bf ATLAS} Collaboration, G.~Aad {\em et~al.}, {\it {Constraints on the
  off-shell Higgs boson signal strength in the high-mass $ZZ$ and $WW$ final
  states with the ATLAS detector}},  {\em Eur. Phys. J.} {\bf C75} (2015),
  no.~7 335, [\href{http://xxx.lanl.gov/abs/1503.01060}{{\tt 1503.01060}}].

\bibitem{Khachatryan:2016ctc}
{\bf CMS} Collaboration, V.~Khachatryan {\em et~al.}, {\it {Search for Higgs
  boson off-shell production in proton-proton collisions at 7 and 8 TeV and
  derivation of constraints on its total decay width}},  {\em JHEP} {\bf 09}
  (2016) 051, [\href{http://xxx.lanl.gov/abs/1605.02329}{{\tt 1605.02329}}].

\bibitem{Khachatryan:2014aep}
{\bf CMS} Collaboration, V.~Khachatryan {\em et~al.}, {\it {Search for a
  standard model-like Higgs boson in the $\mu^+\mu^−$ and $e^+e^−$ decay
  channels at the LHC}},  {\em Phys. Lett.} {\bf B744} (2015) 184--207,
  [\href{http://xxx.lanl.gov/abs/1410.6679}{{\tt 1410.6679}}].

\bibitem{Aaboud:2017ojs}
{\bf ATLAS} Collaboration, M.~Aaboud {\em et~al.}, {\it {Search for the dimuon
  decay of the Higgs boson in $pp$ collisions at $\sqrt{s}$ = 13 TeV with the
  ATLAS detector}},  {\em Phys. Rev. Lett.} {\bf 119} (2017), no.~5 051802,
  [\href{http://xxx.lanl.gov/abs/1705.04582}{{\tt 1705.04582}}].

\bibitem{Barr:1990vd}
S.~M. Barr and A.~Zee, {\it {Electric Dipole Moment of the Electron and of the
  Neutron}},  {\em Phys. Rev. Lett.} {\bf 65} (1990) 21--24. [Erratum: Phys.
  Rev. Lett.65,2920(1990)].

\bibitem{Leigh:1990kf}
R.~G. Leigh, S.~Paban, and R.~M. Xu, {\it {Electric dipole moment of
  electron}},  {\em Nucl. Phys.} {\bf B352} (1991) 45--58.

\bibitem{Gunion:1990ce}
J.~F. Gunion and R.~Vega, {\it {The Electron electric dipole moment for a CP
  violating neutral Higgs sector}},  {\em Phys. Lett.} {\bf B251} (1990)
  157--162.

\bibitem{Chang:1990sf}
D.~Chang, W.-Y. Keung, and T.~C. Yuan, {\it {Two loop bosonic contribution to
  the electron electric dipole moment}},  {\em Phys. Rev.} {\bf D43} (1991)
  R14--R16.

\bibitem{Kao:1992jv}
C.~Kao and R.-M. Xu, {\it {Charged Higgs loop contribution to the electric
  dipole moment of electron}},  {\em Phys. Lett.} {\bf B296} (1992) 435--439.

\bibitem{Ilisie:2015tra}
V.~Ilisie, {\it {New Barr-Zee contributions to $\mathbf{(g-2)_\mu}$ in
  two-Higgs-doublet models}},  {\em JHEP} {\bf 04} (2015) 077,
  [\href{http://xxx.lanl.gov/abs/1502.04199}{{\tt 1502.04199}}].

\bibitem{Inoue:2014nva}
S.~Inoue, M.~J. Ramsey-Musolf, and Y.~Zhang, {\it {CP-violating phenomenology
  of flavor conserving two Higgs doublet models}},  {\em Phys. Rev.} {\bf D89}
  (2014), no.~11 115023, [\href{http://xxx.lanl.gov/abs/1403.4257}{{\tt
  1403.4257}}].

\bibitem{Altmannshofer:2015qra}
W.~Altmannshofer, J.~Brod, and M.~Schmaltz, {\it {Experimental constraints on
  the coupling of the Higgs boson to electrons}},  {\em JHEP} {\bf 05} (2015)
  125, [\href{http://xxx.lanl.gov/abs/1503.04830}{{\tt 1503.04830}}].

\bibitem{Bian:2014zka}
L.~Bian, T.~Liu, and J.~Shu, {\it {Cancellations Between Two-Loop Contributions
  to the Electron Electric Dipole Moment with a CP-Violating Higgs Sector}},
  {\em Phys. Rev. Lett.} {\bf 115} (2015) 021801,
  [\href{http://xxx.lanl.gov/abs/1411.6695}{{\tt 1411.6695}}].

\bibitem{Jung:2013hka}
M.~Jung and A.~Pich, {\it {Electric Dipole Moments in Two-Higgs-Doublet
  Models}},  {\em JHEP} {\bf 04} (2014) 076,
  [\href{http://xxx.lanl.gov/abs/1308.6283}{{\tt 1308.6283}}].

\end{thebibliography}

\end{document}